\def\Title{Uncertainty Relations for
Noise and Disturbance in Generalized Quantum 
Measurements}

\documentclass[aps,twocolumn,groupedaddress,%
draft,showpacs]{revtex4}
\usepackage{amsmath}
 
\newcommand{\beq}{\begin{equation}}
  \newcommand{\beql}[1]{\begin{equation}\label{eq:#1}}
  \newcommand{\beqa}{\begin{eqnarray}}
  \newcommand{\eeqa}{\end{eqnarray}}
  \newcommand{\beqas}{\begin{eqnarray*}}
  \newcommand{\eeqas}{\end{eqnarray*}}
  \newcommand{\qed}{{\em QED}}

  \newcommand*{\R}{\mathbf{R}}
  
  \newcommand*{\bA}{\mathbf{A}}
  
  \newcommand*{\bC}{\mathbf{C}}

  \newcommand*{\bP}{\mathbf{P}}
  \newcommand*{\bR}{\mathbf{R}}
  \newcommand*{\bS}{\mathbf{S}}

  \newcommand*{\bx}{\mathbf{x}}
  \newcommand*{\by}{\mathbf{y}}
  
  \newcommand*{\cB}{\mathcal{B}}

  \newcommand*{\cE}{\mathcal{E}}

  \newcommand*{\cH}{\mathcal{H}}
  \newcommand*{\cI}{\mathcal{I}}
  
  \newcommand*{\cK}{\mathcal{K}}
  \newcommand*{\cL}{\mathcal{L}}
  \newcommand*{\cM}{\mathcal{M}}

  \newcommand*{\cS}{\mathcal{S}}
  
  \newcommand*{\cU}{\mathcal{U}}
  
  \newcommand*{\cW}{\mathcal{W}}

  \newcommand*{\hA}{{\hat A}}

  \newcommand*{\hH}{{\hat H}}

  \newcommand*{\hX}{{\hat X}}

  \newcommand*{\hp}{\hat{p}}
 \newcommand*{\hq}{\hat{q}}
  \newcommand*{\hx}{\hat{x}}

  \newcommand*{\al}{\alpha}
  \newcommand*{\be}{\beta} 
  \newcommand*{\ch}{\chi}
  \newcommand*{\da}{\dagger}
  \newcommand*{\de}{\delta}
  \newcommand*{\ep}{\epsilon}
  \newcommand*{\et}{\eta}
  \newcommand*{\ga}{\gamma}

  \newcommand*{\la}{\lambda}
  \newcommand*{\mb}{\mbox}
  \newcommand*{\nn}{\nonumber}
  
  \newcommand*{\ph}{\phi}
  \newcommand*{\ps}{\psi} 
  \newcommand*{\rh}{\rho}
  \newcommand*{\si}{\sigma} 
  \newcommand*{\ta}{\tau}

  \newcommand*{\ve}{\varepsilon}

  \newcommand*{\De}{\Delta}                                          
  
  \newcommand*{\Eq}[1]{Eq.~(\ref{eq:#1})}

  \newcommand*{\La}{\Lambda}

  \newcommand*{\Ps}{\Psi}

  \newcommand*{\Then}{\Rightarrow}
                                            
  \newcommand*{\Tr}{\mbox{\rm Tr}}

  \newcommand*{\eq}[1]{(\ref{eq:#1})}
\newcommand*{\bra}[1]{\langle#1|}
\newcommand*{\ket}[1]{|#1\rangle}

\newcommand*{\bracket}[1]{\langle#1\rangle}

\newcommand{\ot}{\otimes}
\renewcommand{\hA}{A}
\newcommand{\bq}{{\mathbf q}}
\renewcommand{\hH}{H}
\renewcommand{\hX}{X}
\newcommand{\In}{{\rm in}}
\newcommand{\Out}{{\rm out}}
\newcommand{\bz}{{\mathbf z}}
\newtheorem{Theorem}{Theorem}[section]
\newtheorem{Corollary}[Theorem]{Corollary}
\newenvironment{Proof}{\begin{trivlist}
 \item[\hskip \labelsep {\em \indent Proof.}]}{\qed\end{trivlist}}
\newcommand{\dom}{{\rm dom}}

\newcommand{\tc}{\tau c}
\newcommand{\<}{\langle}
\renewcommand{\>}{\rangle}
\newcommand{\llabel}{\label}
\begin{document}

\title{\Title}
\author{Masanao Ozawa}
\affiliation{Graduate School of Information Sciences,
T\^{o}hoku University, Aoba-ku, Sendai,  980-8579, Japan}

%\date{\today}

\begin{abstract}
Heisenberg's uncertainty relation for measurement noise and
disturbance states that any position measurement with noise
$\ep$ brings the momentum disturbance not less than 
$\hbar/2\ep$. 
This relation holds only for restricted class of measuring apparatuses.
Here, Heisenberg's uncertainty relation is generalized to 
a relation that holds for all the possible quantum measurements,
from which conditions are obtained for measuring apparatuses
to satisfy Heisenberg's relation.
In particular, every apparatus with the noise and the disturbance
statistically independent from the measured object is proven
to satisfy Heisenberg's relation.
For this purpose, all the possible quantum measurements are
characterized by naturally acceptable axioms.
Then, a mathematical notion of the distance between probability operator valued
measures and observables is introduced and the basic properties are explored.
Based on this notion, the measurement noise and disturbance are 
naturally defined for any quantum measurements in a model independent
formulation.  
Under this formulation, various uncertainty relations are also derived
for apparatuses with independent noise, independent disturbance, unbiased 
noise, and unbiased disturbance as well as noiseless apparatuses and 
nondisturbing apparatuses. 
Two models of position measurements are discussed to show that Heisenberg's
relation can be violated  even by approximately repeatable position
measurements.
\end{abstract}
\maketitle

\section{Introduction}

Quantum mechanics introduced the intrinsic discreteness of some physical 
quantities represented by polynomials of continuous quantities.
This discrepancy was resolved by the non-commutativity of the canonical 
observables, the canonical commutation relation (CCR), found by Heisenberg.
Another basic feature of quantum mechanics is that 
every measurement 
introduces an unavoidable and uncontrollable disturbance on the measured 
object.
Heisenberg's uncertainty relation interprets the physical content of the
non-commutativity as the limitation to our ability of observation by
quantifying the amount of unavoidable disturbance caused by
measurement. 

According to the celebrated paper by Heisenberg \cite{Hei27} in 1927,  
Heisenberg's uncertainty relation can be formulated as follows: {\em For
every measurement of the position $Q$ of a mass with  root-mean-square
error $\ep(Q)$, the root-mean-square disturbance $\et(P)$ of the
momentum $P$ of the mass caused  by the interaction of this
measurement always satisfies the  relation
\begin{equation}\label{eq:Heisenberg}
\ep(Q)\et(P)\ge \frac{\hbar}{2},
\end{equation}
where $\hbar$ is Planck's constant divided by $2\pi$.}
Here, we use the lower bound $\hbar/2$ for the consistency
with the modern treatment.

Heisenberg \cite{Hei27} not only explained the physical intuition 
underlying the above relation by discussing the famous $\gamma$ 
ray microscope thought experiment, but also claimed that this 
relation is a straightforward mathematical consequence of the 
CCR, $QP-PQ=i\hbar$.
Heisenberg's argument runs as follows.
He assumed that the mass state just after the measurement of
position $Q$ with root-mean-square error $\ep(Q)$ is
represented by a Gaussian wave function $\ps$ with the
spread $Q_{1}=\ep(Q)$.  Then, by Fourier transform of $\ps$,
he showed that the momentum spread $P_{1}$ in this state satisfies
the relation
\beqa\label{eq:Heisenberg2}
Q_{1}P_{1}\ge\frac{\hbar}{2}.
\eeqa
He identified the momentum disturbance $\et(P)$ with 
the momentum spread $P_{1}$ in the state just after the 
measurement, and concluded \Eq{Heisenberg}
(see \cite{03HUR} for the detained discussion).

The mathematical part of his proof  leading to \Eq{Heisenberg2} was 
refined by introducing the notion of standard deviation shortly afterward by
Kennard \cite{Ken27}.
He explicitly defined the spreads $Q_{1}$ and $P_{1}$
to be the standard deviations of position and momentum, $\si(Q)$
and $\si(P)$.
Using Fourier analysis, he proved 
\begin{equation}\label{eq:Kennard}
\si(Q)\si(P)\ge \frac{\hbar}{2}
\end{equation}
in an arbitrary wave function $\ps$.

Kennard's relation above was later generalized to arbitrary 
pair of observables by Robertson \cite{Rob29} as follows.
{\em For any pair of observables $A$ and $B$, their standard 
deviations, $\si(A)$ and $\si(B)$, satisfy the relation
\begin{equation}\label{eq:Robertson}
\si(A)\si(B)\ge \frac{1}{2}|\bracket{\ps,[A,B]\ps}|
\end{equation}
in any state $\ps$ with $\si(A), \si(B)<\infty$.}
In the above, $[A,B]$ stands for the commutator
$[A,B]=AB-BA$, and the standard deviation is defined as
$\si(A)=(\bracket{\ps,A^{2}\ps}-\bracket{\ps,A\ps}^{2})^{1/2}$,
where $\bracket{\cdots,\cdots}$ denotes the inner product;
in this paper, every state vector is assumed normalized and
the domain of the commutator $[A,B]$ is considered extended
appropriately.

Robertson proved the above relation using a simple application of
the Schwarz inequality without using Fourier analysis.
Thus, it was made clear that Heisenberg's
relation \Eq{Heisenberg2} is a straightforward mathematical
consequence of the CCR.
However, Heisenberg's argument that leads to \Eq{Heisenberg}
from \Eq{Heisenberg2} has not been justified for more than
75 years since then.

In fact, Heisenberg himself appears to have changed his position 
from 1927 to 1929. 
Around this time, it was already known that an EPR type thought
experiment violates \Eq{Heisenberg}. 
In this case, by the effect of entanglement between two masses, 1 and 2,
the position of mass 1 at a time $t$ can be indirectly measured very precisely 
by measuring the entangled mass 2 without disturbing any observables 
of mass 1, and hence if the momentum of mass 1 is measured directly 
just after the position measurement, the momentum at the time $t$ can 
also be measured very precisely.
Heisenberg's  response to this criticism appears that he considered 
the uncertainty relation to be \Eq{Kennard} rather than \Eq{Heisenberg}.
He stated,  for instance, that ``every experiment destroys some of the 
knowledge of the system which was obtained by previous experiments.     
This formulation makes it clear that the uncertainty relation does not 
refer to  the past''  (p. 20, Ref.~\cite{Hei30}).    
Heisenberg's response means that even if we can measure 
both the position and the momentum at the past time $t$ very precisely, 
after the momentum measurement the mass no longer has 
definite position so that \Eq{Kennard} is not violated at any time. 

Thus, in a few years after its first appearance, Heisenberg's uncertainty 
relation might turn to be a more formal relation like the CCR than what
Heisenberg claimed in 1927 as the fundamental limit on our ability of 
measurements.  
However, this change paid a high price of confusions among physicists
as well as a broad scientific community.
In fact, many text books have associated the formal expression of  
``Heisenberg's uncertainty relation'' to \Eq{Kennard}, 
but also associated the physical meaning of 
``Heisenberg's uncertainty relation'' to \Eq{Heisenberg}
by illustrating many examples including the $\gamma$ ray
microscope.
Such a view has been accepted for many 
\cite{vN32,Boh49,Boh51,Mes59a,Bra74,CTDSZ80,BVT80},
whereas the universal validity of \Eq{Heisenberg} has been also
criticized in many ways 
\cite{EPR35,AK65,Bal70,Yue83,Kra87,AG88,88MS,89RS,%
HV90,MM90,91QU,Ish91,MM92,BK92,App98,01CQSR,02KB5E}.
Thus, it is still a significant problem to clarify the limitation
of \Eq{Heisenberg} and to generalize it to obtain relations
that hold for every measurement.

Just as Heisenberg \cite{Hei27} argued that the mathematical relation 
\Eq{Heisenberg2} or \Eq{Kennard} concludes the physical assertion
expressed by \Eq{Heisenberg}, 
Robertson's relation \Eq{Robertson}
suggests the following physical
assertion: {\em If an apparatus measures
an observable $A$ in a state $\ps$ with the root-mean-square 
error $\ep(A)$,
the apparatus disturbs an observable $B$ with the root-mean-square
disturbance $\et(B)$ satisfying the relation
\begin{equation}\label{eq:URMD}
\ep(A)\et(B)\ge\frac{1}{2}|\bracket{\ps,[A,B]|\ps}|.
\end{equation}}
We refer to the above relation as {\em Heisenberg's noise-disturbance
uncertainty relation}.
Many text books of quantum mechanics have regarded
Robertson's relation \Eq{Robertson} as the rigorous formalization 
of Heisenberg's noise-disturbance uncertainty relation, even
though without claiming the universal validity of the latter.
The purpose of the present paper is to clarify the limitation
of Heisenberg's noise-disturbance uncertainty relation
and to generalize this relation to a universally valid
relations for the noise and the disturbance.

It is not a much surprising fact that we have not obtained even a precise
formulation of Heisenberg's noise-disturbance uncertainty relation.
Robertson's relation is a universal statement for any states and 
any pair of observables.
However, Heisenberg's noise-disturbance uncertainty relation
is a universal statement for any states, any pair of observables,
and, in addition to those, for any measurements or any measuring
apparatuses.
Since von Neumann's axiomatization of quantum mechanics 
\cite{vN32} published in 1932,
we have definite answers to questions as to  
what are general states and what are general observables. 
However, the question was left unanswered for long time
as to what are general measurements.

Towards this problem, Davies and Lewis (DL) \cite{DL70}
introduced the mathematical formulation of the 
notion of  ``instrument'' as normalized positive map valued measures,
to which we shall refer as DL instruments,
and showed that this notion quite generally describes the  
statistical properties of a general measurement,
so that for any sequence of measurements the joint
probability distribution of those measurements are
determined by their corresponding DL instruments.

However, the question left open for some time
as to whether every DL instrument corresponds to a possible 
measuring apparatus \cite{Yue87}.
In order to solve this question, the present author 
\cite{83CR,84QC} introduced a general class of mathematical models of
measuring processes and showed that the statistical properties 
given by any such model is described by a normalized completely 
positive map valued measure, to be referred to as a CP instrument,
and conversely that any CP instrument arises in this way.  
Thus, we naturally conclude that measurements are represented by
CP instruments, just as states are represented by density operators and
observables are represented by self-adjoint operators.

In this paper, we start with presenting the above characterization of
measurements in more accessible form.
In Section \ref{se:II}, we propose the two axioms 
for general measuring apparatuses, 
the mixing law (of joint output probability) 
and the extendability axiom, which characterize the statistical 
description of general measuring apparatuses.
In Section \ref{se:MP}, we introduce mathematical models of
measuring processes, called indirect measurement models,
and pose the realizability postulate.
Under the above three axioms, we show that (i) every
apparatus corresponds to a unique CP instrument
that describes the statistical properties of that apparatus, 
(ii) conversely, every CP instrument corresponds to at least one apparatus,
(iii) the correspondence is a one-to-one correspondence
up to statistical equivalence of apparatuses, and (iv) any 
apparatus has a statistically equivalent apparatus which is
described by an indirect measurement model.
Thus, we establish the notion of  ``instrument'' as the function
of a measuring apparatus by the mathematical notion
``CP instrument'' that represents the statistical equivalence 
class of a measuring apparatus.
In the above sections, we are also devoted to explain 
how the notion of CP instruments integrates such notions as effects,
operations, probability operator valued measures (POVMs), and 
trace-preserving completely positive maps,  
widely accepted in the field of quantum information \cite{NC00}.
It should be also pointed out that 
since in an indirect measurement model, the measuring interaction 
is described purely quantum mechanically, the above results
provide a useful approach to explore statistical properties of
general quantum measurements using quantum mechanical laws.

In Section \ref{se:Noise},
a mathematical notion of the distance between 
POVMs and observables is introduced
and the basic properties are explored. 
Then, we formulate the notion of measurement
noise and obtain the basic properties.  
In particular,  we clarify the meaning of noise in the
indirect measurement model and show that this notion is
equivalent to the distance of the POVM of the apparatus 
from the observable to be measured, and hence the noise
is independent of particular models but depend only on 
the POVM of the apparatus.
In Section \ref{se:Disturbance},
we formulate the notion of disturbance 
caused by a measurement and we obtain the basic properties.  
Disturbance is rather straightforward notion for
indirect measurement models, while it is not clear whether
it is model independent.
We show that the disturbance in a given observable 
is determined only by the trace-preserving completely positive map
that describes the nonselective operation of the apparatus.
In Section \ref{se:Uncertainty}, under the formulation provided as above,
Heisenberg's noise-disturbance uncertainty relation is generalized to 
a relation that holds for any measuring apparatuses, from which
conditions are obtained for measuring apparatuses to satisfy 
Heisenberg's relation. 
In particular, every apparatus with the noise and the disturbance
statistically independent from the measured object is proven
to satisfy Heisenberg's relation.
Under this formulation, various uncertainty relations are also derived
for apparatuses with independent noise, independent disturbance, unbiased 
noise, and unbiased disturbance as well as noiseless apparatuses and 
nondisturbing apparatuses.
In Section \ref{se:MBHUP}, we examine von Neumann's model of
position measurement to show that this model typically satisfies
Heisenberg's relation.
Then, we examine the position measurement
model that was introduced in Ref.~\cite{88MS} and show
that this model violates Heisenberg's relation uniformly.
The above model was shown in Ref.~\cite{88MS} to realize
Yuen's contractive state measurement \cite{Yue83} and
to break the standard quantum limit for monitoring free-mass
position claimed by Braginsky and collaborators
\cite{Bra74,CTDSZ80,BVT80} as a consequence of Heisenberg's
relation. 
An experimental proposal was given in Ref.~\cite{03UVR} for realizing the
above model in an equivalent linear optical setting. 
In Section \ref{se:repeatability}, based on the above model 
we show that Heisenberg's relation can be violated even by approximately
repeatable position measurements. 
Some discussions in the final section conclude the present paper.

\section{Statistics of general quantum measurements\label{se:II}}

\subsection{Postulates for quantum mechanics\label{ss:IIA}}

Throughout this paper, we assume the following postulates 
introduced by von Neumann \cite{vN32} for 
non-relativistic quantum mechanics without any superselection
rules.

{\em Postulate I.} (Representations of states and observables)
Any quantum system $\bS$ is associated with a  unique 
separable Hilbert space
$\cH_{\bS}$, called the {\em state space} of $\bS$.
Any {\em state} of $\bS$ is represented in
one-to-one correspondence by a positive operator $\rh$ with unit
trace, called a {\em density operator } on $\cH_{\bS}$.  
Under a fixed unit system, any {\em observable} 
of $\bS$ is represented in one-to-one correspondence by a 
self-adjoint operator $A$ (densely defined) on $\cH_{\bS}$.

{\em Postulate II.} (Schr\"{o}dinger equation)
If system $\bS$ is isolated in a time interval
$(t,t')$, there is a unitary operator $U$, called the 
{\em time evolution operator}, such that if $\bS$ is in state $\rh$ 
at time $t$ then $\bS$ is in state $\rh'=U\rh U^{\da}$ at time $t'$.

{\em Postulate III.}  (Born statistical formula)
Any observable $A$ can be precisely measured in any state $\rh$
in such a way that $A$ takes the value in a Borel set $\De$ with
probability $\Tr[E^{A}(\De)\rh]$,  
where $E^{A}(\De)$
is the spectral projection of $A$  corresponding to Borel set $\De$.

{\em Postulate IV.}  (Composition rule)
The state space of the composite system 
$\bS+\bS'$ of two systems $\bS$ and $\bS'$ is the tensor product 
 $\cH_{\bS}\otimes\cH_{\bS'}$ of their state spaces.
An observable $A$ in $\bS$ and an observable $B$ in $\bS'$
are identified with the observables $A\otimes I$ and 
$I\otimes B$, respectively, in the system $\bS+\bS'$.

For any unit vector $\ps$, the state 
$\rh=\ket{\ps}\bra{\ps}$ is called a {\em vector state} 
represented by $\ps$.  
In this case, $\ps$ is called a {\em state vector} representing 
the state $\rh$.

Let $\rh_{1}, \ldots,\rh_{n}$ be a sequence of density operators and let
$p_{1},\ldots,p_{n}$ be a probability distribution on $\{1,\ldots,n\}$,
i.e., $p_{j}\ge 0$ for all $j$ and $\sum_{j}p_{j}=1$.
We say that a system $\bS$ is a {\em random sample} from the ensemble
$(\{\rh_{j}\},\{p_{j}\})$, iff system $\bS$ is in state $\rh_{j}$
with probability $p_{j}$.
In this case, an observable $A$ takes, in a precise measurement, 
the value in a Borel set $\De$ with probability 
\beqa
P(\De)=\sum_{j}p_{j}\Tr[E^{A}(\De)\rh_{j}].
\eeqa
Let $\rh=\sum_{j}p_{j}\rh_{j}$.  By linearity of trace, the density operator
$\rh$ satisfies
\beqa
P(\De)=\Tr[E^{A}(\De)\rh].
\eeqa
Since $A$ and $\De$ are arbitrary,  from Postulate III 
(Born statistical formula) 
we conclude the following.

\begin{Theorem}\label{th:mixture}
Any random sample from ensemble $(\{\rh_{j}\},\{p_{j}\})$ is in the state 
$\rh=\sum_{j}p_{j}\rh_{j}$.
\end{Theorem}

The above theorem has established the interpretation 
of the mixture of states that the system 
$\bS$ is in the state $p\rh_{1}+(1-p)\rh_{2}$, 
if it is in state $\rh_{1}$ with probability $p$ 
and in state $\rh_{2}$ with probability $1-p$.

The notion of precise measurements of observables is determined 
solely by Postulate III (Born statistical formula) 
without assuming any further conditions 
on the state change caused by measurement 
such as the projection postulate stating that the measurement 
projects the state onto the eigenspace corresponding 
to the eigenvalue to be obtained. 

Postulate III (Born statistical formula) does not assume that the observable has a certain
unknown value in the state just before the measurement
that is reproduced by a precise measurement, but only requires that
the precise measurement statistically reproduces the postulated probability. 

A Hilbert space is separable if and only if its dimension is at most
countable infinite.  Throughout this paper, only separable Hilbert spaces
are considered and simply called {\em Hilbert spaces}.

Throughout this paper, the word ``Borel set'' can be 
safely replaced by the word ``interval'' only
with some modifications on mathematical technicality. 
Readers not familiar with measure theory are recommended to
read with such replacements.

The relation between the present formulation based on spectral
projections due to von Neumann \cite{vN32} 
and Dirac's formulation \cite{Dir58} is as follows.
If the observable $A$ has the Dirac type spectral decomposition
\beqas
A=\sum_{\mu}\sum_{\nu}\mu\ket{\mu,\nu}\bra{\mu,\nu}
+\int_{\R}\sum_{\nu}\lambda\ket{\lambda,\nu}\bra{\lambda,\nu}\,d\lambda,
\eeqas
where $\mu$ varies over the discrete eigenvalues, $\lambda$ varies over
the continuous eigenvalues, and $\nu$ is the degeneracy parameter,
then we have
\beqas
E^{A}(\Delta)=\sum_{\mu\in\Delta}\sum_{\nu}\ket{\mu,\nu}\bra{\mu,\nu}
+\int_{\Delta}\sum_{\nu}\ket{\lambda,\nu}\bra{\lambda,\nu}\,d\lambda.
\eeqas
In this case, we have
\beqas
\lefteqn{\Tr[E^{A}(\De)\rh]}\quad\nn\\
&=&\sum_{\mu\in\De}\sum_{\nu}\bracket{\mu,\nu|\rh|\mu,\nu}
+\int_{\De}\sum_{\nu}
\bracket{\lambda,\nu|\rh|\lambda,\nu}\,d\lambda.
\eeqas
%\beqas
%\Tr[E^{A}(\De)\rh]
%=\sum_{\nu}\sum_{\mu\in\De}\bracket{\mu,\nu|\rh|\mu,\nu}
%+\sum_{\nu}\int_{\De}
%\bracket{\lambda,\nu|\rh|\lambda,\nu}\,d\lambda.
%\eeqas

We do not allow unnormalizable states such as the one
described by Dirac's delta function, since they by no means define the
normalized probability distribution of the output of every measurement 
consistent with the probability theory axiomatized by Kolmogorov
\cite{Kol33}.

\subsection{Output probability distributions}

Every {\em measuring apparatus} has a macroscopic 
{\em output variable} that takes the output of each instance of
measurement.
The output variable is a random variable, in the sense of
classical probability theory \cite{Kol33},
the probability distribution of which depends only on
the  {\em input state}, the state of the system to be measured
at the instant just before the measurement.

Let ${\bf S}$ be a quantum system, to be referred to the {\em object}, 
with state space ${\cal H}$.
Let $\bA(\bx)$ be a measuring apparatus with output
variable $\bx$ to measure the {object} $\bS$.
We assume that $\bx$ takes values in the real line $\R$.
For any Borel set $\De$ in $\R$, we shall denote by ``$\bx\in\De$'' 
the probabilistic event that the output $\bx$ takes a value in $\De$.
The event ``$\bx\in\De$''  is called the {\em outcome} of measurement.
The probability distribution of $\bx$ on {\em input
state} $\rh$ is denoted by $\Pr\{\bx\in\De\|\rh\}$,
where $\De$ varies over all Borel subsets of the real line, and 
called the {\em output  probability distribution} of $\bA(\bx)$.  
We shall write $\Pr\{\bx\in\De\|\rh\}=\Pr\{\bx\in\De\|\ps\}$,
if $\rh=\ket{\ps}\bra{\ps}$.

In this paper, any probability distribution is required to satisfy 
the positivity, the countable additivity, and the normalization condition
\cite{Kol33}, 
so that the output probability distribution satisfies the following conditions.

(i) (Positivity) $\Pr\{\bx\in\De\|\rh\}\ge0$ for any Borel set $\De$.
 
(ii) (Countable additivity)
\beqa
\Pr\{\bx\in\De\|\rh\}=\sum_{j}\Pr\{\bx\in\De_{j}\|\rh\}
\eeqa 
for any disjoint sequence of Borel sets 
$\De_{1}, \De_{2},\ldots$ with $\De=\bigcup_{j}\De_{j}$.

(iii) (Normalization condition)  $\Pr\{\bx\in\R\|\rh\}=1$.

In addition to the above, it is natural to require that the output probability
distribution should satisfy the following postulate.

{\bf Mixing law of output probability:}
{\em For any apparatus $\bA(\bx)$, the
function $\rh\mapsto\Pr\{\bx\in\De\|\rh\}$ is
an affine function of density operators $\rh$ for every
Borel set $\De$, i.e.,  
\beqa
\lefteqn{\Pr\{\bx\in\De\|p\rh_{1}+(1-p)\rh_{2}\}}\nn\\
&=&p\Pr\{\bx\in\De\|\rh_{1}\}+(1-p)\Pr\{\bx\in\De\|\rh_{2}\},
\quad
\eeqa
where $\rh_{1}$ and $\rh_{2}$ are density operators and $0<p<1$. }

The above postulate is justified as follows.
If the system $\bS$ is a random sample from the ensemble
$(\{\rh_{1},\rh_{2}\},\{p,1-p\})$, then the event ``$\bx\in\De$''
occurs with probability
$p\Pr\{\bx\in\De\|\rh_{1}\}+(1-p)\Pr\{\bx\in\De\|\rh_{2}\}$.
On the other hand, from Theorem \ref{th:mixture}
in this case the system $\bS$ is in the state
$p\rh_{1}+(1-p)\rh_{2}$, so that the above equality should hold.

\subsection{Probability operator valued measures}

In order to characterize the output probability distributions, we
need a mathematical definition:  A mapping $\Pi :\De \mapsto
\Pi (\De)$  of the collection $\cB(\R)$ of Borel subsets in
$\R$ into the space  $\cL(\cH)$ of bounded operators on $\cH$ is called a
{\em probability operator valued measure (POVM)}, if the following
conditions are satisfied: 

(i) (Positivity) $\Pi (\De)\ge 0$ for all
$\De\in\cB(\R)$. 

(ii)  (Countable additivity) For any disjoint sequence
$\De_{1},
\De_{2},\ldots$ of Borel sets with $\De=\bigcup_{j}\De_{j}$, 
we have 
\beqa
\Pi (\De)=\sum_{j}\Pi (\De_{j}),
\eeqa
where the summation is convergent in the weak operator topology,
i.e., we have $\bracket{\ps|\Pi (\De)|\ps}=\sum_{j}\bracket{\ps|\Pi (\De_{j})|\ps}$
for every state vector $\ps$.

(iii) (Normalization condition) $\Pi (\R)=I$, where  $I$ is the identity
operator on
$\cH$.
 
For mathematical properties of POVMs we refer to Berberian \cite{Ber66}.
One of important consequences from the 
mixing law of output probability is
the following characterization of output probability 
distributions given in Ref.~\cite{80OG}.

\begin{Theorem}\label{th:POVM}
The mixing law of output probability is equivalent to
the following requirement:

For any apparatus $\bA(\bx)$ there exists  a unique
POVM $\Pi $ satisfying
\begin{equation}\label{eq:Born-POVM}
\Pr\{\bx\in\De\|\rh\}
=
\Tr[\Pi (\De)\rh]
\end{equation}
for any Borel set $\De$ and  density operator $\rh$.
\end{Theorem}

A sketch of the proof runs as follows.
It is easy to check that \Eq{Born-POVM} defines the output 
probability distribution satisfying the mixing law of output probability.
Conversely, suppose that the output probability distribution satisfies
the mixing law of output probability.
Recall that every trace class operator $\rh$ can be decomposed
as 
\beqa
\rh=\sum_{j=1}^{4}\al_{j}\rh_{j},
\eeqa
where $\rh_{1},\ldots,\rh_{4}$ are density operators and
$\al_{1},\ldots,\al_{4}$ are complex numbers; one of the decompositions
can be easily found from the spectral decomposition.
By the above decomposition, for every trace class operator $\rh$ and
every Borel set $\De$, we can define a complex number $\Pi (\De,\rh)$ by
\begin{equation}\label{eq:DL instruments}
\Pi (\De,\rh)=\sum_{j=1}^{4}\al_{j}\Pr\{\bx\in\De\|\rh_{j}\}.
\end{equation}
The mixing law of output probability ensures that for every decompositions of the
same $\rh$, the above equation defines the unique value, and
moreover that the function $\Pi $ is linear in $\rh$.
Let $\ket{1},\ket{2},\ldots$ be an orthonormal basis of $\cH$.
Then, we can define an operator $\Pi (\De)$ by
\beqa
\Pi (\De)=\sum_{n,m}\Pi (\De,\ket{m}\bra{n})\ket{n}\bra{m}.
\eeqa
Then, by linearity of $\Pi (\De,\rh)$ in $\rh$, we have 
\beqa
\Tr[\Pi (\De)\rh]&=&
\sum_{n,m}\Pi (\De,\ket{m}\bra{n})\bra{m}\rh\ket{n}\nn\\
&=&
\Pi \left(\De,\sum_{n,m}\bra{m}\rh\ket{n}\ket{m}\bra{n}\right)\nn\\
&=&
\Pi (\De,\rh)\nn\\
&=&
\Pr\{\bx\in\De\|\rh\}.\label{eq:output-POVM}
\eeqa
Thus, $\Pi (\De)$ is a unique operator satisfying \Eq{Born-POVM}.
Now, conditions (i)--(iii) for $\Pi $ follow easily from \Eq{Born-POVM},
and hence $\Pi $ is a POVM.  This completes the proof.

The POVM $\Pi $ defined by \Eq{Born-POVM} is called the {\em POVM of
apparatus} $\bA(\bx)$.  
The operator $\Pi (\De)$ is called the {\em effect of apparatus
$\bA(\bx)$} associated with the outcome $\bx\in\De$.
For the general notion of effects, we refer to Kraus \cite{Kra83}.
For applications of POVMs to quantum measurement, quantum
estimation, and quantum information, we refer the reader to Helstrom
\cite{Hel76}, Davies \cite{Dav76}, Holevo \cite{Hol82},
Peres \cite{Per93}, and Nielsen-Chuang \cite{NC00}. 

%\subsection{The Born statistical formula}

Let $A$ be an observable of system $\bS$.
Postulate III (Born statistical formula) naturally leads to the following
definition.
We say that apparatus $\bA(\bx)$ {\em satisfies the Born statistical
formula (BSF)} for observable $A$ on input state $\rh$, if we have
\begin{equation}\label{eq:BSF}
%\label{eq:010530a}\label{eq:precise-measurement}\label{eq:BSF}
\Pr\{\bx\in\De\|\rh\}=\Tr[E^{A}(\De)\rh]
\end{equation}
for every Borel set $\De$.
The mapping $E^{A}$ that maps every Borel set $\De$
to the spectral projection  $E^{A}(\De)$ of $A$
corresponding  to $\De$ is called the
the   {\em spectral measure of $A$}.
For mathematical theory of spectral measures, we refer to Halmos \cite{Hal51}.
From Postulate III (Born statistical formula), apparatus $\bA(\bx)$ precisely
measures an observable $A$ if and only if
$\bA(\bx)$ satisfies the BSF for observable $A$ on every input state, 
and moreover for every observable $A$ of $\bS$ 
there is at least one apparatus that precisely measures $A$. 
From Eqs.~\eq{Born-POVM} and \eq{BSF}, {\em apparatus
$\bA(\bx)$   precisely measures observable $A$ 
if and only if the POVM $\Pi $ of $\bA(\bx)$ is
the spectral measure $E^{A}$, i.e., 
\beqa
\Pi =E^{A}.
\eeqa}

\subsection{Quantum state reductions}

We have shown that every apparatus is associated with a POVM which
determines the output probability distribution.
However, POVMs of apparatuses do not determine the joint probability
distributions of outputs from successive measurements using several apparatuses.
In the following, we introduce the notion of quantum state reduction
to determine such joint probability distributions.

Depending on the input state $\rh$ and the outcome $\bx\in\De$, 
let $\rh_{\{\bx\in\De\}}$ be the state just after the measurement 
conditional upon the outcome $\bx\in\De$.
We assume that for any Borel set $\De$ with $\Pr\{\bx\in\De\|\rh\}>0$  
the state $\rh_{\{\bx\in\De\}}$ is uniquely determined.
If $\Pr\{\bx\in\De\|\rh\}=0$, the state $\rh_{\{\bx\in\De\}}$ is taken to be
indefinite and the notation $\rh_{\{\bx\in\De\}}$ denotes an arbitrary state.
The state $\rh_{\{\bx\in\De\}}$ is called the {\em output
state} given the outcome $\bx\in\De$ on input state $\rh$.

The state change from the input state to the output state 
is generally called the {\em quantum state reduction};
while the transformation from the input state to the output
probability distribution, namely the state of the macroscopic meter, 
is called the {\em objectification} or the {\em objective state reduction}. 
Those two different notions have been mixed up for long time
\cite{03QSR}.   
 
Two apparatuses are called {\em statistically equivalent}, 
if they have the same objective state reduction and quantum state reduction, 
or they have the same output probabilities and the same output states
for any outcomes and any input states.

\subsection{Mixing law}

For notational convention, we distinguish apparatuses
by their output variables.  For instance, symbols $\bA(\bx)$, 
$\bA(\by)$, and $\bA(\bz)$ denote three apparatuses with output
variables $\bx$, $\by$, and $\bz$, respectively.

The operational meaning of the state $\rh_{\{\bx\in\De\}}$ is given as follows.
Suppose that a measurement using the apparatus $\bA(\bx)$ on input state $\rh$ 
is immediately followed by a measurement using another apparatus $\bA(\by)$.
Then, the joint probability distribution $\Pr\{\bx\in\De,\by\in\De'\|\rh\}$ of the
output variables $\bx$ and $\by$ is given by
\beqa\label{eq:c}
\lefteqn{\Pr\{\bx\in\De,\by\in\De'\|\rh\}}\quad\nn\\
&=&\Pr\{\by\in\De'\|\rh_{\{\bx\in\De\}}\}\Pr\{\bx\in\De\|\rh\},
\eeqa
since the event $\bx\in\De$ occurs with probability $\Pr\{\bx\in\De\|\rh\}$ 
and then the event $\by\in\De'$ occurs with probability
$\Pr\{\by\in\De'\|\rh_{\{\bx\in\De\}}\}$.
We shall call the above joint probability distribution the {\em joint output
probability distribution} of $\bA(\bx)$ and $\bA(\by)$.

Thus, the joint probability distribution of outputs of  successive
measurements depends only on the input state of the first
measurement and should satisfy the following postulate.
%\vskip\topsep

{\bf Mixing law (of joint output probability): }
{\em For any apparatuses $\bA(\bx)$ and $\bA(\by)$, 
the function $\rh\mapsto \Pr\{\bx\in\De,\by\in\De'\|\rh\}$
is an affine function of density operators $\rh$ for every pair of Borel sets 
$\De,\De'$, i.e.,  
\beqa
\lefteqn{\Pr\{\bx\in\De,\by\in\De'\|p\rh_{1}+(1-p)\rh_{2}\}}\nn\\
&=&p\Pr\{\bx\in\De,\by\in\De'\|\rh_{1}\}\nn\\
& &\mb{}+(1-p)\Pr\{\bx\in\De,\by\in\De'\|\rh_{2}\},
\eeqa
where $\rh_{1}$ and $\rh_{2}$ are density operators and $0<p<1$. }

This requirement is justified as follows.
The successive applications of two apparatuses $\bA(\bx)$
and $\bA(\by)$ to a single system $\bS$
can be considered as an application of one
apparatus $\bA(\bx,\by)$ with two output variables $\bx$ and $\by$.  
Thus, the above postulate follows from the mixing law of
output probability (generalized to apparatuses with two
output variables).

By substituting $\De'=\R$ in \Eq{c} and using the normalization condition
$\Pr\{\by\in\R\|\rh_{\{\bx\in\De\}}\}=1$, we have
\beqa
\Pr\{\bx\in\De,\by\in\R\|\rh\}=\Pr\{\bx\in\De\|\rh\}
\eeqa
for any $\De$ and $\rh$.  Thus, we conclude that {\em the mixing law
of joint output probability implies the mixing law of 
output probability.}
From now on, the mixing law of joint output probability will be simply
referred to as the {\em mixing law}.

Consider the case where $\De=\R$.
The symbol $\rh$ in $\Pr\{\bx\in\R,\by\in\De'\|\rh\}$
refers to the state just before $\bA(\bx)$ measurement,
while the symbol $\rh$ in $\Pr\{\by\in\De'\|\rh\}$
refers to the state just before $\bA(\by)$ measurement.
Thus, the above two probabilities are not generally identical.
According to \Eq{c}, we have
\beqa
\Pr\{\bx\in\R,\by\in\De'\|\rh\}
&=&\Pr\{\by\in\De'\|\rh_{\{\bx\in\R\}}\}.
\eeqa
The above relation characterizes the state $\rh_{\{\bx\in\R\}}$.

If $\Pr\{\bx\in\De\|\rh\}=1$, by the additivity of probability,
we have
\beqa
\Pr\{\bx\in\De,\by\in\De'\|\rh\}&=&\Pr\{\bx\in\R,\by\in\De'\|\rh\},
\eeqa
and hence \Eq{c} leads to
\beqa
\Pr\{\by\in\De'\|\rh_{\{\bx\in\De\}}\}
&=&\Pr\{\by\in\De'\|\rh_{\{\bx\in\R\}}\}.
\eeqa
Since apparatus $\bA(\by)$ is arbitrary, 
we have $\rh_{\{\bx\in\De\}}=\rh_{\{\bx\in\R\}}$.
Thus, the condition $\bx\in{\De}$ makes no selection. 
In this case, the state change $\rh\mapsto\rh_{\{\bx\in\De\}}$ is 
called the {\em nonselective state change}.
%If $0<\Pr\{\bx\in\De\|\rh\}<1$, the state change
%$\rh\mapsto\rh_{\{\bx\in\De\}}$ is called the  {\em
%selective state change}.

From \Eq{c}, the conditional probability distribution of $\by$
given $\bx\in\De$ is determined as
\beqa\label{020603a}
\Pr\{\by\in\De'|\bx\in\De\|\rh\}=\Pr\{\by\in\De'\|\rh_{\{\bx\in\De\}}\}.
\eeqa
provided that $\Pr\{\bx\in\De\|\rh\}>0$.
In particular, we have
\begin{equation}\label{eq:030607b}
\Pr\{\by\in\De'|\bx\in\De\|\rh\}=
\Tr[E^{B}(\De')\rh_{\{\bx\in\De\}}],
\end{equation}
if $\bA(\by)$ precisely measures an observable $B$.
The above relation uniquely determines the output
state $\rh_{\{\bx\in\De\}}$.

\subsection{Operational distributions}

In 1970, Davies and Lewis \cite{DL70} 
introduced the following mathematical notion for unified 
description of statistical properties of measurements.
A mapping $\cI:\De\mapsto\cI(\De)$ of $\cB(\R)$ 
into the space $\cL(\tc(\cH))$ of bounded linear transformations
on the space $\tc(\cH)$ of trace-class operators on $\cH$ is called an
{\em DL instrument}, iff the following conditions  are satisfied.

(i) (Positivity)
$\cI(\De)$ is a positive linear transformation of $\tc(\cH)$ for every 
$\De\in\cB(\R)$.

(ii) (Countable additivity) For any disjoint sequence $\De_{1},
\De_{2},\ldots$ of Borel sets with $\De=\bigcup_{j}\De_{j}$, 
we have
\beqas
\cI(\De)=\sum_{j}\cI(\De_{j}),
\eeqas
where the summation is convergent in the strong operator topology of
$\cL(\tc(\cH))$, i.e., 
$\lim_{n\to\infty}\|\cI(\De)\rh-\sum_{j=1}^{n}\cI(\De_{j})\rh\|_{\tc}=0$
for any $\rh\in\tc(\cH)$,
where $\|X\|_{\tc}=\Tr\sqrt{X^{\da}X}$ for any $X\in\tc(\cH)$.

(iii) (Normalization condition) $\cI(\R)$ is trace-preserving, i.e., 
\beqa
\Tr[\cI(\R)]=\Tr\rh
\eeqa
for any $\rh\in\tc(\cH)$.

For mathematical properties of DL instruments we refer to Davies \cite{Dav76}.
One of important consequence from the mixing law
is the following unified characterization of output probability distribution
and quantum state reduction given in Ref.~\cite{00MN,01QI}. 

\begin{Theorem}
\label{th:DL instrument}
The mixing law is equivalent to the
following requirement:

For any apparatus $\bA(\bx)$ there exists a unique 
DL instrument $\cI$ 
satisfying
\begin{equation}\label{eq:DL instruments2}
\cI(\De)\rh=
\Pr\{\bx\in\De\|\rh\}\rh_{\{\bx\in\De\}}
\end{equation}
for any Borel set $\De$ and  density operator $\rh$.
\end{Theorem}

A sketch of the proof runs as follows.
Let $\bA(\bx)$ be an apparatus.
For any state $\rh$ and Borel set $\De$, 
we define an operator $\cI(\De,\rh)$ by
\begin{equation}\label{eq:DL instruments3}
\cI(\De,\rh)=\Pr\{\bx\in\De\|\rh\}\rh_{\{\bx\in\De\}}.
\end{equation}
If $\Pr\{\bx\in\De\|\rh\}=0$, then $\cI(\De,\rh)=0$, so that $\cI(\De,\rh)$
is determined definitely, despite that $\rh_{\{\bx\in\De\}}$ is indefinite
in this case.  
Then, for any apparatus $\bA(\by)$ to precisely measure
an observable $B$, we have
\beqas
\lefteqn{\Pr\{\bx\in\De,\by\in\De'\|\rh\}}\quad\\
&=&
\Tr[E^{B}(\De')\rh_{\{\bx\in\De\}}]\Pr\{\bx\in\De\|\rh\}\\
&=&\Tr[E^{B}(\De')\cI(\De,\rh)].
\eeqas
Thus, by the mixing law, we have
\beqa\label{eq:030608a}
\lefteqn{\cI(\De, p\rh_{1}+(1-p)\rh_{2})}\qquad\nn\\
&=&p\cI(\De,\rh_{1})+(1-p)\cI(\De,\rh_{2}),
\eeqa
where $\rh_{1}$ and $\rh_{2}$ are density operators and $0<p<1$. 
Thus, the definition of $\cI(\De,\rh)$ can be extended to all trace
class operators $\rh$ by the relation
\beqa
\cI(\De,\rh)
&=&\sum_{j=1}^{4}\al_{j}\cI(\De,\rh_{j}),
\eeqa
for any density operators $\rh_{1},\ldots,\rh_{4}$ and complex
numbers $\al_{1},\ldots,\al_{4}$ such that 
$\rh=\sum_{j=1}^{4}\al_{j}\rh_{j}$.
Since every density operator has at least one such decomposition,
and since \Eq{030608a} ensures the uniqueness of 
extension, the operator $\cI(\De,\rh)$ is well-defined for all
Borel sets $\De$ and all trace class operators $\rh$.  
Then, we can see that the mapping that maps $\rh$ to $\cI(\De,\rh)$
is a bounded linear transformation of $\tc(\cH)$ for every Borel set
$\De$.  We denote this mapping by $\cI(\De)$.
Then, we define $\cI$ as the mapping that maps $\De$ to $\cI(\De)$.
Now, we have only to show three properties (i)--(iii) above;
however, these are easy consequences from the positivity, countable
additivity, and normalization condition of the probability distribution
$\Pr\{\bx\in\De\|\rh\}$.  For the detail, see Refs.~\cite{97OQ,00MN,01QI}. 

The mapping $\cI(\De)$ given above is called 
the {\em operation of apparatus $\bA(\bx)$}
associated with the outcome $\bx\in\De$. The mapping $\cI$
is called the {\em operational distribution of apparatus $\bA(\bx)$}.
Then, the output probability and the output state can be
expressed by
\beqa
\Pr\{\bx\in\De\|\rh\}&=&\Tr[\cI(\De)\rh],
\label{eq:DL1}\\
\rh_{\{\bx\in\De\}}&=&\frac{\cI(\De)\rh}{\Tr[\cI(\De)\rh]}
\label{eq:DL2},
\eeqa
where the second relation assumes $\Pr\{\bx\in\De\|\rh\}>0$.
Thus, if $\cI_{\bx}$ and $\cI_{\by}$ are the operational distributions
of $\bA(\bx)$ and $\bA(\by)$, respectively, 
then the joint output probability distribution can be expressed by
\begin{equation}\label{eq:JOPD1}
\Pr\{\bx\in\De,\by\in\De'\|\rh\}
=\Tr[\cI_{\by}(\De')\cI_{\bx}(\De)\rh]
\end{equation}
for any state $\rh$ and any Borel sets $\De_{1},\De_{2}$.

Both the output probability distribution and the output states
are determined by the operational distribution.  Thus,
two apparatuses are statistically equivalent if and only if 
they have the same operational distribution.

Let us consider three apparatuses $\bA(\bx),\bA(\by),\bA(\bz)$
with operational distributions $\cI_{\bx},\cI_{\by},\cI_{\bz}$,
respectively,
and suppose that in a state $\rh$, these three apparatuses are
applied to the system $\bS$ successively in this order.
Then, the joint probability distribution of three outputs $\bx,\by,\bz$
are given by
\beqas
\lefteqn{\Pr\{\bx\in\De,\by\in\De',\bz\in\De''\|\rh\}}\quad\nn\\
&=&
\Pr\{\by\in\De',\bz\in\De''\|\rh_{\{\bx\in\De\}}\}\Pr\{\bx\in\De\|\rh\}\\
&=&
\Tr[\cI_{\bz}(\De'')\cI_{\by}(\De')\rh_{\{\bx\in\De\}}]\Pr\{\bx\in\De\|\rh\}\\
&=&
\Tr[\cI_{\bz}(\De'')\cI_{\by}(\De')\cI_{\bx}(\De)\rh].
\eeqas
Thus, by mathematical induction, we obtain the relation
\beqa
\lefteqn{
\Pr\{\bx_{1}\in\De_{1},\ldots,\bx_{n}\in\De_{n}\|\rh\}}\quad\quad\nn\\
&=&
\Tr[\cI_{n}(\De_{n})\ldots\cI_{1}(\De_{1})\rh]
\eeqa
for the joint probability distribution of the output variables $\bx_{1},\ldots,\bx_{n}$
of the successive measurements on the initial input state $\rh$ using apparatuses
$\bA(\bx_{1}),\ldots,\bA(\bx_{n})$ in this order with operational distributions 
$\cI_{1},\ldots,\cI_{n}$, respectively.
Thus, joint probability distribution of the output variables in any successive 
measurements are determined by the operational distributions of apparatuses,
so that statistically equivalent apparatuses are mutually exchangeable
without affecting the joint probability of their outcomes.

In this subsection, under the mixing law,  
we have shown that statistical properties of every apparatus are described 
by a DL instrument and that two apparatuses are statistically equivalent 
if and only if they corresponds to the same DL instrument.

\subsection{Duality}

For any bounded linear transformation $T$ on 
$\tc(\cH)$,
the {\em dual} of $T$ is defined to be the bounded linear
transformation $T^{*}$ on $\cL(\cH)$ satisfying 
\begin{equation}\label{eq:dual}
\Tr[A(T\rh)]=\Tr[(T^{*}A)\rh]
\end{equation}
for any $A\in\cL(\cH)$ and $\rh\in\tc(\cH)$.
The dual  $\cI(\De)^{*}$ of the operation $\cI(\De)$ is called the 
{\em dual operation}  associated with $\bx\in\De$;
by \Eq{dual} it is defined by the relation
\begin{equation}\label{eq:dual-operation}
\Tr[A\cI(\De)\rh]=\Tr\{[\cI(\De)^{*}A]\rh\}
\end{equation}
for any $A\in\cL(\cH)$ and $\rh\in\tc(\cH)$.

The operator $\cI(\De)^{*}I$ obtained by applying
the dual operation $\cI(\De)^{*}$ to the identity operator
is called the {\em effect of operation $\cI(\De)$}.
By \Eq{DL1} and \Eq{dual} we have
\begin{equation}
\Pr\{\bx\in\De\|\rh\}=\Tr\{[\cI(\De)^{*}I]\rh\}.
\end{equation}
Since $\rh$ is arbitrary, comparing with \Eq{Born-POVM},
we have
\begin{equation}
\Pi (\De)=\cI(\De)^{*}I
\end{equation}
for any Borel set $\De$.  Thus, the POVM of $\bA(\bx)$ is
determined by the effects of the operational distribution $\cI$.

Let $\cI_{\bx}$ and $\cI_{\by}$ be the operational distributions of
$\bA(\bx)$ and $\bA(\by)$, respectively, and let $\Pi _{\by}$ be the
POVM of $\bA(\by)$. 
Then, we have
\beqa
\Tr[\cI_{\by}(\De')\cI_{\bx}(\De)\rh]
&=&
\Tr\{[\cI_{\by}(\De')^{*}I][\cI_{\bx}(\De)\rh]\}\nn\\
&=&
\Tr\{[\Pi _{\by}(\De')[\cI_{\bx}(\De)\rh]\}\nn\\
&=&
\Tr\{[\cI(\De)^{*}\Pi _{\by}(\De')]\rh\}.
\eeqa
Thus, the joint output probability distribution can be expressed by
\begin{equation}\label{eq:JOPD2}
\Pr\{\bx\in\De,\by\in\De'\|\rh\}
=\Tr\{[\cI(\De)^{*}\Pi _{\by}(\De')]\rh\}
\end{equation}
for any $\De,\De'\in\cB(\R)$.

Given the operational distribution $\cI$ of an apparatus $\bA(\bx)$,
the operation $T=\cI(\R)$ is called the {\em nonselective operation}
of apparatus $\bA(\bx)$ and $T^{*}=\cI(\R)^{*}$ is called the {\em
nonselective dual operation} of apparatus $\bA(\bx)$.  The nonselective
operation $T$  is trace-preserving, i.e., 
\begin{equation}
\Tr(T\rh)=\Tr\rh
\end{equation}
for any trace-class operator $\rh$, while the nonselective dual
operation $T^{*}$ is unit-preserving,  i.e., 
\begin{equation}
T^{*}I=I.
\end{equation}

\subsection{Individual quantum state reductions}

It is natural to assume that the output variable $\bx$ can be 
read out with arbitrary precision.  It follows that each instance
of measurement has the output value $\bx=x$.
Let $\rh_{\{\bx=x\}}$ be the state of the system $\bS$
at the time just after the measurement on input state $\rh$
provided that the measurement yields the output 
value $\bx=x$.
The {\em individual quantum state reduction} caused by the apparatus
$\bA(\bx)$  is the state change 
$\rh\mapsto\rh_{\{\bx=x\}}$ for any real number $x$.
The state $\rh_{\{\bx=x\}}$ is called the {\em output
state} given the output $\bx=x$ on input state $\rh$.

For distinction, we shall call the previously defined quantum
state reduction $\rh\mapsto\rh_{\{\bx\in\De\}}$ as the {\em collective
quantum state reduction}.

If $\Pr\{\bx\in\{x\}\|\rh\}>0$, 
the state $\rh_{\{\bx=x\}}$ is determined by the relation
\begin{equation}
\rh_{\{\bx=x\}}=\rh_{\{\bx\in\{x\}\}}.
\end{equation}
However, the above relation determines no $\rh_{\{\bx=x\}}$, 
if the output probability is continuously distributed.  
In order to determine states $\rh_{\{\bx=x\}}$, 
the following mathematical notion was introduced in Ref.~\cite{85CA}.
A family $\{\rh_{\{\bx=x\}}|\ x\in{\bf R}\}$ of states is called a {\em family of
posterior states} for a DL instrument $\cI$ and a {\em prior state}
$\rh$, if it satisfies the following conditions.

(i) The function $x\mapsto\rh_{\{\bx=x\}}$ is Borel measurable.

(ii) For any Borel set $\De$, we have
\begin{equation}
{\cI}(\Delta)\rh
=
\int_{\Delta}\rh_{\{\bx=x\}}\Tr[d\cI(x)\rh].
\label{eq:226a}
\end{equation}

It was shown in Ref.~\cite{85CA} that for any DL instrument $\cI$ and
prior state $\rh$, there exists a {\em family of posterior
states} uniquely, where two families are taken {\em identical}, if they differ
only on a set $\De$ such that $\Tr[\cI(\De)\rh]=0$.

We define the individual quantum state reduction to be the correspondence 
from the input state $\rh$ to the family $\{\rh_{\{\bx=x\}}|\ x\in{\bf R}\}$ 
of posterior states for the operational distribution $\cI$ 
of $\bA(\bx)$ and prior state $\rh$.

According to the above definition, the individual quantum state reduction 
and the collective quantum state reduction are related by
\begin{equation}\label{eq:individual-qsr}
\rh_{\{\bx\in\De\}}
=\frac{1}{\Pr\{\bx\in\De\|\rh\}}\int_{\De}\rh_{\{\bx=x\}}\,\Pr\{\bx\in dx\|\rh\}.
\end{equation}
Thus, the individual quantum state reduction 
and the collective quantum state reduction
are equivalent under \Eq{individual-qsr}.

The operational meaning of 
the individual quantum state reduction is given as
follows. Suppose that a measurement using the apparatus $\bA(\bx)$ on
input state $\rh$  is immediately followed by a measurement using
another apparatus $\bA(\by)$. Then, the joint probability distribution
$\Pr\{\bx\in\De,\by\in\De'\|\rh\}$ of the output variables $\bx$ and $\by$
is given by \Eq{c}.
The conditional probability distribution of $\by$ given $\bx\in\De$ is
defined in probability theory by
\beqa\label{eq:cond-prob}
\Pr\{\by\in\De'|\bx\in\De\}
=\frac{\Pr\{\bx\in\De,\by\in\De'\|\rh\}}{\Pr\{\bx\in\De\|\rh\}}.
\eeqa
However, this definition does not cover the conditional probability
distribution of $\by$ given $\bx=x$, since it may happen that 
$\Pr\{\bx\in\{x\}\|\rh\}=0$ for every $x$.
To avoid this difficulty, in probability theory the conditional probability
distribution of $\by$ given $\bx=x$ is defined as the function
$x\mapsto \Pr\{\by\in\De'|\bx=x\|\rh\}$ satisfying
\beqa\label{eq:3624a}
\lefteqn{
\Pr\{\bx\in\De,\by\in\De'\|\rh\}}\quad\nn\\
&=&
\int_{\De}\Pr\{\by\in\De'|\ \bx=x\|\rh\}\Pr\{\bx\in dx\|\rh\}.
\eeqa
From Eqs.~\eq{individual-qsr} and \eq{cond-prob}, we have the 
following characterization of the individual quantum state reduction,
\beqa\label{eq:3622a}
\Pr\{\by\in\De'|\bx=x\|\rh\}=\Pr\{\by\in\De'\|\rh_{\{\bx=x\}}\}.
\eeqa
Thus, the individual quantum state reduction is determined by 
the conditional probability distribution of the output of 
any succeeding measurement conditional upon the 
individual output.

\subsection{Extendability postulate}

In the previous discussions, under a sole hypothesis,
the mixing law,  we have 
shown that statistical  properties of every apparatus are described
by a DL instrument and that two apparatuses are statistically equivalent 
if and only if they corresponds to the same DL instrument. 
Consequently, the set of statistical equivalence classes of apparatuses 
are considered to be a subset of the set of DL instruments. 
In the above sense, ``apparatus'' denotes a physical system for measurement
and ``DL instrument'' is intended to denote the function of an apparatus 
or to mathematically denote the statistical equivalence class of an apparatus;
of course, we consider that two apparatuses have the same function if and only
if they are statistically equivalent.
However, up to this point, some DL instruments represent statistical equivalence 
classes of apparatuses, but some of them may not. 
In this subsection, we shall eliminate physically irrelevant  DL instruments 
by another physically plausible requirement.

We deal with any apparatus $\bA(\bx)$ as a mathematical
description of a physical system which has a macroscopic
variable $\bx$ to measure a quantum system $\bS$.
However, even if we sufficiently specify the physical entity of the 
measuring apparatus described by $\bA(\bx)$, 
an ambiguity still remains as to what is the system to be measured.
For example, let $\bS'$ be another quantum system which
is remote from both $\bS$ and $\bA(\bx)$.
Then, we always make the composite system $\bS+\bS'$.
Since we identify an observable $A$ of $\bS$ with the
observable $A\otimes I$ of $\bS+\bS'$, the physical apparatus 
measuring an observable $A$ of the system $\bS$ is also considered 
as the one measuring the observable $A\otimes I$ of the extended
system $\bS+\bS'$.  Thus, every real apparatus has the property 
that if it is described to measure a system $\bS$, 
then it is also described to measure the trivially extended system $\bS+\bS'$.
The above consideration naturally leads to the following postulate.

{\bf Extendability postulate:}
{\em For any apparatus $\bA(\bx)$ measuring a system $\bS$
and any quantum system $\bS'$ not interacting with $\bA(\bx)$ 
nor $\bS$,  there exists an apparatus $\bA(\bx')$ measuring system
$\bS+\bS'$  with the following statistical properties:
\beqa\label{eq:ex-1}
\Pr\{\bx'\in\De\|\rh\otimes\rh'\}
&=&
\Pr\{\bx\in\De\|\rh\},\\
(\rh\otimes\rh')_{\{\bx'\in\De\}}&=&\rh_{\{\bx\in\De\}}\otimes
\rh'
\label{eq:ex-2}
\eeqa
for any Borel set $\De$, any state $\rh$ of $\bS$, 
and any state $\rh'$ of $\bS'$.}

In order to obtain a mathematical condition characterizing 
the models satisfying the above requirement, we need the following
mathematical notions.
Let $\bC^{n}$ be the Hilbert space of $n$-dimensional vectors.
Since every linear operator on a finite dimensional space
is bounded and of finite trace, we have $\tc(\bC^{n})=\cL(\bC^{n})$.
Then, the space of trace class operators on the tensor product
Hilbert space $\cH\otimes\bC^{n}$ is decomposed as 
$\tc(\cH\otimes\bC^{n})=\tc(\cH)\otimes\cL(\bC^{n})$.
Thus, any linear transformation $T$ on $\tc(\cH)$
can be extended naturally to the linear transformation $T\otimes id_{n}$
on $\tc(\cH\otimes\bC^{n})$ by
\begin{equation}
(T\otimes id_{n})(\sum_{j}\rh_{j}\otimes \rh'_{j})
=\sum_{j}T(\rh_{j})\otimes\rh'_{j}
\end{equation}
for any $\rh_{j}\in\tc(\cH)$ and $\rh'_{j}\in\cL(\bC^{n})$.
Then, $T$ is called 
{\em completely positive (CP)}, if $T\otimes id_{n}$ maps
positive operators in $\tc(\cH\otimes\bC^{n})$ to positive
operators in  $\tc(\cH\otimes\bC^{n})$ for any positive
integer $n$.
A DL instrument $\cI$ is called  a {\em completely positive (CP)
instrument}, if $\cI(\De)$ is CP for every $\De$.
Completely positive maps on C*-algebras were introduced by
Stinespring \cite{Sti55}, CP operations were introduced by Kraus
\cite{Kra71}, and CP instruments were introduced in Ref.~\cite{84QC}.
For the general theory of CP maps we refer to Takesaki \cite{Tak79}.

\begin{Theorem}
Under the mixing law, the extendability
postulate implies the following requirement:

The operational distribution of every apparatus should be a 
CP instrument.
\end{Theorem}

The proof runs as follows.
Let $\bA(\bx)$, $\bS$, and $\bS'$ be those given
in the extendability postulate.
The state space of $\bS$ is denoted by $\cH$ and 
the state space of $\bS'$ is supposed to be $\bC^{n}$.
It has been proven that under the mixing law, 
every apparatus has its own operational distribution.
Let $\cI$ be the operational distribution of $\bA(\bx)$.
By the extendability postulate, 
there is an apparatus $\bA(\bx')$ satisfying
Eqs.~\eq{ex-1} and \eq{ex-2}.
The mixing law ensures the existence of the 
operational distribution $\cI'$ of the apparatus $\bA(\bx')$.
Then, we have
\beqa
\cI'(\De)(\rh\otimes\rh')
&=&\Pr\{\bx'\in\De\|\rh\otimes\rh'\}(\rh\otimes\rh')_{\{\bx'\in\De\}}\nn\\
&=&\Pr\{\bx\in\De\|\rh\}\rh_{\{\bx\in\De\}}\otimes\rh'\nn\\
&=&[\cI(\De)\rh]\otimes \rh'.
\eeqa
It follows that the operation $\cI'(\De)$ of the extended apparatus
$\bA(\bx')$ associated with $\De$ is represented by  
$\cI'(\De)=\cI(\De)\otimes id_{n}$.  
Then, by the positivity of operation $\cI'(\De)$,
$\cI(\De)\otimes id_{n}$ should be a positive linear
transformation on $\tc(\cH\otimes\bC^{n})$.
Since $n$ is arbitrary, we conclude that $\cI(\De)$
is completely positive, so that the proof is completed.

The transpose operation of matrices in a fix basis is 
a typical example of a positive linear map which is not CP \cite{NC00}.
Let $T$ be a transpose operation on $\tc(\cH)=\cL(\cH)$ 
for $\cH=\bC^{m}$,
and let $\mu$ be any probability measure on $\bR$.
Then, the relation
\beqa
\cI(\De)\rh=\mu(\De)T(\rh)
\eeqa
for any Borel set $\De$ and any operator $\rh$
defines a DL instrument.  
However, since $T$ is not CP, the operation
$\cI(\De)$ is not CP, so the $\cI$ is not a CP instrument.
The extendability postulate implies that there is no 
apparatus corresponding to the above DL instrument.

As a consequence of the whole argument of this section,
we have reached the following conclusion:

{\em For any apparatus $\bA(\bx)$ there exists 
a unique CP instrument $\cI$ 
such that the output probability distribution
and the quantum state reduction are described by
\beqa
\Pr\{\bx\in\De\|\rh\}&=&\Tr[\cI(\De)\rh],\\
\rh\mapsto\rh_{\{\bx\in\De\}}
&=&\frac{\cI(\De)\rh}{\Tr[\cI(\De)\rh]}
\eeqa
for any Borel set $\De$ and any input state $\rh$, where
the second equality assumes
$\Tr[\cI(\De)\rh]>0$.} 

We have posed two plausible requirements,
the mixing law and 
the extendability postulate,  as a set of necessary conditions for
every apparatus to satisfy.  
Under these conditions, we have shown that every apparatus corresponds
uniquely to a CP instrument, called the operational distribution, that determines
the output probability distributions and the quantum state reduction. 
Thus, the problem of determining all the possible quantum 
measurements is reduced to the problem as to 
which CP instrument corresponds to an apparatus. 
This problem will be discussed in the next section and it will be shown  that
every CP instrument corresponds to at least one apparatus.
Thus, the statistical equivalence classes of all the possible measuring
apparatuses are described in one-to-one correspondence by the CP instruments.

\section{Measuring processes\label{se:MP}}

\subsection{Indirect measurement models}

The disturbance on the object caused by a measurement 
can be attributed to an interaction, called the {\em
measuring interaction}, between the object and the apparatus.  
In this section, we shall consider {indirect measurement models}
in which the measuring interactions are subject to the equations of
motions in quantum mechanics \cite{00MN,01OD}
and show that even though the indirect measurement models 
are only a subclass of all the possible quantum measurements, 
every measurement is statistically equivalent to one of indirect 
measurement models. 

Let $\bA(\bx)$ be a measuring apparatus with
macroscopic  output variable $\bx$ to measure the object $\bS$.  
The measuring
interaction turns on at time $t$, the {\em time of measurement}, and turns off at
time $t+\De t$ between  object $\bS$ and apparatus $\bA(\bx)$.  
We assume that {\em the object and the apparatus do not 
interact each other before $t$ nor after $t+\De t$ and that 
the composite system $\bS+\bA(\bx)$ is isolated in the
time interval $(t,t+\De t)$}.  The {\em probe} $\bP$ is
defined to be the minimal part of apparatus $\bA(\bx)$ 
such that the composite system $\bS+\bP$ is isolated in the time interval
$(t,t+\De t)$. By minimality, we naturally assume that probe $\bP$ is a 
quantum system represented by a Hilbert space $\cK$.  
Denote by $U$ the unitary operator on $\cH\otimes\cK$ representing 
the time evolution of $\bS+\bP$ for the time interval $(t,t+\Delta t)$.  

At the time of measurement the object is supposed to
be in an arbitrary input state $\rh$
and the probe is
supposed to be prepared in a fixed state $\si$.
Thus, the composite system $\bS+\bP$ is in the state $\rh\otimes\si$
at time $t$ and in the state $U(\rh\otimes\si)U^{\da}$ at time $t+\De t$. 
Just after the measuring interaction, the object is separated from the apparatus,
and the probe is subjected to a local interaction with the subsequent stages of the
apparatus.  
The last process is assumed to measure an observable $M$, called
the {\em probe observable}, of the probe, and the output is
represented by the value of the output variable $\bx$. 
The above measurement of $M$ is assumed to be {\em local},
in the sense that the measuring apparatus for the measurement of $M$ 
interacts with the probe $\bP$ but does not interact with the system
$\bS$ \cite{01OD}.

The measuring process of the apparatus $\bA(\bx)$ described above
is thus modeled by the state space $\cK$ of the probe $\bP$, 
the initial state $\si$ of $\bP$, 
the time evolution operator $U$ of $\bS+\bP$,
and the probe observable $M$.

In order to develop the theory of measuring processes described 
above, we define an {\em indirect measurement
model} to be any quadruple $(\cK,\si,U,M)$ consisting of 
a Hilbert space $\cK$, 
a density operator $\si$ on $\cK$, a unitary operator $U$ on
$\cH\otimes\cK$, and an observable $M$ on $\cK$. 
An apparatus $\bA(\bx)$ is said to be
{\em described by an indirect measurement model $(\cK,\si,U,M)$}, 
if the measuring process of $\bA(\bx)$ admits 
the above description with the state space $\cK$ of the probe,  the initial
state $\si$ of the probe,  the time evolution operator $U$ of the object plus
probe during the measuring interaction, and the probe observable $M$.
Two indirect measurement models $(\cK,\si,U,M)$ and 
$(\cK',\si',U',M')$ are said to be {\em unitarily equivalent},
if there is a unitary transformation $\cU$ from $\cK$ onto $\cK'$
such that $\si=\cU^{\da}\si'\cU$, 
$U=(I\otimes\cU^{\da})\si'(I\otimes\cU)$, and  
$M=\cU^{\da}M'\cU$.
We shall not distinguish two unitarily equivalent models, 
since they may describe the same physical system.
An indirect measurement model $({\cal K},\si,U,M)$
is called {\em pure}, if $\si$ is a pure state; we shall
write $({\cal K},\si,U,M)=({\cal K},\xi,U,M)$, if
$\rh=\ket{\xi}\bra{\xi}$.

In an indirect measurement model, the role of the measuring
interaction is well characterized as a transducer,
and the subsequent stages as an amplifier.
In the Stern-Gerlach measurement of the
$z$-component of spin,
the object system models the spin-degree of freedom of the 
particle, the probe models the orbital-degrees of freedom of the 
particle, and the probe observable corresponds to the $z$-component
of the linear momentum of the particle.  Moreover, the amplification
process models the free orbital-motion plus the 
interaction with a detector, and the output variable corresponds to the
$z$-coordinate of the position of the detector that captures the
particle.

\subsection{Output probability distributions}

Let $\bA(\bx)$ be an apparatus with
indirect measurement model
$(\cK,\si,U,M)$.
Since the outcome
of this measurement is obtained by the measurement of 
the probe observable $M$ at time $t+\De t$, 
by the BSF for observable $M$ on input state $U(\rh\otimes\si)U^{\da}$
the output probability distribution of $\bA(\bx)$ is determined 
by
\begin{equation}\label{eq:215a}
\Pr\{\bx\in \De\|\rh\}
=\Tr\{[I\otimes E^{M}(\De)]U(\rh\otimes\si)U^{\da}\}.
\end{equation}

By linearity of operators and the trace, it is easy to check that
the output probability distribution of $\bA(\bx)$ satisfies the
mixing law of output probability.
Thus, by Theorem \ref{th:POVM} there exists the POVM  $\Pi $ of 
$\bA(\bx)$.
To determine $\Pi $, 
using the partial trace operation $\Tr_{\cK}$ over $\cK$ 
we rewrite \Eq{215a} as
\begin{equation}\label{eq:020604a}
\Pr\{\bx\in \De\|\rh\}
=\Tr[\Tr_{\cK}\{U^{\da}[I\otimes E^{M}(\De)]U(I\otimes\si)\}\rh].
\end{equation}
Since $\rh$ is arbitrary, comparing Eqs.~\eq{Born-POVM} and
\eq{020604a}, POVM of $\bA(\bx)$ is determined  as
\begin{equation}\label{eq:POVM1}
\Pi (\De)=\Tr_{\cK}\{U^{\da}[I\otimes E^{M}(\De)]U(I\otimes\si)\}
\end{equation}
for any Borel set $\De$.

\subsection{Conditional expectation}

Now we shall introduce a convenient mathematical notion
to deal with such formulas as \Eq{POVM1}.

Let $\cH$ and $\cK$ be two Hilbert spaces and 
let $\si$ be a density operator on $\cK$.
For any $C\in\cL(\cH\otimes\cK)$, we define
the operator $\cE_{\si}(C)\in\cL(\cH)$ by the relation
\beqa
\cE_{\si}(C)=\Tr_{\cK}[C(I\otimes\si)].
\eeqa
The operator $\cE_{\si}(C)\in\cL(\cH)$ is called the
{\em conditional expectation} of $C$ in $\si$.

The conditional expectations have the following properties
easily obtained from the properties of partial trace operation.

(i) For $C=\sum_{i}A_{i}\otimes B_{i}\in\cL(\cH\otimes\cK)$,
\beqa
\cE(C)=\sum_{i}\Tr[B_{i}\si]A_{i}.
\eeqa

(ii) For any $C\in\cL(\cH\otimes\cK)$ and $\rh\in\tc(\cH)$,
\beqa\label{eq:cond-ex-dual}
\Tr[\cE_{\si}(C)\rh]=\Tr[C(\rh\otimes\si)].
\eeqa

(iii) For any $A,D\in\cL(\cH)$ and $B,E\in\cL(\cK)$, 
\beqa\label{eq:cond-ex}
\cE_{\si}[(A\otimes B)C(D\otimes E)]
=A\,\cE_{\si}[(I\otimes B)C(I\otimes E)]D
\eeqa

(iv) The transformation $C\mapsto\cE_{\si}(C)$ is a linear
transformation from $\cL(\cH\otimes\cK)$ to $\cL(\cH)$.

(v) If $C\ge 0$, then $\cE_{\si}(C)\ge 0$.

Mathematically, \Eq{cond-ex-dual} shows that the transformation
$C\mapsto\cE_{\si}(C)$ is the dual of the trace-preserving 
CP map $\rh\mapsto\rh\otimes\si$ from $\tc(\cH)$ to $\tc(\cH\otimes\cK)$.
Thus, $C\mapsto\cE(C)$ is a unit-preserving CP map
from $\cL(\cH\otimes\cK)$ to $\cL(\cH)$.

If $\si$ is a vector state such that $\si=\ket{\xi}\bra{\xi}$, 
we shall write $\cE_{\xi}(C)=\cE_{\si}(C)$
and call it the conditional expectation of $C$ in $\xi$.
In this case, we have
\beqa
\bracket{\ps|\cE_{\xi}(C)|\ps}
=\bracket{\ps\otimes\xi|C|\ps\otimes\xi}
\eeqa
for any $\ps\in\cH$.
Thus, we shall also write
\beqa
\cE_{\xi}(C)=\bracket{\xi|C|\xi}.
\eeqa

From \Eq{POVM1}, the POVM of apparatus $\bA(\bx)$ with
indirect measurement model $(\cK,\si,U,M)$ is the conditional 
expectation of the spectral measure of the observable $M(t+\De t)$
in the state $\si$,
i.e., 
\beqa
\Pi (\De)
&=&\cE_{\si}[E^{M(t+\De t)}(\De)]\\
&=&\cE_{\si}\{U^{\da}[I\otimes E^{M}(\De)]U\}.
\eeqa

\subsection{Quantum state reductions}

Since the composite system $\bS+\bP$ is in the state
$U(\rh\otimes\si)U^{\da}$ at time $t+\De t$, 
it is standard  that the object state at time $t+\De t$
is obtained by tracing out the probe part of that state,
and, in fact, this rule is justified by Postulate IV (Composition rule) 
in Subsection
\ref{ss:IIA}.
Thus, the nonselective state change is determined by
\begin{equation}  \rh\mapsto\rh'=
\Tr_{{\cal K}}[U(\rh\otimes\sigma)U^{\dagger}].
\end{equation}
%where $\Tr_{{\cal K}}$ denotes the partial trace over ${\cal K}$.

In order to determine the quantum state reduction, 
suppose that at time $t+\Delta t$
the observer would locally measure
an arbitrary observable $B$
of the same object ${\bf S}$.  
Let $\bA(\by)$ be an apparatus with output variable $\by$ 
to make a precise measurement of $B$ . Since both the $M$
measurement on $\bP$ and the $B$ measurement on $\bS$ 
at time $t+\Delta t$ are local,  
the joint probability distribution of their
outputs   satisfies the joint probability formula for the
simultaneous measurement of $I\otimes M$ and 
$B\otimes I$ in the state 
$U(\rh\otimes\sigma)U^{\dagger}$ \cite{01OD}.

It follows that the joint output probability distribution of 
 $\bA(\bx)$ and $\bA(\by)$ is given by
\begin{eqnarray}
\lefteqn{{\rm Pr}\{\bx\in\De,\by\in\Delta'\|\rh\}}
\nonumber\\
&=&\Tr\{[E^{B}(\Delta')\otimes E^{M}(\Delta)]
U(\rh\otimes\sigma)U^{\dagger}\}.\quad
\label{eq:215d}
\end{eqnarray}
%\begin{eqnarray}
%{\rm Pr}\{\bx\in\De,\by\in\Delta'\|\rh\}
%&=&\Tr\{[E^{B}(\Delta')\otimes E^{M}(\Delta)]
%U(\rh\otimes\sigma)U^{\dagger}\}.
%\label{eq:215d}
%\end{eqnarray}
Thus, using the partial trace $\Tr_{\cK}$ we have
\begin{eqnarray}\lefteqn{{\rm
Pr}\{\bx\in\De,\by\in\Delta'\|\rh\}}
\nonumber\\
&=&\Tr[E^{B}(\Delta') \Tr_{\cK}\{[I\otimes E^{M}(\Delta)]
U(\rh\otimes\sigma)U^{\dagger}\}].\quad\quad
\label{eq:020604b}
\end{eqnarray}
%\begin{eqnarray}
%{\rm Pr}\{\bx\in\De,\by\in\Delta'\|\rh\}
%&=&\Tr[E^{B}(\Delta') \Tr_{\cK}\{[I\otimes E^{M}(\Delta)]
%U(\rh\otimes\sigma)U^{\dagger}\}].
%\label{eq:020604b}
%\end{eqnarray}
On the other hand, from \Eq{c}
the same joint output probability distribution can be represented by
\begin{eqnarray}
\lefteqn{{\rm Pr}\{\bx\in\De,\by\in\Delta'\|\rh\}}
\quad\nonumber\\
&=&
{\Tr}[E^{B}(\Delta')\rh_{\{\bx\in\De\}}]
{\Pr}\{\bx\in\De\|\rh\}\nn\\
&=&
{\Tr}[E^{B}(\Delta'){\Pr}\{\bx\in\De\|\rh\}\rh_{\{\bx\in\De\}}].
\label{eq:215e}
\end{eqnarray}
%\begin{eqnarray}
%{\rm Pr}\{\bx\in\De,\by\in\Delta'\|\rh\}
%&=&
%{\Tr}[E^{B}(\Delta')\rh_{\{\bx\in\De\}}]
%{\Pr}\{\bx\in\De\|\rh\}\nn\\
%&=&
%{\Tr}[E^{B}(\Delta'){\Pr}\{\bx\in\De\|\rh\}\rh_{\{\bx\in\De\}}].
%\label{eq:215e}
%\end{eqnarray}
Since $B$ and $\De'$ are chosen arbitrarily,
comparing Eqs.~\eq{020604b} and \eq{215e}, we have
\begin{equation}\label{eq:020604c}
{\Pr}\{\bx\in\De\|\rh\}\rh_{\{\bx\in\De\}}=
\Tr_{\cK}\{[I\otimes E^{M}(\Delta)]
U(\rh\otimes\sigma)U^{\dagger}\}
\end{equation}
From \Eq{215a}, the state 
$\rh_{\{\bx\in\De\}}$ is uniquely determined as
%\lefteqn{\rh_{\{\bx\in\De\}}}\quad\nonumber\\
%&=&
%\frac{\Tr_{{\cal K}}\{[I\otimes E^{M}(\Delta)]
%U(\rh\otimes\sigma)U^{\dagger}\}}
%     {\Tr\{[I\otimes
%E^{M}(\Delta)]U(\rh\otimes\sigma)U^{\dagger}\}}
%\label{eq:215f}
%\end{eqnarray}
\begin{eqnarray}
\rh_{\{\bx\in\De\}}
&=&
\frac{\Tr_{{\cal K}}\{[I\otimes
E^{M}(\Delta)]U(\rh\otimes\sigma)U^{\dagger}\}}
     {\Tr\{[I\otimes
E^{M}(\Delta)]U(\rh\otimes\sigma)U^{\dagger}\}}
\label{eq:215f}
\end{eqnarray}
for any Borel set $\Delta$ with ${\rm Pr}\{\bx\in\De\|\rh\}>0$.

The above formula was obtained in Ref.~\cite{84QC}.
It should be noted that Eq.~(\ref{eq:215f}) does not assume such an
illegitimate use of the projection postulate as assuming 
that the composite system ${\bf S}+{\bf P}$ with the outcome
$\bx\in\De$ is in the state
%\begin{eqnarray}
%\lefteqn{\rho^{\bS +\bP}_{\{\bx\in\De\}}}
%\nonumber\\
%&=&
%\frac{[I\otimes E^{M}(\Delta)]U(\rh\otimes\sigma)U^{\dagger}
%[I\otimes E^{M}(\Delta)]}
%{\Tr\{[I\otimes
%E^{M}(\Delta)]U(\rh\otimes\sigma)U^{\dagger}\}}\quad
%\label{eq:215g}
%\end{eqnarray}
\begin{eqnarray}
\rho^{\bS +\bP}_{\{\bx\in\De\}}
&=&
\frac{[I\otimes E^{M}(\Delta)]U(\rh\otimes\sigma)U^{\dagger}
[I\otimes E^{M}(\Delta)]}
{\Tr\{[I\otimes E^{M}(\Delta)]U(\rh\otimes\sigma)U^{\dagger}\}}
\qquad\label{eq:215g}
\end{eqnarray}
just after the measurement.
It is true that the state
$\rho^{\bS +\bP}_{\{\bx\in\De\}}$ leads to the same 
conclusion by defining $
\rh_{\{\bx\in\De\}}=\Tr_{\cK}[\rho^{\bS +\bP}_{\{\bx\in\De\}}]$,
but such an assumption is by no means correct,
since for any partition $\Delta=\Delta'\cup\Delta''$ the state 
$\rho^{{\bf S}+{\bf P}}_{\{\bx\in\De\}}$ should be 
a mixture of
$\rho^{{\bf S}+{\bf P}}_{\{\bx\in\De'\}}$ and 
$\rho^{{\bf S}+{\bf P}}_{\{\bx\in\De''\}}$ 
but this is not the case
for Eq.~(\ref{eq:215g}).
It is a significant merit of our derivation of Eq.~(\ref{eq:215f})
to make no assumptions on the state of 
the composite system $\bS+\bP$ after the measurement.
%of probe observable.
 
\subsection{Operational distributions}

In the previous subsection, we have confined our attention
to the case where the measurement using 
$\bA(\bx)$ is followed
by a precise measurement of an {\em observable}.
Now, we generally suppose that at time $t+\Delta t$
the observer would  locally measure the same system $\bS$
by an arbitrary apparatus  $\bA(\by)$.
We shall show that the joint output probability distribution of 
$\bA(\bx)$ and $\bA(\by)$ satisfies the mixing law.
Let $\Pi _{\by}$ be the POVM of $\bA(\by)$.
Under the condition that the measurement of $\bA(\bx)$ leads to
the outcome $\bx\in\De$, the state at time $t+\De t$ 
is  $\rh_{\{\bx\in\De\}}$.  
It follows from \Eq{020604c} that 
the joint output probability distribution is given by
\begin{eqnarray}
\lefteqn{{\rm Pr}\{\bx\in\De,\by\in\Delta'\|\rh\}}
\nonumber\\
&=&\Pr\{\by\in\De'\|\rh_{\{\bx\in\De\}}\}\Pr\{\bx\in\De\|\rh\}
\nn\\
&=&\Tr[\Pi _{\by}(\De')\Pr\{\bx\in\De\|\rh\}\rh_{\{\bx\in\De\}}]
\nn\\
&=&
\Tr\{[\Pi _{\by}(\Delta')\otimes E^{M}(\Delta)]
U(\rh\otimes\sigma)U^{\dagger}\}.
\label{eq:020522a}
\end{eqnarray}
%\begin{eqnarray}
%{\rm Pr}\{\bx\in\De,\by\in\Delta'\|\rh\}
%&=&\Pr\{\by\in\De\|\rh_{\{\bx\in\De\}}\}\Pr\{\bx\in\De\|\rh\}
%\nn\\
%&=&\Tr[\Pi _{\by}(\De')\Pr\{\bx\in\De\|\rh\}\rh_{\{\bx\in\De\}}]
%\nn\\
%&=&
%\Tr\{[\Pi _{\by}(\Delta')\otimes E^{M}(\Delta)]
%U(\rh\otimes\sigma)U^{\dagger}\}.
%\label{eq:020522a}
%\end{eqnarray}

By linearity of operators and the trace, it is easy to check that
the joint output probability distribution of $\bA(\bx)$ and
$\bA(\by)$ satisfies the mixing law.  
Thus, by Theorem \ref{th:DL instrument} there exists 
the operational distribution $\cI$ of $\bA(\bx)$.

By \Eq{020604c} the operational distribution $\cI$
is determined  by
\begin{equation}\label{eq:020609a}
\cI(\De)\rh=\Tr_{{\cal K}}\{[I\otimes E^{M}(\Delta)]
U(\rh\otimes\sigma)U^{\dagger}\}
\end{equation}
for any Borel set $\De$ and any state $\rh$.
From the above relation, it is easy to see that
$\cI$ satisfied the complete positivity;
as an alternative characterization, 
it is well-know that a linear transformation $T$ on
$\tc(\cH)$ is completely positive if and only if 
\begin{equation}
\sum_{ij}
\bracket{\xi_{i}|T(\rh_{i}^{\da}\rh_{j})|\xi_{j}}\ge 0
\end{equation}
for any finite sequences $\xi_{1},\ldots,\xi_{n}\in\cH$ and 
$\rh_{1},\ldots,\rh_{n}\in\tc(\cH)$ \cite{Tak79}.
Thus, we conclude that 
{\em the operational distribution of any apparatus
with indirect measurement  model
$({\cal K},\sigma,U,M)$ is a CP instrument}.

Let $\cI$ be the operational distribution of an apparatus $\bA(\bx)$
with indirect measurement model $(\cK,\si,U,M)$.
From \Eq{020609a}, the nonselective operation $T=\cI(\R)$ 
is represented as
\begin{equation}\label{eq:020609b}
T\rh=\Tr_{{\cal K}}[U(\rh\otimes\sigma)U^{\dagger}].
\end{equation}
For any bounded operator $A\in\cL(\cH)$, 
trace-class operator $\rh\in\tc(\cH)$,
and Borel set $\De\in\cB(\R)$, we have
\beqa
\lefteqn{\Tr[A\cI(\De)\rh]}\nn\\
&=&
\Tr(A\Tr_{{\cal K}}\{[I\otimes E^{M}(\Delta)]
U(\rh\otimes\sigma)U^{\dagger}\})\nn\\
\label{eq:020609b1}
&=&
\Tr\{[A\otimes E^{M}(\Delta)]U(\rh\otimes\sigma)U^{\dagger}\}\nn\\
\label{eq:020609b2}
&=&
\Tr\{U^{\dagger}[A\otimes E^{M}(\Delta)]U(\rh\otimes\sigma)\}\nn\\
\label{eq:020609b3}
&=&
\Tr(\Tr_{{\cal K}}\{U^{\dagger}[A\otimes E^{M}(\Delta)]
U(I\otimes\sigma)\}\rh).\quad
\label{eq:020609b4}
\eeqa
Thus, from Eqs.~\eq{dual-operation} and \eq{020609b4} we have
\beqa\label{eq:020609c}
\lefteqn{
\Tr\{[\cI(\De)^{*}A]\rh\}}\nn\\
&=&\Tr(\Tr_{{\cal K}}\{U^{\dagger}[A\otimes
E^{M}(\Delta)] U(I\otimes\sigma)\}\rh)\nn\\
&=&\Tr(\cE_{\si}\{U^{\dagger}[A\otimes
E^{M}(\Delta)] U\}\rh).\quad
\eeqa
Since $\rh$ is arbitrary, we have
\begin{equation}
\cI(\De)^{*}A
=\cE_{\sigma}\{U^{\dagger}[A\otimes E^{M}(\Delta)]U\}
\end{equation}
for  any bounded operator $A\in\cL(\cH)$.
In particular, the POVM $\Pi $ of $\bA(\bx)$ satisfies
\begin{equation}
\Pi (\De)=\cE_{\sigma}\{U^{\dagger}[I\otimes E^{M}(\Delta)]U\}
\end{equation}
for any Borel set $\De$
and the nonselective dual operation $T^{*}$ satisfies
\begin{equation}\label{eq:dual-nonselective-operation}
T^{*}A=\cE_{\sigma}\{U^{\dagger}[A\otimes I]U\}
\end{equation}
for any bounded operator $A$.

\subsection{Canonical Measurements\label{se:5}}

In this section, we shall consider a model which has been considered to
describe a typical  measuring process for an arbitrary observable
\cite{Boh51,93CA}. Let $\bA(\bx)$ be an apparatus to measure an
observable  
$A$ of the object $\bS$ described by a Hilbert space $\cH$.
The measuring process of $\bA(\bx)$ is described by an
indirect measurement model $(\cK,\si,U,M)$ as follows.
The probe is modeled by a mass of one degree of freedom
with position $\hq$ and momentum $\hp$,
so that the Hilbert space $\cK$ is the $L^{2}$ space
of wave functions on $\R$, i.e., $\cK=L^{2}(\R)$.
The measuring interaction is turned on in the time
interval $(t,t+\De t)$ that couples $A$ and $\hp$,
so that the total Hamiltonian in the time interval $(t,t+\De t)$
is given by
\beqa
H_{\bS+\bP}=H_{\bS}\otimes I+I\otimes H_{\bP}
+KA\otimes\hp,
\eeqa
where $H_{\bS}$ and $H_{\bP}$ are free Hamiltonians 
of the object and the probe, respectively, and $K$ is the
coupling constant. 
We suppose that the coupling is so strong that we can neglect
the free evolutions and the duration $\De t$ is so small as 
to satisfy $K\De t=1$.
Thus, the time evolution of the composite system $\bS+\bP$
during the measuring interaction is  given by
\beqa
U=\exp\left(\frac{-i}{\hbar}A\otimes\hp\right).
\eeqa
After the measuring interaction, the apparatus makes a precise
measurement of the position $\hq$ of the probe to output 
the measurement result.
Let $\xi$ be the state vector of the probe at the time of the 
measurement.  Then, the above apparatus is modeled by
the indirect measurement model $\cM(A,\xi)$ defined by
\beqas
\cM(A,\xi)=\left(
L^{2}(\R),\xi,\exp(-iA\otimes\hp/\hbar),\hq
\right).
\eeqas
We shall call this model the {\em canonical model} with observable
$A$ and probe state $\xi$.
In what follows,
we shall denote its operational distribution
by $\cI_{\xi}$ and its POVM by
$\Pi _{\xi}$, respectively.  

The Schr\"odinger equation for the wave function  
$\Ps_{t+\ta}\in\cH\otimes L^{2}(\bR)$ in
the time interval $t<t+\ta<t+\De t$ becomes
\beqa  
\frac{\partial \Psi _{t+\ta}(q) }{\partial  \ta}
=  - K{\hA}\frac{\partial  \Psi _{t+\ta}(q)}{\partial q}.
\llabel{eq:5.2}
\eeqa
Now assume the initial condition
\beqa     \Psi _{t}(q) = \xi(q)\ps ,     \llabel{eq:5.3}
\eeqa
where $\ps\in{\cH}$ is a state vector of the measured system,
and the vector valued function $\xi(q)\ps$ represents
the tensor product $\ps\otimes\xi$ in $\cH\otimes
L^{2}(\bR)$.
The solution of the Schr\"odinger equation is given by
\beqa     \Ps_{t+\ta}(q)=\xi(qI-\ta K\hA)\ps,\llabel{eq:5.5}
\eeqa
where $I$ is the identity operator on $\cH$.
For $\ta = \De t$, we have
\beqa\label{eq:3620a}
     \Psi _{t+\De t }(q) = \xi (qI - {\hA})\ps .
\llabel{eq:5.6}
\eeqa

In order to determine the operational distribution of this
measurement, we first obtain the following useful general result.

\begin{Theorem}\label{th:3627a}
For any unitary operator $U$ on $\cH\otimes L^{2}(\R)$,
the indirect measurement model $(L^{2}(\R),\xi,U,\hq)$ has
the operational distribution $\cI$ determined by
\beqa\label{eq:3627a}
\cI(\De)\ket{\ps}\bra{\ps} 
= \int_{\De}
\ket{U(\ps\ot  \xi)(q)}\bra{U(\ps \ot \xi)(q)}\,dq
\eeqa
for any input state $\ps$ and Borel set $\De$.
\end{Theorem}

A formal proof using the Dirac notation runs as follows.
Let $\ps \in {\cH}$ and $\De \in
{\cB}({\bR} )$.  Then, we have
\begin{eqnarray*} 
\lefteqn{\cI(\De)\ket{\ps}\bra{\ps}}\quad\\ 
&=& \Tr_{\cK}\{[I\ot  E^{\hq}(\De)]
|U(\ps\ot  \xi)\> \< U(\ps \ot \xi)|\} \\
&=&
\Tr_{\cK}\{[I\ot \int_{\De}\ket{q}\bra{q}\,dq]
|U(\ps\ot  \xi)\> \< U(\ps \ot \xi)|\}\\
&=& \int_{\De}
\Tr_{\cK}\{I\ot\ket{q}\bra{q}
|U(\ps\ot  \xi)\> \< U(\ps \ot \xi)|\}\,dq\\
&=& \int_{\De}
\bracket{q|U(\ps\ot  \xi)}\bracket{U(\ps \ot \xi)|q}\,dq\\
&=&
\int_{\De}
\ket{U(\ps\ot  \xi)(q)}\bra{U(\ps \ot \xi)(q)}\,dq.
\end{eqnarray*}
Thus, we obtain \Eq{3627a}.

     The statistics of the canonical model $\cM(A,\xi)$ is determined by the 
operational distribution $\cI_{\xi}$. 
From \Eq{3620a}, 
\begin{eqnarray*} 
\lefteqn{
\ket{U(\ps\ot  \xi)(q)}\bra{U(\ps\ot  \xi)(q)}}\quad\nn\\
&=&
\ket{\xi (qI - {\hA})\ps}\bra{\xi (qI - {\hA})\ps}\\
&=&
\xi (qI - {\hA})\ket{\ps}\bra{\ps}\xi (qI - {\hA})^{\da}
\end{eqnarray*}
for any $\ps \in {\cH}$, and hence from \Eq{3627a},
\beqa
     {\cI}_\xi (\De)\ket{\ps}\bra{\ps}  
= \int _\De \xi (qI -
{\hA})\ket{\ps}\bra{\ps} 
\xi (qI - {\hA})^{\da}\,dq.
\eeqa
By linearity and continuity, we obtain
\beqa
     {\cI}_\xi (\De)\rho = \int _\De \xi (qI - {\hA})\rho
\,\xi (qI - {\hA})^{\da}\,dq.
\llabel{eq:5.8}
\eeqa
Then, the
dual operational distribution is given by
\beqa\cI_{\xi}(\De)^{*}X = \int _\De \xi (qI - {\hat
A})^{\da}X\xi (qI - {\hA}) \,dq,
\llabel{eq:5.7}
\eeqa
where $X$ is an arbitrary bounded operator. 
The associated POM $\Pi _{\xi}$ is given by
\beqa
\Pi _{\xi}(\De)=\int_{\De}|\xi(qI-\hA)|^{2}\,dq.
\llabel{eq:5.9}
\eeqa
It follows that the output probability distribution is given by
\beqa
\Pr\{\bq\in\De\|\rh\} = \int_{\De}\Tr[|\xi(qI-\hA)|^{2}\rh]\,dq
\eeqa
for any input state $\rh$. 
From \Eq{5.8}, the output state given the output $\bq=q$
is obtained as 
 \beqa
\rh_{\{\bq=q\}}=\frac{\xi(qI-\hA)\rh\,\xi(qI-\hA)^{\da}}
        {\Tr[|\xi(qI-\hA)|^{2}\rh]}.          
\eeqa
If the input state is a vector state $\rh=\ket {\ps}\bra{\ps}$,
we also have
\beqa
\Pr\{\bq\in\De\|\ps\} = \int_{\De}\|\xi(qI-\hA)\ps\|^{2}\,dq
\eeqa
and the output state is also 
the vector state $\rh_{\{\bq=q\}}=
\ket{\ps_{\{\bq=q\}}}\bra{\ps_{\{\bq=q\}}}$
such that
\beqa
\ps_{\{\bq=q\}}
=\frac{\xi (qI - {\hA})\ps}{\|\xi (qI - {\hA})\ps\|}.
\eeqa

By the function calculus of self-adjoint operator $A$, we have
\beqa
\lefteqn{\int_{\De}|\xi(qI-\hA)|^{2}\,dq}\quad\nn\\
&=&
\int_{\R}dE^{A}(\la)\int_{\R}\ch_{\De}(q)|\xi(q-\la)|^{2}\,dq,
\eeqa
where $\ch_{\De}$ is the characteristic function of the 
Borel set $\De$, i.e., $\ch_{\De}(x)=1$ if $x\in\De$ and
$\ch_{\De}(x)=0$  if $x\not\in\De$.
Let $f(q)=|\xi (-q)|^2$.
From \Eq{5.9}, we have
\beqa  \Pi _{\xi }(\De) = \int _{\bR} (f*\ch_{\De})(\la)\,dE^{A}(\la),    
\llabel{eq:5.20}
\eeqa
where $f*\ch_\De$ is the convolution, i.e., 
\beqa    
(f*\ch_\De)(\la) = \int _{\bR}f(q)\ch_{\De}(\la-q)\,dq.
\eeqa
Thus, if the initial state $\xi$ of the probe goes to the position
eigenstate $\ket{\hq=0}$, the initial position density function
$|\xi(q)|^{2}$ and $f(q)$ approaches to the Dirac delta function,
so that the effect $\Pi _{\xi}(\De)$ approaches to the spectral 
projection $E^{A}(\De)$ of the observable $A$.
Similarly, if $\xi$ goes to the position eigenstate, the output
state $\rh_{\{\bq=q\}}$ goes to the eigenstate of the observable
$A$ corresponding to the output of measurement.
Thus, the model $\cM(A,\xi)$ describes an approximately
precise measurement of $A$ that leaves the object
in an approximate eigenstate of $A$ corresponding to
the output. 
For the notion of approximate eigenvectors,
see Halmos \cite{Hal51}.
For the detailed discussion on the statistical properties of
the model $\cM(A,\xi)$, we refer to Ref.~\cite{93CA}. 

 \subsection{Realizability postulate}

In the preceding section, we have considered the requirements
that every measuring apparatus should satisfy.
However, no postulates were posed as to
what measuring apparatus exists, 
except for Postulate II (Schr\"{o}dinger equation) 
requiring that for any observable there is
at least one apparatus to make a precise measurement of that
observable. 

Here, we introduce a postulate that allows to construct another
measuring apparatus from the apparatus allowed by Postulate II
(Schr\"{o}dinger equation).

{\bf Realizability postulate.}
{\em For any indirect measurement model $(\cK,\si,U,M)$, 
there is an apparatus $\bA(\bx)$ described by $(\cK,\si,U,M)$.}

From the above postulate and 
the statistics of an apparatus with indirect measurement model,
we conclude that {\em for any indirect measurement model
$(\cK,\si,U,M)$, there is an apparatus with the operational
distribution $\cI$ such that 
\begin{equation}%\label{eq:020609a}
\cI(\De)\rh=\Tr_{{\cal K}}\{[I\otimes E^{M}(\Delta)]
U(\rh\otimes\sigma)U^{\dagger}\}
\end{equation}
for any input state $\rh$.}

The above postulate is justified by the assumption
that  our quantum systems obey no superselection rules.
The argument runs as follows.
By our assumption, 
every observable $A$ admits a precise measurement, 
so that we can assume that there is at least one plausible 
model of a measuring apparatus for the measurement of $A$.
Although this is related to a long standing controversy on
the measurement problem, for the simplicity of the coupling, 
the model $\cM(\xi,A)$ has been considered be the first one 
to be plausible \cite{vN32,Boh51}.
Now, we shall argue that the realizability of any other indirect 
measurement model is, in principle, as feasible 
as the realizability of $\cM(\xi,A)$.
From the negation of any nontrivial superselection rules, any self-adjoint
operator corresponds to an observable and any density operator
corresponds to a state, so that we can prepare $\bP$ in $\si$ 
and measure $M$ within a given experimental error limit.
Thus, we have only to show that the unitary operator $U$ is
realizable.  Since any unitary operator can be represented by an
exponential of some observable, we can find an observable $A$
and a parameter $\theta$ such that $U=e^{-i\theta A}$.
Then, in order to realize the model $(\cK,\si,U,M)$, we can
follow the following steps:

(i) Prepare the probe $\bP$ in the state $\si$ at time $t$.

(ii) Prepare the model $\cM(A,\xi)$ in the state $\xi$ near
the momentum eigenstate $\ket{\theta}=\ket{\hp=\theta}$ at time $t$.

(iii) Couple the composite system $\bS+\bP$ to the model
$\cM(A,\xi)$.

(iv) At the time $t+\De t$, the coupling with the model $\cM(A,\xi)$
is turned off, and the observer measures the probe observable $M$.

Now, let $\rh$ be the input state to the model $(\cK,\si,U,M)$.
If the state $\xi$ were such that $\xi=\ket{\theta}$,
the coupling between the composite system $\bS+\bP$ and the model
$\cM(A,\xi)$ changes the state of the $\bS+\bP$ from
$\rh\otimes\si$ to
$U(\rh\otimes\si)U^{\da}$, by the relation
\beqas
\lefteqn{
\exp(-iA\otimes\hp)
(\rh\otimes\si\otimes\ket{\theta}\bra{\theta})
\exp(-iA\otimes\hp)^{\da} }\quad\nn\\
&=&
\exp(-i\theta A)(\rh\otimes\si)\exp(-i\theta
A)^{\da}\otimes\ket{\theta}\bra{\theta}\\
&=&
U(\rh\otimes\si) U^{\da}\otimes\ket{\theta}\bra{\theta}.
\eeqas
Thus, the above procedure realizes the model $(\cK,\si,U,M)$
within the given error limit, if the state preparation is 
sufficiently near to the eigenstate $\ket{\hp=\theta}$.
Thus, any indirect measurement model can be realized, in principle,
by a physical system under a given unit system and in a given
experimental error limit.
This supports the realizability postulate.

In the preceding subsection, we concluded that 
the operational distribution of any apparatus
with indirect measurement  model
$({\cal K},\sigma,U,M)$ is a CP instrument.  
The converse of this assertion was proven by Ref.~\cite{83CR,84QC}
as follows.

\begin{Theorem}[Realization Theorem]
\label{th:realization}
For any CP instrument $\cI$ there exists a pure indirect measurement
model $(\cK,\xi,U,M)$ satisfying 
\beqa
\cI(\De)\rh&=&\Tr_{{\cal K}}\{[I\otimes E^{M}(\Delta)]
U(\rh\otimes\ket{\xi}\bra{\xi})U^{\dagger}\}\quad\\
\cI(\De)^{*}X
&=&\cE_{\xi}\{U^{\dagger}[X\otimes E^{M}(\Delta)]U\}
\eeqa
for any state $\rh$ and observable $X$.
\end{Theorem}

The above theorem has the following two significant corollaries.

\begin{Theorem}
For any POVM $\Pi$ there exists a pure indirect measurement
model $(\cK,\xi,U,M)$ satisfying 
\begin{equation}\label{eq:realization-POVM}
\Pi(\De)=\cE_{\xi}\{U^{\dagger}[I\otimes E^{M}(\Delta)]U\}
\end{equation}
for any Borel set $\De$.
\end{Theorem}
Proof runs as follows.
Note that given POVM $\Pi$ and any fixed state $\rh_{0}$,
the relation 
\beqa
\cI(\De)\rh=\Tr[\Pi(\De)\rh]\rh_{0}.
\eeqa
defines a CP instrument $\cI(\De)$ with 
\beqa
\cI(\De)^{*}X=\Pi(\De)\rh_{0}(X).
\eeqa
Then, by the relation $\Pi(\De)=\cI(\De)^{*}I$,
the assertion follows immediately from Theorem \ref{th:realization}.

\begin{Theorem}
For any trace-preserving CP map $T$ on $\tc(\cH)$ 
there exists a pure indirect measurement model $(\cK,\xi,U,E)$ 
with projection $E$ satisfying 
\beqa
T\rh&=&\Tr_{{\cal K}}\{U(\rh\otimes\ket{\xi}\bra{\xi})U^{\dagger}\},\\
T^{*}X&=&\cE_{\xi}[U^{\dagger}(X\otimes I)U]
\eeqa
for any Borel set $\De$.
\end{Theorem}
This representation was also given by Kraus \cite{Kra83} independently.
Proof runs as follows.
Note that given trace-preserving CP map $T$ and any fixed fixed probability measure
$\mu$,
the relation 
\beqa
\cI(\De)\rh=\mu(\De)T(\rh).
\eeqa
defines a CP instrument $\cI(\De)$ with 
\beqa
\cI(\De)^{*}X=\mu(\De)T^{*}(X).
\eeqa
Then, by the relation $T=\cI(\bR)$,
the assertion follows immediately from Theorem \ref{th:realization}.

We summarize the results.

\begin{Theorem}
The operational distribution of any apparatus
with indirect measurement model is a CP instrument,
and conversely every CP instrument is obtained in this
way with a pure indirect measurement model.
\end{Theorem}

From the realization theorem,  every CP instrument $\cI$ 
has a pure indirect measurement model $(\cK,\xi,U,M)$.
In this case, 
from Eqs.~\eq{dual-operation} and \eq{020609b3} we have
\beqas
\lefteqn{\bracket{\ps|\cI(\De)^{*}A|\ps}}\nn\\
&=&
\Tr\{[\cI(\De)^{*}A]\ket{\ps}\bra{\ps}\}\\
&=&
\Tr\{U^{\dagger}[A\otimes
E^{M}(\Delta)]
U(\ket{\ps}\bra{\ps}\otimes\ket{\xi}\bra{\xi})\}\\
&=&
\bracket{\ps\otimes\xi|U^{\dagger}[A\otimes
E^{M}(\Delta)]U|\ps\otimes\xi}.
\eeqas
Since $\ps\in\cH$ is arbitrary, we have
\begin{equation}
\cI(\De)^{*}A=\bracket{\xi|U^{\dagger}[A\otimes
E^{M}(\Delta)]U|\xi}.
\end{equation}
Let $V$ be the linear transformation from $\cH$ to
$\cK$ defined by
\begin{equation}
V\ps=U(\ps\otimes\xi)
\end{equation}
for all $\ps\in\cH$.
Then, we have 
\begin{equation}
V^{\da}V=I.
\end{equation} 
Now, we have the following useful representation applied to every
CP instruments:
\begin{equation}\label{eq:CP instrument}
\cI(\De)^{*}A=V^{\dagger}[A\otimes
E^{M}(\Delta)]V
\end{equation}
for any $A\in\cL(\cH)$ and $\De\in\cB(\De)$.

Under the realizability postulate, any CP instrument 
has a corresponding apparatus.
From the three postulates discussed above,
we conclude that
the set of statistical equivalence classes of apparatuses is in
one-to-one correspondence with the set of CP instruments.

We have also another useful conclusion: {\em Any apparatus
is statistically equivalent to an apparatus with indirect measurement
model}.  Thus, when we discuss statistical properties of all the 
possible measurements, we can assume without any loss of generality
that the apparatus under consideration has an indirect measurement
model.

\section{Noise in Measurements\label{se:Noise}}

\subsection{Measurements of observables} 

Let $\bA(\bx)$ be a measuring apparatus 
%for the object $\bS$ described by
with indirect measurement model $(\cK,\si,U,M)$.
Let $A$ be an observable of the object $\bS$.
As defined previously, $\bA(\bx)$ precisely measures
$A$ if and only if $\bA(\bx)$ satisfies the BSF
for observable $A$ on every input state.

In order to clarify the meaning of the above definition, 
let us examine the case where observable $A$ has a complete
orthonormal basis of eigenvectors.  Then,  we 
can write
\begin{equation}\label{eq:discrete}
A=\sum_{n,\nu}a_{n}\ket{a_{n},\nu}\bra{a_{n},\nu},
\end{equation}
where $a_{n}$ varies over all eigenvalues and $\nu$ is the
degeneracy parameter. 
From \Eq{BSF}, it is obvious that if $\bA(\bx)$ 
precisely measures $A$, then $\bA(\bx)$ outputs $a_{n}$ with probability one 
on input state $\ket{a_{n},\nu}$.
In what follows, we shall show that the converse is also true.
Suppose that $\bA(\bx)$ outputs $a_{n}$ with probability one on input state
$\ket{a_{n},\nu}$ for all $n$ and $\nu$. 
Then we have
\begin{equation}
\Pr\{\bx=a_{n}\|\,\ket{a_{m},\nu}\}=\de_{m,n},
\end{equation}
so that the  POVM $\Pi $ of $\bA(\bx)$ satisfies
\begin{equation}
\bracket{a_{m},\nu|\Pi \{a_{n}\}|a_{m},\nu}=\de_{m,n}.
\end{equation}
Consequently,
\begin{equation}
\Pi \{a_{n}\}\ket{a_{m},\nu}=\de_{m,n}\ket{a_{m},\nu}.
\end{equation}
Hence, we have
\begin{equation}
\Pi \{a_{n}\}=\sum_{\nu}\ket{a_{n},\nu}\bra{a_{n},\nu}
\end{equation}
for all $n$.
Thus, we conclude 
\begin{equation}
\Pi (\De)=E^{A}(\De)
\end{equation}
for all Borel set $\De$, so that apparatus $\bA(\bx)$ 
precisely measures observable $A$.
Thus, {\em apparatus $\bA(\bx)$ precisely measures 
observable $A$ if and only if
it outputs the value of $A$ whenever the object has a
definite value of $A$ just before the measurement.}

\subsection{Noise in direct measurements}

Measurement noise should be defined to be the difference between
the true value of the quantity to be measured and the output from
the measuring apparatus.
This is meaningful in classical mechanics, 
but there is a difficulty in quantum mechanics,  
since we cannot always expect that the definite true value exists.
However, this does not mean that we cannot define the
average amount of noise of the measurement in a given state.
In fact, if we can identify the noise with a physical quantity,
we can describe the statistical properties of the noise even in quantum
mechanics . 
We shall call this physical quantity the noise operator.

In this subsection, we consider a case where the noise operator can
be determined easily.
We suppose that in order to measure an observable $A$ 
in a given state, 
the observer actually make a precise measurement of
another observable $X$, the meter observable, 
in the same state.  
In this case, it is natural to define the noise operator to be the
observable
\beqa
N(A)=X-A.
\eeqa
Accordingly, the root-mean-square (rms) noise of this measurement in the
input state $\ps$ should be defined to be
\beqa
\ep(A)=
\bracket{\ps|N(A)^{2}|\ps}^{1/2}=
\bracket{\ps|(X-A)^{2}|\ps}^{1/2}.
\eeqa
The above formula is easily rewritten as
\beqa\label{eq:geometric1}
\ep(A)=\|X\ket{\ps}-A\ket{\ps}\|,
\eeqa
and hence the rms noise $\ep(A)$ has properties of
distance between two vectors $A\ket{\ps}$ and $X\ket{\ps}$. 

If the observable $A$ has a definite value $a$ in the state $\ps$,
i.e., $A\ket{\ps}=a\ket{\ps}$, we have
\beqa\label{eq:definite}
\ep(A)&=&\bracket{\ps|(X-a)^{2}|\ps}^{1/2}\\
&=&
\left(\int_{\R}(x-a)^{2}\bracket{\ps|dE^{X}(x)|\ps}\right)^{1/2},
\eeqa
and hence $\ep(A)$ is the root-mean-square of the difference
between the output $x$ and the true value $a$.

Let $\bracket{A}$, $\bracket{X}$, $\si(A)$, and $\si(X)$ be 
the means and the standard deviations of observables
$A$ and $X$, respectively, in state $\ps$. 
Then, we have $\si(A)=\|A\ket{\ps}-\bracket{A}\ket{\ps}\|$
and so on.
From the triangular inequality for the distance between vectors,
we have 
\beqa
\|X\ket{\ps}-\bracket{X}\ket{\ps}\|
&\le&\|X\ket{\ps}-A\ket{\ps}\|+
\|A\ket{\ps}-\bracket{A}\ket{\ps}\|\nn\\
& &+\|\bracket{A}\ket{\ps}-\bracket{X}\ket{\ps}\|
\label{eq:triangular}
\eeqa
Thus, the geometric inequality \Eq{triangular} implies 
the statistical inequality
\beqa\label{eq:triangular2}
\si(X)&\le&\ep(A)+\si(A)+|\bracket{X}-\bracket{A}|.
\eeqa
From an analogous inequalities for vectors, we have
\beqa\label{eq:triangular3}
\si(A)&\le&\ep(A)+\si(X)+|\bracket{X}-\bracket{A}|,\\
\ep(A)&\le&\si(A)+\si(X)+|\bracket{X}-\bracket{A}|.
\label{eq:triangular4}
\eeqa
From the above, we have
\beqa\label{eq:triangular5}
|\si(X)-\si(A)|\le\ep(A)+|\bracket{X}-\bracket{A}|,
\eeqa
and hence the increase and decrease of the standard deviation
of the output from the standard deviation of the measured 
observable in the input state is bounded from above
by the rms noise plus the bias, the difference
of their means.

If $A$ has a definite value and the output is unbiased, i.e., 
$A\ket{\ps}=\bracket{X}\ket{\ps}$, from Eqs.~\eq{triangular2} and
\eq{triangular4} we have
\beqa\label{eq:definite-unbias}
\ep(A)=\si(X).
\eeqa
Thus, the rms noise in this case is identical with the fluctuation of 
the meter observable.

If the output is constant, i.e., $X=x_{0} I$, from \Eq{triangular4}
we have
\beqa\label{eq:definite-unbias2}
\ep(A)\le\si(A)+|x_{0}-\bracket{A}|.
\eeqa
This inequality already shows that Heisenberg's 
noise-disturbance uncertainty relation does not cover
all the possible ways of measuring the same observable $A$.
In fact, suppose that in order to measure the position observable 
$Q$ in a state $\ps$,
the observer actually make a  precise measurement of a constant
observable $X=x_{0}I$.
Then, this measurement can be done without disturbing 
any observables, in particular, the momentum $P$.
However, the rms noise of this measurement
is bounded by the finite number $\si(Q)+|x_{0}-\bracket{Q}|$
for any state $\ps$ with $\|Q\ket{\ps}\|<\infty$.
Thus, for this measurement the product of the root-mean-square
noise and the root-mean-square disturbance vanishes uniformly
over all states $\ps$ in the domain of the operator $Q$.

\subsection{Noise in indirect measurements}

Let  $\bA(\bx)$  be an apparatus with indirect measurement
model $(\cK,\si,U,M)$. 
We suppose that the apparatus $\bA(\bx)$ is used for
measuring an observable $A$ in the state $\rh$ at time $t$.
In the Heisenberg picture with the original state $\rh\otimes\si$, 
we write 
$A^{\In}=A\otimes I$,
$M^{\In}=I\otimes M$,
$A^{\Out}=U^{\da}(A\otimes I)U$,
and $M^{\Out}=U^{\da}(I\otimes M)U$.
In this subsection,
for any observable $C$ of $\bS+\bP$, the mean value 
and the standard deviation of $C$ in
state $\rh\otimes\si$ is denoted by $\bracket{C}$
and $\si(C)$, respectively, i.e.,
\beqa
\bracket{C}&=&\Tr[C(\rh\otimes\si)],\\
\si(C)&=&\Tr[(C-\bracket{C})^{2}(\rh\otimes\si)]^{1/2}.
\eeqa
The above definition can be rewritten as
\beqa\label{eq:SD-HS}
\si(C)
=\|(C-\bracket{C})\sqrt{\rh\otimes\si}\|_{HS},
\eeqa
where $\|\cdots\|_{HS}$ is the Hilbert-Schmidt norm defined by
\beqa
\|X\|_{HS}=\sqrt{\Tr X^{\da}X}
\eeqa
for any Hilbert-Schmidt class operator $X$, i.e.,
$\Tr X^{\da}X<\infty$.
Then, a simple application of the Schwarz inequality for
the inner product $\bracket{X,Y}=\Tr X^{*}Y$ on 
Hilbert-Schmidt class operators, we have
\beqa
\si(C)\si(D)&\ge&|\Tr[(C-\bracket{C})(D-\bracket{D})\rh\otimes\si]|\nn\\
&\ge&\frac{1}{2}|\Tr([C,D]\rh\otimes\si)|
\eeqa
for any observables $C,D$ with $\si(C),\si(D)<\infty$.
We shall refer to the last inequality as the 
{\em Heisenberg-Robertson uncertainty relation for standard deviations}
or {\em Heisenberg-Robertson relation}, for short.

In order to quantify the noise, we introduce the noise operator
$N(A)$ of $\bA(\bx)$ for measuring $A$.
According to the measuring process described in Section \ref{se:MP},
this measurement can be described as follows:
in order to measure the observable $A^{\In}$ in the state 
$\rh\otimes\si$ the observer actually make a precise measurement 
of the observable $M^{\Out}$ in the same state.
It follows that we can apply the definition of the noise operator 
given in the preceding section.
Thus, we define the {\em noise operator} $N(A)$ of $\bA(\bx)$ for
measuring $A$ by
\beqa
N(A)&=&M^{\Out}-A^{\In}\label{1.1a}\\
&=&U^{\da}(I\otimes M)U-A\otimes I.
\eeqa
The {\em root-mean-square (rms) noise} $\ep(A,\rh)$, 
or denoted by $\ep(A)$ for short,
of  $\bA(\bx)$ for measuring $A$ on input state $\rh$ is, then,  defined by 
\begin{equation}\label{eq:noise}
\ep(A,\rh)=\bracket{N(A)^{2}}^{1/2}.
\end{equation}
Using the Hilbert-Schmidt norm, 
the above definition can be rewritten as
\beqa\label{eq:noise-HS}
\ep(A,\rh)
=\|M^{\Out}\sqrt{\rh\otimes\si}-A^{\In}\sqrt{\rh\otimes\si}\|_{HS}.
\eeqa
We shall write $\ep(A,\rh)=\ep(A,\ps)$, if $\rh=\ket{\ps}\bra{\ps}$. 

In order to clarify the meaning of the above definition, 
suppose that the probe preparation is a pure state $\si=\ket{\xi}\bra{\xi}$
and let us consider the observable $A$ in \Eq{discrete}.
Suppose that the input state is $\ket{a_{n},\nu}$.
Then, we have
\begin{equation}
N(A)\ket{a_{n},\nu}\ket{\xi}
=(M^{\Out}-a_{n})\ket{a_{n},\nu}\ket{\xi}
\end{equation}
and 
\begin{equation}
\ep(A,\ket{a_{n},\nu})=\bracket{(M^{\Out}-a_{n})^{2}}^{1/2}.
\end{equation}
Thus, $\ep(A,\ket{a_{n},\nu})$ stands for the root-mean-square
difference between the experimental output $M^{\Out}$ and 
the true value $a_{n}$ of observable $A$.

If $\ep(A,\ket{a_{n},\nu})=0$, we have
\begin{equation}
M^{\Out}\ket{a_{n},\nu}\ket{\xi}=a_{n}\ket{a_{n},\nu}\ket{\xi},
\end{equation}
so that $\bA(\bx)$ outputs $a_{n}$ with probability one.  Thus,
we have shown that {\em if $\ep(A,\ps)=0$ for any eigenstates $\ps$ of 
a purely discrete observable $A$, then
$\bA(\bx)$ precisely measures $A$.}

For a general observable $A$,
if the observable $A$ has a definite value $a$ in the state $\rh$,
i.e., $A\rh=a\rh$, we have
\beqa\label{eq:definite2}
\ep(A)&=&
\left(\int_{\R}(x-a)^{2}\bracket{dE^{M^{\Out}}(x)}\right)^{1/2},
\eeqa
and hence $\ep(A)$ is the root-mean-square of the difference
between the output $x$ and the true value $a$.

From the similar argument leading to Eqs.~\eq{triangular2}--\eq{triangular4},
we have
\beqa\label{eq:triangular2-M}
\si(M^{\Out})
&\le&\ep(A)+\si(A^{\In})+|\bracket{M^{\Out}}-\bracket{A^{\In}}|,\\
\si(A^{\In})
&\le&\ep(A)+\si(M^{\Out})+|\bracket{M^{\Out}}-\bracket{A^{\In}}|,
\label{eq:triangular3-M}\\
\ep(A)
&\le&\si(A^{\In})+\si(M^{\Out})+|\bracket{M^{\Out}}-\bracket{A^{\In}}|.
\qquad\label{eq:triangular4-M}
\eeqa
From the above, we also have
\beqa\label{eq:triangular5-M}
|\si(M^{\Out})-\si(A^{\In})|
\le\ep(A)+|\bracket{M^{\Out}}-\bracket{A^{\In}}|.
\eeqa

\subsection{Distance of POVMs from observables\label{se:Distance}}

In the preceding subsection, we have defined the root-mean-square 
noise of measurement
using the associated indirect measurement model.
Thus, this amount of noise apparently depends on the model;
for example, two different models with different boundaries
between the apparatus and the observer describing the physically 
identical apparatus might have different amounts of noise.
In the next subsection, we shall show that this is only apparently the case. 
The root-mean-square noise depends only on the POVM of the apparatus and hence
statistically equivalent apparatuses have the same amount of noise.
In this subsection, we shall generally introduce a notion of distance
between a POVM and an observable, which will play an important role in
the study of quantum noise and disturbance in measurements.

Let $\Pi $ be a POVM on a Hilbert space $\cH$.
Let $f(x)$ be a real Borel function on $\R$. 
Denote by $\int f(x) d\Pi(x)$, or $\int f d\Pi$ for short,
the symmetric operator defined by 
\begin{equation}
\bracket{\xi|\int f(x) d\Pi(x)|\et} 
= \int_\R f(x)\,d\<\xi|\Pi (x)|\et\>
\end{equation}
for any $\xi,\et\in\dom(\int f(x)d\Pi(x))$, where the domain is defined by
\beqa
\lefteqn{{\dom\left(\int f(x) d\Pi(x)\right) }}\nn\\
&=& \left\{\xi\in\cH \mid \int_\R f(x)^2 \,d\<\xi|\Pi (x)|\xi\> <
\infty
\right\}.\quad
\eeqa
The {\em first and the second moment operators of $\Pi $}, denoted by 
$O(\Pi )$ and $O^{(2)}(\Pi )$, are defined by
\beqa
O(\Pi )=\int x\, d\Pi(x),\label{eq:moment1}\\
O^{(2)}(\Pi )=\int_{\R}x^{2}\, d\Pi(x).\label{eq:moment2}
\eeqa

By the Naimark theorem \cite{RN55}, 
there is a Hilbert space
$\cW$, an isometry $V:\cH\to\cW$,
and a self-adjoint operator $C$ such that 
\beqa\label{eq:Naimark}
\Pi (\De)=V^{\da}E^{C}(\De)V
\eeqa 
for every Borel set $\De$.
We shall call any triple $(\cW,V,C)$ satisfying \Eq{Naimark}
a {\em Naimark extension} 
of $\Pi $.  By integrating the both sides of \Eq{Naimark}, we have
\beqa
O(\Pi )=V^{\da}CV,\label{eq:Naimark1}\\
O^{(2)}(\Pi )=V^{\da}C^{2}V.\label{eq:Naimark2}
\eeqa
Since $V^{\da}C^{2}V\ge V^{\da}CVV^{\da}CV$, we have 
\beqa
O^{(2)}(\Pi )\ge O(\Pi )^{2}.
\eeqa
Let $A$ and $\rh$ be an observable and a density operator on $\cH$.
We define the {\em distance} $d_{\rh}(\Pi ,A)$ 
of POVM $\Pi $ from observable $A$ in $\rh$ by
\beqa
\lefteqn{d_{\rh}(\Pi ,A)}\nn\\
&=&\Tr\{[O^{(2)}(\Pi )-O(\Pi )^{2}+(O(\Pi )-A)^{2}]\rh\}^{1/2}\quad
\label{eq:distance}\\
&=&\Tr[O^{(2)}(\Pi )-O(\Pi )A-AO(\Pi )+A^{2})\rh]^{1/2}.\quad
\label{eq:distance2}
\eeqa
We shall abbreviate $d_{\ket{\ps}\bra{\ps}}$ as $d_{\ps}$
for a vector state $\ps$.

In the case where $\Pi $ is the spectral measure of an observable $X$, i.e.,
$\Pi =E^{X}$, we have
\beqa
O(E^{X})&=&\int x\, dE^{X}(x)=X,\\
O^{(2)}(E^{X})&=&\int_{\R}x^{2}\, d\Pi(x)=X^{2}.
\eeqa
Consequently, we have
\beqa\label{eq:distance4}
d_{\rh}(E^{X},A)
&=&\Tr[(X-A)^{2}\rh]^{1/2}\nn\\
&=&\|X\sqrt{\rh}-A\sqrt{\rh}\|_{HS}.
\eeqa
Thus, the distance $d_{\rh}$ generalizes the distance of two observables
given by $\|X\sqrt{\rh}-A\sqrt{\rh}\|_{HS}$.
%\beqa
%d_{\rh}(X,A)=\|X\sqrt{\rh}-A\sqrt{\rh}\|_{HS}.
%\eeqa

Now, we have the following properties of the distance $d_{\rh}$.

\begin{Theorem}
\label{th:Naimark}
Let $A$ and $\rh$ be an observable 
and a density operator on $\cH$.
For any Naimark extension $(\cW,V,C)$ of a POVM $\Pi $ on $\cH$,
we have
\beqa
d_{\rh}(\Pi ,A)=\|CV\sqrt{\rh}-VA\sqrt{\rh}\|_{HS}.
\eeqa
\end{Theorem}
\begin{Proof}
The assertion follows from Eqs.~\eq{Naimark1}, \eq{Naimark2}, 
\eq{distance2}, and the relations
\beqas
\lefteqn{\|CV\sqrt{\rh}-VA\sqrt{\rh}\|_{HS}^{2}}\quad\\
&=&
\Tr[(CV\sqrt{\rh}-VA\sqrt{\rh})^{\da}(CV\sqrt{\rh}-VA\sqrt{\rh})]\\
&=&
\Tr[(V^{\da}C^{2}V-V^{\da}CVA-AV^{\da}CV+A^{2})\rh].
\eeqas

\end{Proof}

\begin{Theorem}
\label{th:zero-distance}
A POVM $\Pi $ on $\cH$ is a spectral measure of an observable $A$
on $\cH$, i.e, $\Pi =E^{A}$ if and only if $d_{\ps}(\Pi ,A)=0$ for any state
vector $\ps\in\cH$.
\end{Theorem}
\begin{Proof}
From \Eq{distance4}, if $\Pi =E^{A}$, we have $d_{\rh}(\Pi ,A)=0$
for any $\rh$. Conversely, suppose that $d_{\ps}(\Pi ,A)=0$ for all state
vector $\ps\in\cH$.
Let $(\cW,V,C)$ be a Naimark extension of $\Pi $. 
From Theorem \ref{th:Naimark}, we have
\beqa
CV\ket{\ps}\bra{\ps}=VA\ket{\ps}\bra{\ps}
\eeqa
for all $\ps\in\cH$.  
Thus, we have
\beqa\label{eq:030603}
CV=VA,
\eeqa
and hence $CVV^{\da}=VAV^{\da}$.
By taking the adjoint of the both sides, we have
$CVV^{\da}=VV^{\da}C$. 
Since $VV^{\da}$ is a projection, it follows that 
all the spectral projections $E^{C}(\De)$ commutes with
$VV^{\da}$.
Since $V$ is isometry, i.e., $V^{\da}V=I$, we have 
\beqa
V^{\da}E^{C}(\De)VV^{\da}E^{C}(\De)V
&=&V^{\da}E^{C}(\De)V.
\eeqa
Thus, $\Pi (\De)=V^{\da}E^{C}(\De)V$ is projection valued.
From \Eq{030603}, we have also
\beqa
A=V^{\da}CV=\int \la d\Pi(\la).
\eeqa
By the uniqueness of the spectral decomposition, we conclude
that $\Pi $ is the spectral measure of $A$, i.e., $\Pi =E^{A}$.
\end{Proof}

\begin{Corollary}\label{th:zero-distance-1}
For any POVM $\Pi $ on $\cH$ and any observable $A$ on $\cH$,
the following conditions are equivalent.

(i) $\Pi =E^{A}$.

(ii) $d_{\rh}(\Pi ,A)=0$ for any state $\rh$.

(iii) $d_{\rh}(\Pi ,A)=0$ for a faithful state $\rh$.

(iv) $d_{\ket{n}}(\Pi ,A)=0$ for any $\ket{n}$ in 
an orthonormal basis $\{\ket{n}\}$.

(v) $d_{\ps}(\Pi ,A)=0$ for any state vector $\ps\in\cH$.

\end{Corollary}
\begin{Proof}
The implication (i) $\Then$ (ii) is 
an immediate consequence of \Eq{distance4}, and the implication
(ii) $\Then$ (iii) is obvious, since a faithful state exists on any
separable Hilbert space.
To show the implication (iii)$\Then$(iv), assume
that $d_{\rh}(\Pi ,A)=0$ for a faithful state $\rh$.
Let $(\cW,V,C)$ be a Naimark extension of $\Pi $. 
From Theorem \ref{th:Naimark}, we have
\beqa\label{eq:0300603b}
(CV-VA)\rh=0.
\eeqa
Let $\ket{1},\ket{2},\ldots$ be an orthonormal  basis consisting
of eigenvectors of $\rh$.  Then, we have  $\rh\ket{n}=p_{n}\ket{n}$
with $p_{n}>0$ for all $n$.  Thus, applying the both sides of 
\Eq{0300603b} to the vector $p_{n}^{-1}\ket{n}$, we have
\beqa
(CV-VA)\ket{n}=0.
\eeqa
By Theorem \ref{th:Naimark}, we have $d_{\ket{n}}(\Pi ,A)=0$
for all $\ket{n}$, and (iii)$\Then$(iv) has been shown.
To show the implication (iv)$\Then$(v), assume
that $d_{\ket{n}}(\Pi ,A)=0$ for an orthonormal basis 
$\ket{1},\ket{2},\ldots$.
From Theorem \ref{th:Naimark}, we have
\beqa\label{eq:0300603c}
(CV-VA)\ket{n}=0.
\eeqa
By linearity, it follows easily that for any state vector
$\ps\in\cH$, we have
\beqa\label{eq:0300603d}
(CV-VA)\ket{\ps}=0.
\eeqa
Thus, we conclude $d_{\ps}(\Pi ,A)=0$ for any state $\ps$,
so that (iv)$\Then$(v) has shown.
Since the implication (v)$\Then$(i) has been proven in the
proof of Theorem \ref{th:zero-distance}, this completes the proof.
\end{Proof}

\begin{Theorem}
\label{th:distance}
Let $C$ be an observable on Hilbert space $\cH\otimes\cK$
and let $\si$ be a density operator on $\cK$.
If $\Pi _{C}$ is a POVM defined by 
\begin{equation}\label{eq:POVM1'}
\Pi _{C}(\De)=\cE_{\si}[E^{C}(\De)]
\end{equation}
for any $\De\in\cB(\R)$, then we have
\begin{equation}\label{eq:distance3}
d_{\rh}(\Pi _{C},A)
=\|C\sqrt{\rh\otimes\si}-A\otimes I\sqrt{\rh\otimes\si}\|_{HS}.
\end{equation}
\end{Theorem}
\begin{Proof}
By integrating the both sides of \Eq{POVM1'}, we have
\beqa
O(\Pi _{C})
&=&
\Tr_{\cK}[C(I\otimes\si)],\\
O^{(2)}(\Pi _{C})
&=&
\Tr_{\cK}[C^{2}(I\otimes\si)].
\eeqa
Thus, by the properties of the partial trace, we have
\beqa
\Tr[O(\Pi _{C})A\rh]
&=&\Tr[C(A\otimes I)(\rh\otimes\si)],\\
\Tr[AO(\Pi _{C})\rh]
&=&\Tr[(A\otimes I)C(\rh\otimes\si)],\\
\Tr[O^{(2)}(\Pi _{C})\rh]
&=&\Tr[C^{2}(\rh\otimes\si)].
\eeqa
Thus, we have
\beqas
\lefteqn{\|C\sqrt{\rh\otimes\si}-A\otimes
I\sqrt{\rh\otimes\si}\|_{HS}^{2}}\nn\quad\\
%&=&
%\Tr[(C-A\otimes I)^{2}(\rh\otimes\si)]\\
&=&
\Tr[C^{2}(\rh\otimes\si)]-\Tr[C(A\otimes I)(\rh\otimes\si)]\\
& &\mb{}-\Tr[(A\otimes I)C(\rh\otimes\si)]-\Tr[(A^{2}\otimes
I)(\rh\otimes\si)]\\ &=&
\Tr[(O^{(2)}(\Pi _{C})-O(\Pi _{C})A-AO(\Pi _{C})+A^{2})\rh].
\eeqas
Thus, \Eq{distance3} follows from \Eq{distance2}.
\end{Proof}

%\widetext
\subsection{Model independent definition of noise}

The following theorem shows that the root-mean-square 
noise of an apparatus is
determined only by its POVM, and hence
statistically equivalent apparatuses have the same amount of noise.

\begin{Theorem} 
\label{th:noise}
Let $\bA(\bx)$ be an apparatus with indirect measurement
model $(\cK,\si,U,M)$.
Then, the rms noise $\ep(A,\rh)$ is determined by the POVM  $\Pi $ 
of $\bA(\bx)$ as
\beqa
\ep(A,\rh)&=&d_{\rh}(\Pi ,A).
\eeqa
\end{Theorem}
\begin{Proof}
The assertion follows from \Eq{noise-HS} and Theorem \ref{th:distance}
for $C=U^{\da}(I\otimes M)U$.
\end{Proof}

We define the {\em root-mean-square (rms) noise $\ep(A,\rh)$ of apparatus
$\bA(\bx)$ for measuring observable $A$ in state $\rh$} to be the 
distance $d_{\rh}(\Pi ,A)$
of the POVM $\Pi $ of $\bA(\bx)$ 
from observable $A$ in state $\rh$.
As above, this definition is consistent with the definition 
for apparatuses with indirect measurement models.

The following theorem asserts that apparatuses precisely measuring $A$
and apparatuses with numerically zero rms noise for $A$ are equivalent notions.

%{\em Theorem 1.
\begin{Theorem}
\label{th:noise-1}
An apparatus $\bA(\bx)$  precisely measures an observable
$A$ if and only if $\ep(A,\rh)=0$ on any input state $\rh$.
%}
\end{Theorem}

%A proof is given in Appendix \ref{se:ap-noise-1}.
%\section{Proof of Theorem 1}
%\label{se:A}
\begin{Proof}
Let $\Pi $ be the POVM of an apparatus $\bA(\bx)$.
Then,  $\bA(\bx)$  precisely measures $A$ if and only if
$\Pi =E^{A}$.
Thus, the assertion follows immediately from Theorem
\ref{th:zero-distance}.
\end{Proof}

Let $\bracket{\bx}$ and $\si(\bx)$ be the mean and the standard
deviation of the output variable $\bx$ of the apparatus $\bA(\bx)$ in
state $\rh$.  Then, we have
\beqa
\bracket{\bx}&=&\int_{\R}x\,\Pr\{\bx\in dx\|\rh\},
\label{eq:def-mean}\\
\si(\bx)
&=&\left(\int_{\R}(x-\bracket{\bx})^{2}\,\Pr\{\bx\in dx\|\rh\}\right)^{1/2}.
\label{eq:sd-output1}
\eeqa
From Eqs.~\eq{Born-POVM}, \eq{moment1},
and \eq{moment2}, we have 
\beqa
\bracket{\bx}&=&\Tr[O(\Pi )\rh],\label{eq:output-mean}\\
\si(\bx)
&=&\left(\Tr[O^{(2)}(\Pi )\rh]-\Tr[O(\Pi )\rh]^{2}\right)^{1/2}.
\label{eq:sd-output2}
\eeqa

From Eqs.~\eq{triangular2-M}--\eq{triangular4-M},
we have
\beqa\label{eq:triangular2-A}
\si(\bx)
&\le&\ep(A)+\si(A)+|\bracket{\bx}-\bracket{A}|,\\
\si(A)
&\le&\ep(A)+\si(\bx)+|\bracket{\bx}-\bracket{A}|,
\label{eq:triangular3-A}\\
\ep(A)
&\le&\si(A)+\si(\bx)+|\bracket{\bx}-\bracket{A}|.
\label{eq:triangular4-A}
\eeqa
In particular, we have
\beqa\label{eq:triangular5-A}
|\si(\bx)-\si(A)|
&\le&\ep(A)+|\bracket{\bx}-\bracket{A}|.
\eeqa

In this subsection, we have shown that the rms noise of an apparatus is
defined independent of a particular model to describe the measuring
process of the apparatus.  This suggests that the rms noise can be 
statistically estimated from the experimental data.  
In fact, this can be done as follows.
Let $\Pi $ be a POVM and let $A$ be an observable.
By the relation
\beqa
O(\Pi )A+AO(\Pi )&=&(A+I)O(\Pi )(A+I)-AO(\Pi )A\nn\\
& &\mb{}-O(\Pi ),
\eeqa
we have
\beqa\label{eq:data}
d_{\ps}(\Pi ,A)^{2}
&=&\bracket{\ps|A^{2}|\ps}+\bracket{\ps|O^{(2)}(\Pi )|\ps}\nn\\
& &\mb{}+\bracket{\ps|O(\Pi )|\ps}+\bracket{A\ps|O(\Pi )|A\ps}
\mb{  }\nn\\
& &\mb{}-\bracket{(A+I)\ps|O(\Pi )|(A+I)\ps}.
\eeqa
In the above, $\bracket{\ps|A^{2}|\ps}$ is the theoretical mean value
of $A^{2}$ in state $\ps$, $\bracket{\ps|O^{(2)}(\Pi )|\ps}$ is the 
mean of the squared output $\bx^{2}$ in state $\ps$, 
and the other terms are the means of the output $\bx$ in the respective
input states.  Thus, the error $\ep(A,\ps)$ can be 
statistically estimated,
in principle, from experimental data of the measurements
in states $\ps$, $A\ps/\|A\ps\|$, and $(A+I)\ps/\|(A+I)\ps\|$.

\subsection{Relations to other approaches}

In Refs.~\cite{88MS,89RS,91QU} the notion of rms noise was previously
introduced for a  restricted class of measurements.
In what follows, we shall show that those definitions are equivalent
to the general definition introduced above.

Let $A$ be an observable of $\bS$.  A POVM $\Pi $ of $\bS$ is said to be
{\em compatible with $A$}, or {\em $A$ compatible} for short, 
if it satisfies the relation
\begin{equation}\label{eq:020605a} 
[\Pi (\De_{1}),E^{A}(\De_{2})] = 0
 \end{equation}                                   
for all $\De_{1}, \De_{2} \in \cB(\R)$.

Let $\rh$ be a state.  For an $A$-compatible
POM $\Pi $, the joint probability distribution 
of $\Pi $ and $A$ in state $\rh$ is defined by
\begin{equation}\label{eq:020606b}     
\mu^{(\Pi ,A)}_{\rh}(\De_{1}\times\De_{2})
= \Tr[\Pi (\De_{1})E^{A}(\De_{2})\rh]
\end{equation}
for any 
$\De_{1},\De_{2} \in \cB(\R)$.
By \Eq{020605a} it is easy to see that \Eq{020606b} defines a
unique Borel measure on $\R^{2}$.
As a notational convention,  we shall write
\beqa
\lefteqn{
\iint_{\R^2}
f(x,y)\,d\mu^{(\Pi ,A)}_{\rh}(x,y)}\nn\quad\\
&=& \iint_{\R^2}
f(x,y)\Tr[d\Pi(x)dE^{A}(y)\rh]
\eeqa
%\beqa
%\iint_{\R^2}
%f(x,y)\,d\mu^{(\Pi ,A)}_{\rh}(x,y)
%&=& \iint_{\R^2}
%f(x,y)\,\Tr[d\Pi(x)dE^{A}(y)\rh]
%\eeqa
for a  Borel function $f(x,y)$ on $\R^{2}$.
If $f(x)g(y)$ is a $\mu^{(\Pi ,A)}_{\rh}(x,y)$-integrable function
on $\R^{2}$, then we have
%\beqa
%\iint_{\R^{2}}f(x)g(y)\Tr[d\Pi(x)dE^{A}(y)\rh]
%&=&
%\Tr[(\int fd\Pi) g(A)\rh].
%\eeqa
\beqa
\lefteqn{\iint_{\R^{2}}f(x)g(y)\Tr[d\Pi(x)dE^{A}(y)\rh]}\nn\quad\\
&=&
\Tr\left[\left(\int f\,d\Pi\right) g(A)\rh\right].
\eeqa

Now, let us assume that the POVM $\Pi $ of an apparatus $\bA(\bx)$
is compatible with an observable $A$.
Then, we have
\beqas
\lefteqn{\iint_{\R^{2}}\,(x - y)^{2}\,\Tr[d\Pi(x)dE^{A}(y)\rh]}\nn\quad\\
&=&
\iint_{\R^{2}}\,x ^{2}\,\Tr[d\Pi(x)dE^{A}(y)\rh]\\
& &
\mb{}-2\iint_{\R^{2}}\,xy\,\Tr[d\Pi(x)dE^{A}(y)\rh]\\
& &
\mb{}+\iint_{\R^{2}}\,y^{2}\,\Tr[d\Pi(x)dE^{A}(y)\rh]\\
&=&
\Tr[O^{(2)}(\Pi )\rh]-2\Tr[O(\Pi )A\rh]+\Tr[A^{2}\rh]\\
&=&
\Tr\{[O^{(2)}(\Pi )-O(\Pi )^{2}+(X-A)^{2}]\rh\}\\
&=&
d_{\rh}(\Pi ,A).
\eeqas
Thus, by Theorem \ref{th:noise}, we have
\begin{equation}\label{eq:(3.5)}
\ep(A,\rh)^{2} =
\iint_{\R^{2}}\,(x - y)^{2}\,\Tr[d\Pi(x)dE^{A}(y)\rh].
\end{equation} 
The above relation shows that the rms noise $\ep(A,\rh)$ represents the
root-mean-square deviation of the output $\bx$ of the measurement
using $\bA(\bx)$ from the output $\by$ of an precise
$A$ measurement using another apparatus $\bA(\by)$, when these two
were made simultaneously in the state $\rh$.
In Ref.~\cite{91QU}, the rms noise of an apparatus with $A$-compatible
POVM was introduced by \Eq{(3.5)}.

Let us consider the case where the object $\bS$ is a one-dimensional
mass and the observable to be measured is the position $\hx$ of the mass.
Suppose that the POVM $\Pi $ of apparatus $\bA(\bx)$ to measure
$\hx$  is compatible with $\hx$, i.e.,
\begin{equation}
[\Pi (\De_{1}),E^{\hx}(\De_{2})]=0
\end{equation}
for all Borel sets $\De_{1},\De_{2}$.
 Under this condition, there is a
kernel function $G(a,x)$  called the {\em resolution kernel}, 
which
may be a distribution or a generalized function, such that
\begin{equation}
\Pi (\De)=\int_{\De}da\int_{\R}G(a,x)\,dE^{\hx}(x)
\end{equation}
or
 \begin{equation}
 d\Pi(a) = da \int_{\R} \,G(a,x)|x\rangle \langle x|\, dx           
\label{eq:4.1.5}
 \end{equation}
in the Dirac notation.
 Even if the apparatus $\bA(\bx)$ measures position $\hx$ 
approximately, the output probability distribution
$\Pr\{\bx\in da\|\psi \}$ on input state represented by a
wave function $\ps(x)$ 
is expected to be related to the position
distribution $|\psi (x)|^2$ --- from \Eq{4.1.5}, this relation is
expressed in the following form
 \begin{equation}
  \Pr\{\bx\in da\|\psi \}= da \int_{\R}  dx\,G(a,x)|\psi (x)|^2.
                                                  \label{eq:4.1.6}
 \end{equation}
 Note that $G(a,x)$ is independent of a particular wave
function $\psi (x)$. 
Obviously, $\bA(\bx)$ precisely measures $\hx$, i.e.,  
\begin{equation}
\Pr\{\bx\in da\|\psi \} =|\psi (a)|^2\,da
\end{equation}
for all $\psi $, if and only if $G(a,x) =
\delta (x-a)$.  
Roughly speaking, $G(a,x)$ is the conditional
probability density of the output $\bx=a$, given that
the mass is in the position $\hx=x$ at the time of
measurement; hence the rms noise $\epsilon (\hx,\ket{x})$ of 
the  apparatus $\bA(\bx)$ on input state $\ket{x}$ 
should satisfy
 \begin{equation}
 \ep (\hx,\ket{x})^2 = \int_{\R}  da\,(a - x)^2G(a,x).     
\label{eq:4.1.7}
 \end{equation}
Since our definition of the rms noise excludes the case where the input
state is an unnormalizable state like $\ket{x}$, \Eq{4.1.7}
cannot be justified.  However, if the input 
mass state is a normalized wave function $\psi (x)$,
the rms noise $\ep(\hx,\psi )$ should satisfy
\begin{equation}
 \ep(\hx,\psi )^2 = \int_{\R}  \ep(\hx,\ket{x})^2|\psi (x)|^2\, dx
\end{equation}
or equivalently
\begin{equation}\label{eq:4.1.8}
\ep(\hx,\psi )^2 = \iint_{\R^{2}}  da\,(a - x)^2G(a,x)|\psi
(x)|^2\,dx.
\end{equation}
The following computations show that \Eq{4.1.8} is actually
derived from our general definition.
For $\rh=\ket{\ps}\bra{\ps}$, we have
\beqas
\lefteqn{\Tr[\Pi (\De_{1})E^{\hx}(\De_{2})\rh]}\nn\quad\\
&=&
\bracket{\ps|\Pi (\De_{1})E^{\hx}(\De_{2})|\ps}\\
&=&
\int_{\De_{1}}da
\int_{\R}G(a,x)\bracket{\ps|dE^{\hx}(x)E^{\hx}(\De_{2})|\ps}\\
&=&
\int_{\De_{1}}da
\int_{\De_{2}}G(a,x)\bracket{\ps|dE^{\hx}(x)|\ps}.
\eeqas
Thus, by properties of Lebesgue integral,
we have
\beqa
\lefteqn{\iint_{\R^{2}}(x-a)^{2}\Tr[d\Pi(a)dE^{\hx}(x)\rh]}\nn\quad\\
&=&
\iint_{\R^{2}}(x-a)^{2}\,da\,G(a,x)|\ps(x)|^{2}dx.
\eeqa
Therefore, 
from \Eq{(3.5)}
we conclude that \Eq{4.1.8} actually holds.

\section{Disturbance in measurement\label{se:Disturbance}}

\subsection{Nondisturbing measurements}

Let $\bA(\bx)$ be an apparatus with indirect measurement model
$(\cK,\si,U,M)$.
%(971224)\\
We should generally say that apparatus $\bA(\bx)$ {\em does
not disturb} an observable $B$ of $\bS$, if the nonselective state 
change does not
perturb the dynamical evolution of the probability distribution of $B$, i.e., 
\beqa\label{1224a}
\lefteqn{
\Tr\{[E^{B}(\De)\otimes I]U(\rh\otimes \si)U^{\da}\}}\nn\\
&=&
\Tr[E^{B}(\De)e^{-iH\De t/\hbar}\rh e^{iH\De t/\hbar}]
\eeqa 
%\beqa\label{1224a}
%\Tr\{[E^{B}(\De)\otimes I]U(\rh\otimes \si)U^{\da}\}
%&=&
%\Tr[E^{B}(\De)e^{-iH\De t/\hbar}\rh e^{iH\De t/\hbar}]
%\eeqa 
for any Borel set $\De$ and any input state $\rh$, 
where $H$ is the Hamiltonian of the system $\bS$. 
In this paper, we assume that the apparatus carries out
{\em instantaneous} measurements in the sense that the time duration
$\De t$ is very small and the coupling between $\bS$ and $\bP$
is very large so that the free evolution of $\bS$ in the
time interval $(t,t+\De t)$ can be neglected. 
In this case, we say that apparatus
$\bA(\bx)$ {\em does not change the probability distribution} of
an observable $B$ of $\bS$ on input state $\rh$, if 
\begin{equation}\label{eq:10}
\Tr\{[E^{B}(\De)\otimes I]U(\rh\otimes\si)U^{\da}\}
=\Tr[E^{B}(\De)\rh],
\end{equation}
or in the Heisenberg picture,
\begin{equation}\label{eq:829e}
\bracket{E^{B^{\Out}}(\De)}=\bracket{E^{B^{\In}}(\De)}
\end{equation}
for every Borel set  $\De$, 
where we write $B^{\In}=B\otimes I$ and $B^{\Out}=U^{\dagger}(B\otimes
I)U$.   We say that apparatus $\bA(\bx)$ {\em does not disturb} observable
$B$, or $\bA(\bx)$ is called {\em $B$-nondisturbing}, 
if apparatus $\bA(\bx)$
does not disturb the probability distribution of  observable $B$ on 
{\em any} input state $\rh$
\cite{01OD}. 

The next theorem shows that nondisturbing measurements 
are characterized by nonselective operations, so that it is independent of the
particular choice of the indirect measurement model associated with
the apparatus.

\begin{Theorem}
\label{th:nondisturbing2}
An apparatus $\bA(\bx)$ with indirect measurement model
$(\cK,\si,U,M)$ does not disturb an observable $B$ if and only if 
we have
\begin{equation}\label{eq:nondisturbing}
T^{*}E^{B}(\De)=E^{B}(\De)
\end{equation}
for any Borel set $\De$, where $T$ is the nonselective operation
of $\bA(\bx)$.
\end{Theorem}
\begin{Proof}
By the property of the partial trace, we have
\beqa
\lefteqn{\Tr\{[E^{B}(\De)\otimes I]U(\rh\otimes\si)U^{\da}\}}\quad\nn\\
&=&\Tr\left(\Tr_{\cK}\{U^{\da}[E^{B}(\De)\otimes
I]U(I\otimes\si)\}\rh\right).
\label{eq:108e}
\eeqa
%\beqa
%\Tr\{[E^{B}(\De)\otimes I]U(\rh\otimes\si)U^{\da}\}
%&=&\Tr\left(\Tr_{\cK}\{U^{\da}[E^{B}(\De)\otimes
%I]U(I\otimes\si)\}\rh\right).
%\label{eq:108e}
%\eeqa
Thus, \Eq{10} is equivalent to
\begin{equation}
\Tr(\Tr_{\cK}\{U^{\da}[E^{B}(\De)\otimes I]U(I\otimes\si)\}\rh)
=\Tr[E^{B}(\De) \rh].
\end{equation} 
Since $\rh$ is arbitrary, $\bA(\bx)$ does not disturb $B$ if and only
if
\begin{equation}\label{eq:11} %\label{829f}
\Tr_{\cK}\{U^{\da}[E^{B}(\De)\otimes I]U(I\otimes\si)\}
=E^{B}(\De) 
\end{equation} 
for any Borel set $\De$.
Thus, by \Eq{dual-nonselective-operation} we conclude that
$\bA(\bx)$ does not disturb $B$ if and only
if \Eq{nondisturbing} holds for any Borel set $\De$.
\end{Proof}

\subsection{Joint measurements with nondisturbing apparatuses}

The relation between simultaneous measurements and nondisturbing
measurements were investigated in Ref.~\cite{01OD} and it was proven that
any apparatus $\bA(\bx)$ precisely measuring an observable $A$
does not disturb observable $B$ if and only if 
successive precise measurements of observables $A$ and $B$, 
using $\bA(\bx)$ for the $A$ measurement, 
satisfies the joint probability formula for
simultaneous measurements in the first input state.
Here, we shall generalize the above result
for apparatuses which do not necessarily make a 
precise measurement.

\begin{Theorem}
\label{th:020609e} 
Suppose that an apparatus $\bA(\by)$ precisely measures an
observable $B$ immediately after a measurement using an apparatus
$\bA(\bx)$ with POVM $\Pi _{\bx}$.  
Then, apparatus $\bA(\bx)$ does not disturb 
observable $B$ if and only if their joint output probability distribution
satisfies
\beqa\label{eq:joint-nondisturbing}
\Pr\{\bx\in\De,\by\in\De'\|\rh\}
&=&\Tr[\Pi _{\bx}(\De)E^{B}(\De')\rh]
\eeqa 
for any input state  $\rh$ and any Borel sets $\De$ and $\De'$. 
In this case, $\Pi $ is necessarily compatible with $B$.
\end{Theorem}
\begin{Proof}
By the realization theorem, we can assume without any
loss of generality that the apparatus $\bA(\bx)$ has a pure
indirect measurement model $(\cK,\xi,U,M)$.  

Since the apparatus $\bA(\by)$ precisely measures $B$,
the POVM $\Pi _{\by}$ of $\bA(\by)$ is such that
$\Pi _{\by}=E^{B}$.  Thus, from \Eq{JOPD2}, we have
\begin{equation}
\Pr\{\bx\in\De,\by\in\De'\|\rh\}
=\Tr\{\cI_{\bx}(\De)^{*}[E^{B}(\De')]\rh\}.
\end{equation}
By \Eq{CP instrument} we have
\begin{equation}\label{eq:CP instrument1}
\cI_{\bx}(\De)^{*}E^{B}(\De')=V^{\dagger}[E^{B}(\De')\otimes
E^{M}(\Delta)]V
\end{equation}
for any $\De,\De'\in\cB(\R)$, where $V$ is such that
$V\ps=U(\ps\otimes\xi)$ for all $\ps\in\cH$.
Suppose that apparatus $\bA(\bx)$ does not disturb observable $B$.
Then, we have
\begin{equation}
\cI_{\bx}(\R)^{*}E^{B}(\De')=E^{B}(\De'),
\end{equation}
and hence 
\begin{equation}
E^{B}(\De')=V^{\dagger}[E^{B}(\De')\otimes I]V.
\end{equation}
Thus, we have
\begin{equation}
|VE^{B}(\De')-[E^{B}(\De')\otimes I]V|^{2}
=
0.
\end{equation}
Consequently,
\begin{equation}
VE^{B}(\De')=[E^{B}(\De')\otimes I]V.
\end{equation}
By \Eq{CP instrument1}, we have
\beqa
\cI_{\bx}(\De)^{*}E^{B}(\De')
&=&
V^{\dagger}[E^{B}(\De')\otimes E^{M}(\Delta)]V\nn\\
&=&
V^{\dagger}[I\otimes E^{M}(\Delta)]VE^{B}(\De')\nn\\
&=&
\Pi _{\bx}(\De)E^{B}(\De').
\eeqa
Therefore, \Eq{joint-nondisturbing} follows.
Conversely, suppose that \Eq{joint-nondisturbing} holds
for for any input state  $\rh$ and any Borel sets $\De$ and $\De'$. 
Let $\De=\R$.  We have
\begin{equation}
\Tr\{\cI_{\bx}(\R)^{*}[E^{B}(\De')]\rh\}.
=\Tr[E^{B}(\De')\rh]
\end{equation}
for any state $\rh$.  Thus, we conclude 
\begin{equation}
\cI_{\bx}(\R)^{*}[E^{B}(\De')]=E^{B}(\De')
\end{equation}
for any Borel set $\De'$, and the assertion follows from
\Eq{nondisturbing}.
\end{Proof}

%\section{Disturbance in measurement}
%\subsection{Model dependent definition of disturbance}
\subsection{Disturbance in indirect measurement models}

In order to
quantify the disturbance, we introduce the {\em disturbance operator}
$D(B)$ of apparatus $\bA(\bx)$ for observable $B$ defined by 
\beqa
D(B)
&=&B^{\Out}-B^{\In}
\label{disturbance_operator}\\
&=&U^{\da}(B\otimes I)U-B\otimes I.
\label{eq:disturbance_op}
\eeqa
The {\em root-mean-square (rms) disturbance}
 $\et(B,\rh)$ of observable $B$ by
apparatus $\bA(\bx)$ on input state $\ps$ is, then,  defined by 
%\begin{equation}\label{eq:disturbance2}
%\et(B,\rh)=\Tr\{[U^{\da}(B\otimes I)U-B\otimes
%I]^{2}U(\rh\otimes\si)\}^{1/2},
%\end{equation}
%or in the Heisenberg picture
\begin{equation}\label{eq:disturbance}
\et(B,\rh)=\bracket{D(B)^{2}}^{1/2}.
\end{equation}
We shall write $\et(B,\rh)=\et(B,\ps)$ if $\rh=\ket{\ps}\bra{\ps}$.
%The disturbance $\et(B)$ represents the root-mean-square 
%deviation of the observable $B$ before and after the measuring
%interaction. 
The above definition can be rewritten as
\beqa\label{eq:disturbance-HS}
\et(B,\rh)
=\|B^{\Out}\sqrt{\rh\otimes\si}-B^{\In}\sqrt{\rh\otimes\si}\|_{HS}.
\eeqa

From \Eq{disturbance_op} we have
\begin{equation}
D(B)=U^{\da}[B\otimes I,U].
\end{equation}
Thus, we have
\begin{equation}
\et(B,\rh)=\bracket{|[B\otimes I,U]|^{2}}^{1/2},
\end{equation}
and $\et(B,\rh)=0$ if and only if
$[B\otimes I,U]\rh\otimes\si=0$.

\subsection{Model independent definition of disturbance}

In the preceding subsection, we have defined the rms disturbance
of apparatus using the associated indirect measurement model.
In what follows, we shall show that the rms disturbance is 
determined by the nonselective operation of the apparatus
and hence depends only on the statistical equivalence class of the
apparatus.

The following theorem shows that the  rms disturbance of an apparatus 
determined only by its nonselective operation.

\begin{Theorem} 
\label{th:disturbance}
Let $\bA(\bx)$ be an apparatus with indirect measurement
model $(\cK,\si,U,M)$. 
Then, the rms disturbance $\et(B,\rh)$ is determined by the 
nonselective operation $T$ as 
\beqa
\et(B,\rh)&=&d_{\rh}(T^{*}E^{B},B),
\eeqa
where $T^{*}E^{B}$ stands for the POVM defined by
\begin{equation}
(T^{*}E^{B})(\De)=T^{*}[E^{B}(\De)]
\end{equation}
for any $\De\in\cB(\R)$.
\end{Theorem}
\begin{Proof}
Let $\Pi $ be the POVM defined by
\begin{equation}
\Pi (\De)=\cE_{\si}\{U^{\da}[E^{B}(\De)\otimes I]U\}
\end{equation}
for any $\De\in\cB(\R)$.  Then, by Theorem \ref{th:distance}, we have
\beqa
d_{\rh}(\Pi ,B)=\|U^{\da}(B\otimes I)U\sqrt{\rh\otimes\si}-
B\otimes I\sqrt{\rh\otimes\si}\|_{HS},\nn\\
\eeqa
and hence by \Eq{disturbance-HS}, we have
\begin{equation}\label{eq:th-1}
d_{\rh}(\Pi ,B)=\et(B,\rh)
\end{equation}
On the other hand, by \Eq{dual-nonselective-operation} we have
\begin{equation}\label{eq:th-2}
\Pi (\De)=T^{*}E^{B}(\De)
\end{equation}
for any $\De\in\cB(\R)$.  Thus, the assertion follows from \Eq{th-1}
and \Eq{th-2}.
\end{Proof}

We generally define the {\em root-mean-square (rms) disturbance}
$\et(B,\rh)$ of an observable $B$ 
by any apparatus $\bA(\bx)$ in state $\rh$ to be the  
distance $d_{\rh}(T^{*}E^{B},B)$.
As above, this definition is consistent with the definition 
for apparatuses with indirect measurement models.

One of the fundamental properties of the rms disturbance is that
non-disturbing apparatuses and apparatuses with zero disturbances are
equivalent notions, as ensured by the following theorem.

\begin{Theorem} 
\label{th:nondisturbing}
 The apparatus $\bA(\bx)$ does not disturb observable $B$ if
and only if
$\et(B,\rh)=0$ for any state $\rh$.
\end{Theorem}
\begin{Proof} 
From Theorem \ref{th:disturbance},
$\et(B,\rh)=0$ if and only if
$d_{\rh}(T^{*}E^{B},B)=0$.
Thus, from Theorem \ref{th:zero-distance},
$\et(B,\rh)=0$ for all $\rh$ if and only if 
$T^{*}E^{B}=E^{B}$.
By Theorem \ref{th:nondisturbing2},
the last condition holds if and only if 
$\bA(\bx)$ does not disturb $B$.
The proof is completed.
\end{Proof}

\section{New Formulation of Uncertainty Principle\label{se:Uncertainty}}

\subsection{Universally valid uncertainty relation}

Under the general definitions of rms noise and rms disturbance
introduced in the preceding sections,
we can rigorously investigate the validity of Heisenberg's 
noise-disturbance uncertainty relation.
%, \Eq{URMD}.
For this purpose, let $\bA(\bx)$ be an apparatus with
indirect measurement model $(\cK,\si,U,M)$.
Let $A$ and $B$ be two observables of the object.
Recall that the {noise operator} $N(A)$
and the {disturbance operator} $D(B)$ satisfy
%\begin{subequations}
\begin{eqnarray}
M^{\Out}&=&A^{\In}+N(A),\label{eq:NO}\\
B^{\Out}&=&B^{\In}+D(B).\label{eq:DO}
\end{eqnarray}
%\end{subequations}
Since $M$ and $B$ are observables in different systems, we have
$[M^{\Out},B^{\Out}]=0$, and hence we obtain
the following commutation relation for the noise operator
and the disturbance operator, 
\begin{eqnarray}
& &[N(A),D(B)]+[N(A),B^{\In}]+[A^{\In},D(B)]\nn\\
& &\qquad=-[A^{\In},B^{\In}].
\end{eqnarray}
Taking the moduli of means in the original state $\rh\otimes\si$
of the both sides
 and applying the triangular inequality, we have
\begin{eqnarray}
\lefteqn{
|\bracket{[N(A),D(B)]}|+|\bracket{[N(A),B^{\In}]}
+\bracket{[A^{\In},D(B)]}|}\qquad 
\qquad\qquad\qquad\qquad\qquad\qquad\qquad\qquad\qquad
\nn\\
\ge |\Tr([A,B]\rh)|.\quad& &
\end{eqnarray}
Since the variance is not greater than the mean square,  we have 
\begin{eqnarray}
\ep(A,\rh)&\ge&\si(N(A),\rh\otimes\si),\\
\et(B,\rh)&\ge&\si(D(B),\rh\otimes\si),
\end{eqnarray}
and hence by the Heisenberg-Robertson relation, we have
\begin{equation}
\ep(A,\rh)\et(B,\rh)\ge\frac{1}{2}|\bracket{[N(A),D(B)]}|.
\end{equation}
Thus, we obtain the {\em universally valid noise-disturbance uncertainty
relation} for the pair $(A,B)$,
\begin{eqnarray}\label{eq:UVUR1}
& &\ep(A,\rh)\et(B,\rh)
+\frac{1}{2}|\bracket{[N(A),B^{\In}]}
+\bracket{[A^{\In},D(B)]}|\nn\\
&  &\qquad\ge\frac{1}{2}|\Tr([A,B]\rh)|.
\end{eqnarray}

The above relation immediately gives rigorous conditions on
what  apparatus satisfies Heisenberg's noise-disturbance
uncertainty relation.  Some conditions are listed in the
following.

\begin{Theorem}\label{th:HNDUR}
Let $A$ and $B$ be a pair of observables.
An apparatus $\bA(\bx)$ with indirect measurement
model $(\cK,\si,U,M)$ satisfies Heisenberg's
noise-disturbance uncertainty relation, i.e., 
\beqas
\ep(A,\rh)\et(B,\rh)\ge \frac{1}{2}|\Tr([A,B]\rh)|
\eeqas
for any state $\rh$ for which all the relevant terms are finite,
if one of the following conditions holds:

(i) The noise operator commutes with $B^{\In}$ and
the disturbance operator commutes with $A^{\In}$,
i.e.,  
\beqa
[N(A),B^{\In}]&=&0,\\{}
[D(B),A^{\In}]&=&0.
\eeqa

(ii) The noise operator and the disturbance operator 
belong to the probe system, i.e., there are two observables
$N$ and $D$ on $\cK$ such that 
\beqa
N(A)&=&I\otimes N,\\
D(B)&=&I\otimes D.
\eeqa
\end{Theorem}

\subsection{Model-Independent formulation}

The above characterizations are easily obtained, but depend
on the model.  In order to obtain intrinsic characterizations
of apparatuses satisfying Heisenberg's relation, we
reformulate the universally valid relation in terms of
model independent notions.

Let $\bA(\bx)$ be an apparatus with POVM
$\Pi $ and nonselective operation $T$.
We now introduce the {\em mean noise operator} $n(A)$
for observable $A$ and the {\em mean disturbance operator} $d(B)$
for observable $B$ defined by
\beqa
n(A)&=&O(\Pi )-A,\\
d(B)&=&T^{*}(B)-B
\eeqa
The meaning of the above operators will be clarified in the following 
argument.

By the realization theorem, there is an indirect measurement
model $(\cK,\si,U,M)$ such that 
\beqa\label{eq:3617a}
\Pi (\De)&=&\cE_{\si}[E^{M^{\Out}}(\De)],\\
T^{*}(X)&=&\cE_{\si}[U^{\da}(X\otimes I)U]
\eeqa
for any Borel set $\De$ and any observable $X$ on $\cH$.
Then, we also have
\beqa
O(\Pi )&=&\cE_{\si}(M^{\Out}),\\
T^{*}(B)&=&\cE_{\si}(B^{\Out}).\label{eq:3618c}
\eeqa
Thus,
\beqa
\cE_{\si}[N(A)]
&=&\cE_{\si}[M^{\Out}-A^{\In}]\nn\\
&=&O(\Pi )-A,\label{eq:3619a}
\eeqa
and 
\beqa
\cE_{\si}[D(B)]
&=&\cE_{\si}[B^{\Out}-B^{\In}]\nn\\
&=&T^{*}(B)-B.\label{eq:3619b}
\eeqa

Thus, we have
\beqa
n(A)&=&\cE_{\si}[N(A)],\\
d(B)&=&\cE_{\si}[D(B)].
\eeqa

Note that for any observable $C$ on $\cH\otimes\cK$
and any observable $X$ on $\cH$, we have
\beqa\label{eq:3617b}
\cE_{\si}\{[C,X\otimes I]\}=[\cE_{\si}(C),X].
\eeqa
By the relations,
\beqas
\Tr\{[N(A),B^{\In}]\rh\otimes\si\}
&=&
\Tr\{\cE_{\si}([N(A),B^{\In}])\rh\}\nn\\
&=&
\Tr([\cE_{\si}\{N(A)\},B]\rh),
\eeqas
we have
\beqa\label{eq:3618a}
\bracket{[N(A),B^{\In}]}=\Tr([n(A),B]\rh).
\eeqa
Similarly, we also have
\beqa\label{eq:3618b}
\bracket{[A^{\In},D(B)]}=\Tr([A,d(B)]\rh).
\eeqa
Therefore, by substituting Eqs.~\eq{3618a} and \eq{3618b},
we obtain the {\em model-independent universally valid
noise-disturbance uncertainty relation} as follows.

\begin{Theorem}\label{th:UVUR}
Let $A$ and $B$ be a pair of observables.
Every apparatus $\bA(\bx)$ satisfies the relation
\begin{eqnarray}\label{eq:UVUR2}
\lefteqn{\ep(A,\rh)\et(B,\rh)
+\frac{1}{2}|\Tr\{[n(A),B]\rh\}}\hspace{10em}\nn\\ 
+\Tr\{[A,d(B)]\rh\}|
&\ge&\frac{1}{2}\left|\Tr([A,B]\rh)\right|
\end{eqnarray}
for any state $\rh$ for which all the relevant terms are finite,
where $\Pi $ is the POVM of $\bA(\bx)$ and $T$ is the 
nonselective operation of $\bA(\bx)$.
\end{Theorem}

Before stating the conditions for Heisenberg's relation,
we introduce some terminology.  
Let $A$ and $B$ be observables of the system $\bS$
to be measured.
We say that an apparatus $\bA(\bx)$ makes an {\em unbiased
measurement} of  $A$, if the mean output is equal to the
mean of the observable $A$ in the input state, i.e., 
\beqa
\bracket{\bx}=\bracket{A^{\In}}
\eeqa
for any input state $\rh$.  From \Eq{output-mean}, this is the case
if and only if the first moment operator of $\Pi $ is equal to $A$, 
i.e.,
\beqa
O(\Pi )=A.
\eeqa
We say that an apparatus $\bA(\bx)$ makes an {\em unbiased
disturbance} of  $B$, if $\bA(\bx)$ does not change 
the mean of $B$, i.e., 
\beqa
\bracket{B^{\In}}=\bracket{B^{\Out}}
\eeqa 
for any input state $\rh$.  
Since the state just after the measurement is $T(\rh)$, we have
\beqa\label{eq:post-measurement-mean}
\bracket{B^{\Out}}=\Tr[T^{*}(B)\rh],
\eeqa
by the relation $\Tr[BT(\rh)]=\Tr[T^{*}(B)\rh]$.
The above relation is also obtained from indirect
measurement models.
In fact, if $\bA(\bx)$ has an indirect measurement model
$(\cK,\si,U,M)$, then, from \Eq{3618c} we have
\beqas
\bracket{B^{\Out}}&=&\Tr[B^{\Out}(\si\otimes\rh)]\\
&=&\Tr[\cE_{\si}(B^{\Out})\rh]\\
&=&\Tr[T^{*}(B)\rh].
\eeqas
Since $\bracket{B^{\In}}=\Tr[B\rh]$ and $\rh$ is arbitrary,
we conclude that apparatus $\bA(\bx)$ makes an unbiased
disturbance of  $B$, if and only if
\beqa
T^{*}(B)=B.
\eeqa

We say that $\bA(\bx)$ has {\em statistically independent noise}
for $A$, if the mean noise $\bracket{\bx}-\bracket{A^{\In}}$ does not
depend on the input state $\rh$, or equivalently, if the mean noise
operator $n(A)$ is a constant operator, i.e., $n(A)=r I$ for some
$r\in\R$.
We say that $\bA(\bx)$ has {\em statistically independent  
disturbance} for $B$, if the mean disturbance
$\bracket{B^{\Out}}-\bracket{B^{\In}}$ does not depend 
on the input state $\rh$, or equivalently, 
if the mean disturbance operator $d(B)$ is a constant operator, i.e.,
$d(B)=r I$ for some $r\in\R$.

The model-independent universally valid noise-disturbance
uncertainty relation leads to rigorous conditions on
what  apparatus satisfies Heisenberg's noise-disturbance
uncertainty relation, as follows.
 
\begin{Theorem}\label{th:HUR}
Let $A$ and $B$ be a pair of observables.
An apparatus $\bA(\bx)$ satisfies Heisenberg's
noise-disturbance uncertainty relation, i.e., 
\beqas
\ep(A,\rh)\et(B,\rh)\ge \frac{1}{2}|\Tr([A,B]\rh)|
\eeqas
for any state $\rh$ for which all the relevant terms are finite,
if one of the following conditions holds:

(i) The mean noise operator commutes with $B$ and
the mean disturbance operator commutes with $A$,
i.e.,  
\beqa
[n(A),B]&=&0,\\{}
[d(B),A]&=&0.
\eeqa

(ii) The apparatus $\bA(\bx)$ has both statistically independent noise for $A$
and statistically independent disturbance for $B$.

(iii) The apparatus $\bA(\bx)$ makes both unbiased measurement of $A$
and unbiased disturbance of $B$.
\end{Theorem}

\subsection{Generalized noise-disturbance uncertainty relation}

In order to obtain the trade-off among the rms noise
$\ep(A,\rh)$, the disturbance $\et(B,\rh)$, 
and the pre-measurement uncertainties
$\si(A,\rh)$ and $\si(B,\rh)$, we apply the Heisenberg-Robertson relation
to all terms in the left-hand-side of the universally valid noise-disturbance
uncertainty relation.  Then, we now obtain the {\em
generalized noise-disturbance uncertainty relation} as follows.

\begin{Theorem}\label{th:GUR}
For any apparatus $\bA(\bx)$ and observables $A$ and $B$, we have
the relation
\beqa\label{eq:GUR}
& & \ep(A,\rh)\et(B,\rh)+\ep(A,\rh)\si(B,\rh)+\si(A,\rh)\et(B,\rh)
\nn\\
& &\quad\quad \quad\ge \frac{1}{2}|\Tr([A,B]\rh)|
\eeqa
for any state $\rh$ for which all the relevant terms are finite.
\end{Theorem}

Under the finite energy constraint, i.e., $\si(Q), \si(P)<\infty$,
the above relation excludes the possibility of having both
$\ep(Q)=0$ and $\et(P)=0$.
However,
$\ep(Q)=0$ is possible with $\si(Q)\et(P)\ge\hbar/2$;
and also $\et(P)=0$ is possible with $\ep(Q)\si(P)\ge\hbar/2$.
In particular, even the case where $\ep(Q)=0$ and $\et(P)<\ve$ 
with arbitrarily small $\ve$ is possible for some input state with 
$\si(Q)>\hbar/2\ve$,
and also the case where $\et(P)=0$ and $\ep(Q)<\ve$ is
possible for some input state with $\si(P)>\hbar/2\ve$.
Such extreme cases occur in compensation for large uncertainties
in the input state, while
in the minimum uncertainty state with
$\si(Q)=\si(P)=(\hbar/2)^{1/2}$, we have 
\begin{equation}
\ep(Q)\et(P)+\sqrt{\frac{\hbar}{2}}[\ep(Q)+\et(P)]\ge\frac{\hbar}{2}.
\end{equation}
Even in this case, it is allowed to have $\ep(Q)\et(P)=0$ with 
$\ep(Q)=0$ and $\et(P)\ge(\hbar/2)^{1/2}$ or with $\et(P)=0$ and 
$\ep(Q)\ge(\hbar/2)^{1/2}$.

For the general case, we have the following trade-off relations for precise
$A$ measurements  or $B$-non-disturbing measurements.

\begin{Theorem}\label{th:URND}
For any apparatus $\bA(\bx)$ and observables $A$ and $B$,
if $\bA(\bx)$ does not disturb $B$,
we have
\begin{equation}\label{eq:URND}
\ep(A,\rh)\si(B,\rh)
\ge\frac{1}{2}|\Tr([A,B]\rh)|
\end{equation}
for any state $\rh$ for which all the relevant terms are finite.
\end{Theorem}

\begin{Theorem}\label{th:URPM}
For any apparatus $\bA(\bx)$ and observables $A$ and $B$,
if $\bA(\bx)$ precisely measures $A$, we have
\begin{equation}
\si(A,\rh)\et(B,\rh)
\ge\frac{1}{2}|\Tr([A,B]\rh)|
\end{equation}
for any state $\rh$ for which all the relevant terms are finite.
\end{Theorem}

For physical significance of the generalized noise-disturbance uncertainty relation,
we refer the reader to Ref.~\cite{03UVR,03HUR}.
In the next section, we shall give an indirect measurement model
that satisfies inequalities in Theorems \ref{th:GUR} and \ref{th:URPM}
but does not satisfies Heisenberg's relation in Theorem \ref{th:HUR}
for position measurement noise and momentum disturbance.

\subsection{Uncertainty relations for 
measurements with statistically independent noise}

Let $\bA(\bx)$ be an arbitrary apparatus and let
$A,B$ be a pair of observable of the measured object.
Denote by $\cI$, $T$, and $\Pi $ be its 
operational distribution, nonselective operation, and POVM
respectively.
Recall that the standard deviation of the output $\bx$ on input state $\rh$
is given by
\beqas
\si(\bx,\rh)&=&\bracket{(\bx-\bracket{\bx}^{2})}^{1/2}\\
&=&
(\Tr[O^{(2)}(\Pi )\rh]-\Tr[O(\Pi )\rh]^{2})^{1/2}.
\eeqas
From Eqs.~\eq{triangular2-A}--\eq{triangular4-A}, if $\bA(\bx)$
makes an unbised measurement of $A$, 
i.e., $\bracket{\bx}=\bracket{A}$,
we have
\beqa
|\si(A,\rh)-\ep(A,\rh)|\le\si(\bx,\rh)\le\ep(A,\rh)+\si(A,\rh).
\eeqa
In what follows, we shall show that if $\bA(\bx)$ 
has statistically independent noise or
makes an unbiased measurement of $A$, 
the standard deviation $\si(\bx,\rh)$ obeys
a reciprocal trade-off with the disturbance on any observable $B$. 

Let $(\cK,\si,U,M)$ be an indirect measurement model
statistically equivalent to $\bA(\bx)$.
Now, we shall return to the 
input-output relations, Eqs.~\eq{NO} and \eq{DO},
%\begin{subequations}
%\begin{eqnarray*}
%M^{\Out}&=&A^{\In}+N(A),\label{eq:NO'}\\
%B^{\Out}&=&B^{\In}+D(B),\label{eq:DO'}
%\end{eqnarray*}
%\end{subequations}
from which we have
\beqas
\lefteqn{
[M^{\Out},B^{\Out}]}\quad\\
&=&
[M^{\Out},B^{\In}+D(B)]\\
&=&
[M^{\Out},B^{\In}]+[M^{\Out},D(B)]\\
&=&
[A^{\In},B^{\In}]+[N(A),B^{\In}]+[M^{\Out},D(B)].
\eeqas
By the relation $[M^{\Out},B^{\Out}]=0$, we have
\beqa
[N(A),B^{\In}]+[M^{\Out},D(B)]=-[A^{\In},B^{\In}].
\eeqa
Taking the moduli of the both sides in the original state $\rh\otimes\si$
and applying the triangular inequality
as before, we have
\beqas
|\bracket{[N(A),B^{\In}]}|+|\bracket{[M^{\Out},D(B)]}|
\ge|\Tr([A,B]\rh)|.
\eeqas
By the Heisenberg-Robertson relation and the relation
$\si(M^{\Out})=\si(\bx,\rh)$, we have
\beqa\label{eq:x}
\si(\bx,\rh)\et(B,\rh)\ge\frac{1}{2}|\bracket{[M^{\Out},D(B)]}|.
\eeqa
From Eqs.~\eq{3618a} and (\ref{eq:x}), we have

\begin{Theorem}
Let $A$ and $B$ be a pair of observables.
Every apparatus $\bA(\bx)$ satisfies the relation
\beqa
\si(\bx,\rh)\et(B,\rh)+\frac{1}{2}|\Tr\{[n(A),B]\rh\}|
\ge\frac{1}{2}|\Tr([A,B]\rh)|\nn\\
\eeqa
for any state $\rh$ for which all the relevant terms are finite,
where $n(A)$ is the mean noise operator for $A$.
\end{Theorem}

From the above, we have the following reciprocal
uncertainty relation for 
measurements with statistically independent noise
and unbiased measurements.

\begin{Theorem}
Let $A$ and $B$ be a pair of observables.
An apparatus $\bA(\bx)$ satisfies the relation
\beqa
\si(\bx,\rh)\et(B,\rh)
\ge\frac{1}{2}|\Tr([A,B]\rh)|
\eeqa
for any state $\rh$ for which all the relevant terms are finite,
if one of the following conditions holds:

(i) The mean noise operator commutes with $B$,
i.e.,  $[n(A),B]=0$.

(ii) The apparatus $\bA(\bx)$ has an statistically independent noise for $A$.

(iii) The apparatus $\bA(\bx)$ makes an unbiased measurement of $A$.
\end{Theorem}

\subsection{Uncertainty relations for measurements with 
statistically independent disturbance}

Let $\bA(\bx)$ be an arbitrary apparatus and let
$A,B$ be a pair of observable of the measured object.
Denote by $\cI$, $T$, and $\Pi $ be its 
operational distribution, nonselective operation, and POVM
respectively.
For any input state $\rh$, 
the standard deviation $\si(B,\rh)$ is called the {\em pre-measurement
uncertainty} of $B$ and the standard deviation $\si(B,T\rh)$
of $B$ in the state $T\rh$ is called the 
{\em post-measurement uncertainty} of $B$.
By the definition of the root-mean-square disturbance $\et(B,\rh)$, 
they satisfy the relation
\beqa
|\si(B,\rh)-\et(B,\rh)|\le\si(B,T\rh)\le\et(B,\rh)+\si(B,\rh).\nn\\
\eeqa
If the measurement does not disturb an observable $B$,
the rms noise $\ep(A,\rh)$ is constrained by \Eq{URND} so
that
\beqa
\ep(A,\rh)\ge\frac{|\bracket{[A,B]}|}{2\si(B,\rh)}.
\eeqa 
In what follows, we consider the more general case where
the statistically independent disturbance 
or unbiased disturbance is allowed 
and we shall show that the rms noise
$\ep(A)$ obeys another reciprocal trade-off that is obtained
by replacing the pre-measurement uncertainty $\si(B,\rh)$
by the post-measurement uncertainty $\si(B,T\rh)$.

Let $(\cK,\si,U,M)$ be an indirect measurement model
statistically equivalent to the apparatus $\bA(\bx)$.
From the input-output relations, Eqs.~\eq{NO} and \eq{DO},
we have
\beqas
\lefteqn{
[M^{\Out},B^{\Out}]}\quad\\
&=&
[A^{\In}+N(A),B^{\Out}]\\
&=&
[A^{\In},B^{\Out}]+[N(A),B^{\Out}]\\
&=&
[A^{\In},B^{\In}]+[A^{\In},D(B)]+[N(A),B^{\Out}].
\eeqas
By the relation $[M^{\Out},B^{\Out}]=0$, we have
\beqa
[N(A),B^{\Out}]+[A^{\In},D(B)]=-[A^{\In},B^{\In}].
\eeqa
Taking the moduli of the both sides in the original state $\rh\otimes\si$
and applying the triangular inequality
as before, we have
\beqas
|\bracket{[N(A),B^{\Out}]}|+|\bracket{[A^{\In},D(B)]}|
\ge|\Tr([A,B]\rh)|.
\eeqas
By the Heisenberg-Robertson relation and the relation
$\si(B^{\Out})=\si(B,T\rh)$, we have
\beqa\label{eq:x2}
\ep(A,\rh)\si(B,T\rh)\ge\frac{1}{2}|\bracket{[N(A),B^{\Out}]}|.
\eeqa
From Eqs.~\eq{3618b} and \eq{x2}, we have

\begin{Theorem}
Let $A$ and $B$ be a pair of observables.
Every apparatus $\bA(\bx)$ satisfies the relation
\beqa
\ep(A,\rh)\si(B,T\rh)+\frac{1}{2}|\Tr\{[A,d(B)]\rh\}|
\ge\frac{1}{2}|\Tr([A,B]\rh)|\nn\\
\eeqa
for any state $\rh$ for which all the relevant terms are finite,
where $n(A)$ is the mean noise operator for $A$.
\end{Theorem}

From the above, we have the following reciprocal
uncertainty relation for measurements with statistically 
independent disturbance or 
unbiased disturbance.

\begin{Theorem}
Let $A$ and $B$ be a pair of observables.
An apparatus $\bA(\bx)$ satisfies the relation
\beqa
\ep(A,\rh)\si(B,T\rh)
\ge\frac{1}{2}|\Tr([A,B]\rh)|
\eeqa
for any state $\rh$ for which all the relevant terms are finite,
if one of the following conditions holds:

(i) The mean disturbance operator commutes with $B$,
i.e.,  $[A,d(B)]=0$.

(ii) The apparatus $\bA(\bx)$ has an statistically independent disturbance for $B$.

(iii) The apparatus $\bA(\bx)$ makes an unbiased disturbance of $B$.
\end{Theorem}

\section{The model breaking Heisenberg's relation
\label{se:MBHUP}}

\subsection{Von Neumann's model\label{se:vNm}}

For comparison with the model to be presented later,
we shall start with a canonical position measurement 
proposed by von Neumann \cite{vN32},
which turns out to typically satisfy 
Heisenberg's noise-disturbance uncertainty relation.

Let us consider the case where the object  $\bS$ is a
one-dimensional mass with position  $\hat{x}$,  momentum
$\hat{p}_{x}$ $([\hx,\hp_{x}]=i\hbar)$,
and Hamiltonian $\hH_{\bS}$ 
on the Hilbert space $\cH=L^{2}(\R)$.
Under general definitions given in the previous sections, 
we can rigorously formulate Heisenberg's 
noise-disturbance uncertainty relation
as
\begin{equation}\label{eq:vNm-Heisenberg}
\ep(\hat{x})\et(\hat{p}_{x})\ge\frac{\hbar}{2}.
\end{equation}

Let $\bA(\bq)$ be the apparatus measuring the system $\bS$
described as follows.
The probe $\bP$ of $\bA(\bq)$ is supposed 
to be a one-dimensional system with canonical observables
$\hq$ and $\hp$ $([\hq,\hp]=i\hbar)$, 
and Hamiltonian $\hH_{\bP}$
on the Hilbert space $\cK=L^{2}(\R)$.
The probe observable is designed to be 
the coordinate $\hq$ of $\bP$.
The probe is also designed to be prepared in a state with a 
normalized wave function $\xi(q)$ just before measurement.
Mathematically, we assume that the wave
function is rapidly decreasing, i.e., $\xi(q)\in\cS(\R)$,
so that we have $\si(\hq),\si(\hp)<\infty$ in the state
$\xi$.
The object-probe interaction $\hH$
is turned on from time $t$ to $t+\De t$.
The total Hamiltonian for the object plus probe is taken to be
\begin{equation}
\hH_{\bS+\bP} = \hH_{\bS} + \hH_{\bP} + K\hH,
\end{equation}
where $K$ is the coupling constant.  
We assume that the coupling is so strong, i.e.,  $K \gg 1$, 
that the free Hamiltonians can be neglected and that the duration
$\De t$ of the coupling is chosen so that $K\De t=1$.  

Von Neumann \cite{vN32} introduced the measuring interaction  
\begin{equation}\label{eq:829o} 
\hH=\hat{x}\hat{p}
\end{equation} 
for an approximate position measurement (see also
Refs.~\cite{CTDSZ80,Cav85,93CA}).  
Then, the unitary operator
of the time evolution of $\bS+\bP$ from $t$ to $t+\De t$ is given by
\begin{equation}\label{eq:829p}
U=\exp\left(\frac{-i}{\hbar}\hat{x}\hat{p}\right).
\end{equation}
This measurement is, therefore, described by the indirect measurement
model
\beqa
\cM(\hx,\xi)
=\left(L^{2}(\R),\xi,\exp\left(\frac{-i}{\hbar}\hat{x}\hat{p}\right),
\hq\right),
\eeqa
which has been generally introduced in Subsection \ref{se:5}.
From general results in Subsection \ref{se:5}, the model $\cM(\hx,\xi)$
has the operational measure
\beqa
\cI_{\xi}(\De)\rh=\int_{\De}\xi(qI-\hx)\rh\,\xi(qI-\hx)^{\da}\,dq,
\eeqa
the dual operational measure
\beqa
\cI_{\xi}(\De)^{*}X=\int_{\De}\xi(qI-\hx)^{\da}\rh\,\xi(qI-\hx)\,dq,
\eeqa
the POVM 
\beqa
\Pi _{\xi}(\De)=\int_{\De}|\xi(qI-\hx)|^{2}\,dq,
\eeqa
the output probability distribution
\beqa
\Pr\{\bq\in\De\|\rh\} = \int_{\De}\Tr[|\xi(qI-\hx)|^{2}\rh]\,dq,
\eeqa
and the output state 
 \beqa
\rh_{\{\bq=q\}}=\frac{\xi(qI-\hx)\rh\,\xi(qI-\hx)^{\da}}
        {\Tr[|\xi(qI-\hx)|^{2}\rh]}.         
\eeqa
If the input state is a vector state $\rh=\ket {\ps}\bra{\ps}$,
we also have the output probability distribution
\beqa
\Pr\{\bq\in\De\|\ps\} = \int_{\De}\,dq\int_{\R}|\xi(q-x)|^{2}|\ps(x)|^{2}\,dx
\eeqa
and the output state 
\beqa
\ps_{\{\bq=q\}}(x)
=\frac{\xi(q-x)\ps(x)}
{\displaystyle\left(\int_{\R}|\xi(q-x)|^{2}|\ps(x)|^{2}\,dx\right)^{1/2}}
\eeqa
with $\rh_{\{\bq=q\}}=
\ket{\ps_{\{\bq=q\}}}\bra{\ps_{\{\bq=q\}}}$.

Solving the Heisenberg equations of motion 
\beqa
d\hX(t+\ta)/d\ta=-\frac{i}{\hbar}[\hX(t+\ta),K\hH]
\eeqa
for $t<t+\ta<t+\De t$, where $\hX(t+\ta)$ is any Heisenberg observable of
$\bS+\bP$, we
obtain 
\beqa
\hat{x}(t+\ta)&=&\hat{x}(t),\\
\hat{q}(t+\ta)&=&K\ta\hat{x}(t)+\hat{q}(t),\\
\hat{p}_{x}(t+\ta)&=&\hat{p}_{x}(t)-K\ta\hat{p}(t),\\
\hat{p}(t+\ta)&=&\hat{p}(t).
\eeqa

For $\ta=\De t=1/K$, we have
\beqa
\hat{x}(t+\De t)&=&\hat{x}(t),\\
\hat{q}(t+\De t)&=&\hat{x}(t)+\hat{q}(t),\\
\hat{p}_{x}(t+\De t)&=&\hat{p}_{x}(t)-\hat{p}(t),\\
\hat{p}(t+\De t)&=&\hat{p}(t).
\eeqa

It follows that the noise operator, the disturbance operator,
the mean noise operator, and the mean disturbance operator
are given by
\beqa
N(\hat{x})
&=&\hat{q}(t+\De t)-\hat{x}(t)=\hat{q}(t),\\
D(\hat{p}_{x})
&=&\hat{p}_{x}(t+\De t)-\hat{p}_{x}(t)=-\hat{p}(t),\\
n(\hat{x})
&=&\bracket{\xi|\hat{q}|\xi}I,\\
d(\hat{p}_{x})
&=&-\bracket{\xi|\hat{p}|\xi}I.
\eeqa
Thus, this measurement has statistically independent  
position-measurement noise and statistically independent 
momentum disturbance, so that this measurement 
satisfies Heisenberg's noise-disturbance
uncertainty relation.
In fact, the mean-square position-measurement noise and the
mean-square momentum disturbance are given by
\beqa
\ep(\hat{x})^{2}
&=&\bracket{\hat{q}(t)^{2}}\ge\si(\hq)^{2},\\
\et(\hat{p}_{x})^{2}
&=&\bracket{\hat{p}(t)^{2}}\ge\si(\hp)^{2}.
\eeqa
Therefore, we conclude that the von Neumann model obeys 
Heisenberg's noise-disturbance uncertainty relation,
\begin{equation}
\ep(\hat{x})\et(\hat{p}_{x})
\ge\frac{\hbar}{2},
\end{equation}
as a consequence of the Heisenberg-Kennard relation
\beqa 
\si(\hat{q})\si(\hat{p})\ge\frac{\hbar}{2},
\eeqa
applied to the probe state just before measurement.

This model represents a basic feature of the
$\ga$ ray  microscope on the point that the trade-off between the rms noise
and the disturbance arises from the fundamental physical limitation on
preparing the probe. 
It might be expected that such a basic feature is shared by
every model in a reasonable class of position measurements. 
However, the next model suggests that it is not the case.

\subsection{Time independent Hamiltonian  model}

In what follows, we modify the measuring interaction of the von 
Neumann model to construct a model that violates Heisenberg's
noise-disturbance uncertainty relation.
In this new model, the object, the probe, the probe preparation,
and the probe observable to be actually measured
are the same systems, the same state, and the same observable as
the von Neumann model. 
Instead of \Eq{829o}, the measuring interaction is now taken to be
\cite{88MS}
\begin{equation}\label{eq:829ox}
\hH=\frac{\pi}{3\sqrt{3}}
(2\hat{x}\hat{p}-2\hat{p}_{x}\hat{q}
+\hat{x}\hat{p}_{x}-\hat{q}\hat{p}).
\end{equation}
The measuring interaction $\hH$
is turned on from time $t$ to $t+\De t$.
The total Hamiltonian for the object plus probe is 
\begin{equation}
\hH_{\bS+\bP} = \hH_{\bS} + \hH_{\bP} + K\hH.
\end{equation}
The coupling constant  $K$ and the time duration  $\De t$ are chosen as
before so that $K\gg 1$ and $K\De t=1$.
Then, the time evolution operator  $U$ for the time interval 
$(t,t+\De t)$ is given by 
\begin{equation}\label{eq:829px}
U=\exp\left[\frac{-i\pi}{3\sqrt{3}\hbar}
(2\hat{x}\hat{p}-2\hat{p}_{x}\hat{q}
+\hat{x}\hat{p}_{x}-\hat{q}\hat{p})\right].
\end{equation}
This measurement is, therefore, described by the indirect measurement
model
\beqas
\left(L^{2}(\R),\xi,\exp\left[\frac{-i\pi}{3\sqrt{3}\hbar}
(2\hat{x}\hat{p}-2\hat{p}_{x}\hat{q}
+\hat{x}\hat{p}_{x}-\hat{q}\hat{p})\right]\!,\hq\right).
\eeqas
We shall call this model the $(1,-2,2)$ model,
whereas the von Neumann model will be
called the $(0,0,1)$ model;
for general $(\al,\be,\ga)$ model we refer to
Ref.~\cite{90QP}.

For the time interval $t<t+\ta<t+\De t$,
the wave function $\Ps_{t+\ta}(x,q)$ of the composite
system $\bS+\bP$ satisfies the Schr\"{o}dinger equation
\beqa
i\hbar\frac{\partial \Psi_{t+\ta}(x,q)}{\partial \ta}
=
KH\Psi_{t+\ta}(x,q).
\eeqa
The solution is
 \begin{eqnarray}
 \lefteqn{
\Psi _{t+\ta}(x,q) }\nonumber\\
&=& \Psi _{t}\left(\frac{2}{\sqrt 3}
 \left\{x\sin\frac{\left(1-K\ta\right)\pi }{3}
 +q\sin\frac{K\ta\pi }{3}\right\}\right.,\nn\\
 & &\left.\frac{2}{\sqrt 3}\left\{-x\sin\frac{K\ta\pi }{3}
 +q\sin\frac{\left(1+K\ta\right)\pi }{3}\right\}\right). 
 \end{eqnarray}
For $\ta=\De t=1/K$, we have
\begin{eqnarray}
\Psi _{t+\De t}(x,q) = \Psi _{t}(q,q-x). 
 \end{eqnarray}

Now, suppose that 
at time $t$, just before the coupling is turned on,
the object wave function is $\psi (x)$ with
$\si(\hx),\si(\hp_{x})<\infty$ in the state $\psi(x)$. 
Since the 
the probe is prepared in the wave function $\xi(q)$,
the total wave function is
\beqa
\Psi _{t}(x,q) = \psi (x)\xi (q).
\eeqa
At time $t+\De t$, the end of the interaction, the total wave
function becomes 
\begin{equation}
 \Psi_{t+\De t} (x,q) = \psi (q)\xi (q-x).
 \end{equation}
Compare with \Eq{829o}; as simple as the von Neumann model,
but the statistics is much different.  

In the above state, the probe observable $\hq$ is measured
to obtain the outcome.
Thus the output probability distribution of this measurement
is given by
 \beqa
 \Pr\{\bq\in\De\|\psi \} 
&=& \int_{\De}\,dq\int_{\R} |\Psi_{t+\De t} (x,q)|^{2}\,dx \nn\\
&=& \int_{\De}|\psi (q )|^2\,dq.
 \eeqa
The output probability distribution has the probability
density function $|\psi (q )|^{2}$, which coincides with
the Born probability density of the object $x$ just before
the measurement and shows that this measurement
is precise position measurement.

The object wave function $\psi _{\{\bq=q\}} (x)$ just after this
measurement given the output $\bq=q$ is obtained (up to
normalization) by
 \begin{eqnarray*}
 \psi _{\{\bq=q\}} (x)&=& 
\frac{\Psi_{t+\De t} (x,q)}
{{\displaystyle\left(\int_{\R}|\Psi_{t+\De t}
(x,q)|^{2}\,dx\right)^{1/2}}}\\
             &=& \frac{\psi (q)}{|\psi (q)|}\xi(q - x) \\
             &=& C\,\xi(q - x),
 \end{eqnarray*}
where $C$ ($|C|=1$) is a constant phase factor depending only
on the output $\bq=q$.
The above relation can be also derived from a general result in
Section \ref{se:MP}.  
Let $f(x)$ be the wave function in $\cH$ defined by
\beqa
f(x)=\xi(-x)
\eeqa
for all $x\in\R$.
Then, we have
\beqa
\xi(q-x)&=&f(x-q)\nn\\
&=&\left[\exp({-iq\hp_{x}}/{\hbar})f\right](x).
\eeqa
Thus, from
\beqa\label{eq:ozawa-unitary}
U(\ps\otimes\xi)(x,q)=\psi (q)\xi (q-x),
\eeqa
we have
\beqa
U(\ps\otimes\xi)(q)
=\psi(q)\exp({-iq\hp_{x}}/{\hbar})f
\eeqa
From Theorem \ref{th:3627a}, the operational
distribution $\cI$ satisfies
\beqa\label{eq:3627b}
\lefteqn{
\cI(\De)\ket{\ps}\bra{\ps}}\quad\nn\\
&=&
\int_{\De}
\left\ket{\psi(q)\exp({-iq\hp_{x}}/{\hbar})f\right}
\left
\bra{\psi (q)\exp({-iq\hp_{x}}/{\hbar})f
\right}\,dq\nn\\
&=&
\int_{\De}\exp({-iq\hp_{x}}/{\hbar})\ket{f}
\bra{f}\exp({iq\hp_{x}}/{\hbar}) |\ps(q)|^{2}\,dq
\nn\\
&=&
\int_{\De}\exp({-iq\hp_{x}}/{\hbar})\ket{f}
\bra{f}\exp({iq\hp_{x}}/{\hbar})\nn\\
& &\times
\Tr[dE^{\hq}(q)\ket{\ps}\bra{\ps}]
\eeqa
It follows that the output state given $\bq=q$ is 
\beqa
\lefteqn{
\ket{\ps_{\{\bq=q\}}}\bra{\ps_{\{\bq=q\}}}
}\quad\nn\\
&=&\exp({-iq\hp_{x}}/{\hbar})\ket{f}
\bra{f}\exp({iq\hp_{x}}/{\hbar}),
\eeqa
and hence we have
\beqa
\ps_{\{\bq=q\}}(x)
=\left[\exp({-iq\hp_{x}}/{\hbar})f\right](x)
=\xi(q-x),
\eeqa
up to constant phase factor.

By linearity and continuity, from \Eq{3627b} the operational
distribution of this model is given by
\beqa
\cI(\De)\rh
&=&
\int_{\De}\exp({-iq\hp_{x}}/{\hbar})\ket{f}
\bra{f}\exp({iq\hp_{x}}/{\hbar})\nn\\
& &\times\Tr[dE^{\hq}(q)\rh]
\eeqa

Solving the Heisenberg equations of motion for $t<t+\ta<t+\De t$, we
obtain 
% \beqa
% \lefteqn{\hat{x}(t+\ta)}\quad\nn\\
 %&=&\frac{2}{\sqrt{3}}\hat{x}(t)\sin \frac{(1+K\ta)\pi}{3}
 %+\frac{-2}{\sqrt{3}}\hat{q}(t)\sin \frac{K\ta\pi}{3},\\
 %\lefteqn{\hat{q}(t+\ta)}\quad\nn\\
 %&=&\frac{2}{\sqrt{3}}\hat{x}(t)\sin \frac{K\ta\pi}{3}
 %+\frac{-2}{\sqrt{3}}\hat{q}(t)\sin \frac{(1-K\ta)\pi}{3},\\
 %\lefteqn{\hat{p}_{x}(t+\ta)}\quad\nn\\
 %&=&\frac{-2}{\sqrt{3}}\hat{p}_{x}(t)\sin \frac{(1-K\ta)\pi}{3}
 %+\frac{-2}{\sqrt{3}}\hat{p}(t)\sin \frac{K\ta\pi}{3},\\
 %\lefteqn{\hat{p}(t+\ta)}\quad\nn\\
 %&=&\frac{2}{\sqrt{3}}\hat{p}_{x}(t)\sin \frac{K\ta\pi}{3}
 %+\frac{2}{\sqrt{3}}\hat{p}(t)\sin \frac{(1+K\ta)\pi}{3}.
% \eeqa
\beqas
\hat{x}(t+\ta)
 &=&\frac{2}{\sqrt{3}}\hat{x}(t)\sin \frac{(1+K\ta)\pi}{3}
 -\frac{2}{\sqrt{3}}\hat{q}(t)\sin \frac{K\ta\pi}{3},\\
\hat{q}(t+\ta)
 &=&\frac{2}{\sqrt{3}}\hat{x}(t)\sin \frac{K\ta\pi}{3}
 +\frac{2}{\sqrt{3}}\hat{q}(t)\sin \frac{(1-K\ta)\pi}{3},\\
\hat{p}_{x}(t+\ta)
 &=&\frac{2}{\sqrt{3}}\hat{p}_{x}(t)\sin \frac{(1-K\ta)\pi}{3}
 -\frac{2}{\sqrt{3}}\hat{p}(t)\sin \frac{K\ta\pi}{3},\\
\hat{p}(t+\ta)
 &=&\frac{2}{\sqrt{3}}\hat{p}_{x}(t)\sin \frac{K\ta\pi}{3}
 +\frac{2}{\sqrt{3}}\hat{p}(t)\sin \frac{(1+K\ta)\pi}{3}.
 \eeqas

For $\ta=\De t=1/K$, we have
\beqa
\hat{x}(t+\De t)&=&\hat{x}(t)-\hat{q}(t),\label{eq:ozawa-model-1}\\
\hat{q}(t+\De t)&=&\hat{x}(t),\label{eq:ozawa-model-2}\\
\hat{p}_{x}(t+\De t)&=&-\hat{p}(t),\label{eq:ozawa-model-3}\\
\hat{p}(t+\De t)
&=&\hat{p}_{x}(t)+\hat{p}(t)\label{eq:ozawa-model-4}.
\eeqa

It follows that the noise operator, the disturbance
operator, the mean noise operator, and the mean disturbance
operator are given by
\beqa
N(\hat{x})
&=&\hat{q}(t+\De t)-\hat{x}(t)=0,\\
D(\hat{p}_{x})
&=&\hat{p}_{x}(t+\De t)-\hat{p}_{x}(t)=-\hat{p}-\hat{p}_{x},\\
n(\hat{x})
&=&0,\\
d(\hat{p}_{x})
&=&-\bracket{\xi|\hat{p}|\xi}I-\hat{p}_{x},\\
\eeqa
Thus, the position-measurement noise and the
momentum disturbance are given by
\beqa
\ep(\hat{x})
&=&0,\\
\et(\hat{p}_{x})^{2}
&=&\bracket{[\hat{p}_{x}(t)+\hat{p}(t)]^{2}}\nn\\
&=&\si(\hat{p}_{x})^{2}+\si(\hat{p})^{2}
+[\bracket{\hat{p}_{x}(t)}+\bracket{\hat{p}(t)}]^{2}.\nn\\
\eeqa
Consequently,  we have %\cite{note2}
\begin{equation}
\ep(\hat{x})\et(\hat{p}_{x})=0.
\end{equation}
Therefore, our model obviously violates
% \Eq{Heisenberg}.  
Heisenberg's noise-disturbance uncertainty relation.

If $\bracket{\hat{p}_{x}(t)^{2}}\to 0$
and $\bracket{\hat{p}(t)^{2}}\to 0$ (i.e.,  $\ps$ and $\xi$
tend to the momentum eigenstate with zero momentum) then
we have even $\et(\hat{p}_{x}(t))\to 0$ with $\ep(\hat{x})=0$.
Thus, we can precisely measure position without effectively
disturbing  momentum in a near momentum eigenstate.

Taking advantage of 
the above model, we can refute the argument
that the uncertainty principle generally leads to a general sensitivity 
limit, called the standard quantum limit, for monitoring free-mass
position \cite{Yue83,88MS}. 

\subsection{Time dependent Hamiltonian model}

The interaction of the preceding model \Eq{829ox},
the $(1,-2,2)$ model,
includes the term $\hat{x}\hat{p}_{x}-\hat{q}\hat{p}$,
which cannot be implemented by a simple coupling.
Thus, it seems that this model is far more difficult than
the von Neumann model.
In this section, we shall show, however, that 
if we use time dependent interaction, the 
$(1,-2,2)$ model can be
implemented as feasibly as the von Neumann model.

Now, we shall consider the following model description, 
which will
turn out statistically equivalent to the model discussed in the
preceding subsection. In this model, the object, the probe, 
the probe preparation, and the probe
observable are the same as the previous models. 
The object-probe interaction is turned on from time
$t$ to $t+\De t$. 
For the time interval $t<t+\ta<t+\De t$,
the time dependent total Hamiltonian 
$H_{{\bf S}+{\bf P}}(t+\ta)$
of  ${\bf S}+{\bf P}$ is taken to be
\begin{eqnarray}
H_{{\bf S}+{\bf P}}(t+\ta)
&=&H_{{\bf S}}\otimes I+I\otimes H_{{\bf P}}
-K_{1}(\ta){\hat p}_{x}\otimes {\hat q}\nonumber\\
& &\mbox{ }+K_{2}(\ta){\hat x}\otimes{\hat p},
\label{eq:301a}
\end{eqnarray}
where the strengths of couplings, $K_{1}(\ta)$ and $K_{2}(\ta)$, 
satisfy
\begin{mathletters}
\begin{eqnarray}
K_{1}(\ta)&=&0\quad\mbox{if $\ta\not\in(t,t+\frac{\Delta t}{2})$},\\
K_{2}(\ta)&=&0\quad\mbox{if $\ta\not\in(t+\frac{\Delta t}{2},t+\Delta t)$},
\end{eqnarray}
\begin{equation}
\int_{t}^{t+\frac{\Delta t}{2}}K_{1}(\ta)d\ta=1,\quad
\int_{t+\frac{\Delta t}{2}}^{t+\Delta t}K_{2}(\ta)d\ta=1.
\end{equation}
\end{mathletters}
We assume that $\Delta t$ is so small that the system Hamiltonians 
$H_{{\bf S}}$ and
$H_{{\bf P}}$ can be neglected from $t$ to $t+\Delta t$.
Solving the Schr\"{o}dinger equation,
just as von Neumann model, 
the time evolution of ${\bf S}+{\bf P}$ during the coupling 
is described by the unitary evolution operators
\begin{mathletters}
\begin{eqnarray}
U(t+\frac{\Delta t}{2},t)
&=&\exp \left(\frac{i}{\hbar}{\hat p}_{x}{\hat q}
\right),\\
U(t+\Delta t,t+\frac{\Delta t}{2})
&=&\exp\left( -\frac{i}{\hbar}{\hat x}{\hat p}
\right).
\end{eqnarray}
\end{mathletters}
Then, in the position basis we have
\begin{mathletters}
\begin{eqnarray*}
\langle x,y|U(t+\frac{\Delta t}{2},t)|x',y'\rangle
&=&\langle x+y,y|x',y'\rangle,\\
\langle x,y|U(t+\Delta t,t+\frac{\Delta t}{2})|x',y'\rangle
&=&\langle x,y-x|x',y'\rangle,
\end{eqnarray*}
\end{mathletters}
and hence
\begin{eqnarray}
&\langle x,y|U(t+\Delta t,t+\frac{\Delta t}{2})
U(t+\frac{\Delta t}{2},t)|x',y'\rangle& 
\nonumber\\
&\quad=\langle y,y-x|x',y'\rangle.&
\label{eq:226b}
\end{eqnarray}
Thus,  by \Eq{ozawa-unitary}, we conclude that 
the unitary evolution operator 
\beqa
U=U(t+\Delta t,t+\frac{\Delta t}{2})
U(t+\frac{\Delta t}{2},t)
\eeqa
is the same as the unitary operator of the $(1,-2,2)$ model.
Thus, the above model is identical with the $(1,-2,2)$ model.
In particular, we have obtained the relation
\beqa
\lefteqn{\exp\left[\frac{-i\pi}{3\sqrt{3}\hbar}
(2\hat{x}\hat{p}-2\hat{p}_{x}\hat{q}
+\hat{x}\hat{p}_{x}-\hat{q}\hat{p})\right]}
\nn\quad\\
&=&
\exp\left(-\frac{i}{\hbar}{\hat x}{\hat p}\right)
\exp\left(\frac{i}{\hbar}{\hat p}_{x} {\hat q}\right).
\eeqa 
Thus, we can avoid to implement the term
$\hat{x}\hat{p}_{x}-\hat{q}\hat{p}$,
and only von Neumann type interactions
${\hat x} {\hat p}$ and 
${\hat p}_{x}{\hat q}$ are suffice 
to implement the $(1,-2,2)$ model.

\section{Repeatability and uncertainty principle\label{se:repeatability}}

\subsection{Repeatability hypothesis and the projection postulate}
\label{se:RH}

In formulating the canonical description of the measurement of an
observable, von Neumann required not only that the output probability 
distribution satisfies the Born statistical formula but also that the
quantum state reduction satisfies the following hypothesis abstracted from
the result of  the Compton-Simons experiment \cite{vN32}.

{\bf Repeatability hypothesis.}
{\em If an observable is
measured twice in succession in a system, then we get the same value
each time.}

In what follows, we consider the rigorous formulation of this requirement
for general measuring apparatuses. 
Let $\bA(\bx)$ be an apparatus with output variable $\bx$. 
In order to formalize the repeatability hypothesis, we need to consider 
repeated measurements using the identical apparatuses on the same system.
Since the same apparatus cannot be used twice in succession, 
we assume that immediately after the measurement using $\bA(\bx)$,
another statistically equivalent
apparatus $\bA(\by)$ with output variable $\by$
is used for the succeeding measurement. 
Then, the repeatability hypothesis states that if $\bx=x$ then $\by=x$ for any $x$.  
This condition is well-formulated by the concept of conditional probability
as follows.
The apparatus $\bA(\bx)$ satisfies the repeatability hypothesis
if and only if the conditional probability distribution of $\by$
given $\bx=x$ satisfies 
\begin{equation}\label{eq:122a}
\Pr\{\by\in\De|\bx=x\|\rh\}=\ch_{\De}(x)
\end{equation} 
for all $x$, $\De$ and $\rh$.
%, where $\de_{x}$ stands for the point mass of $x$, i.e., 
%$\de_{x}(\De)=\ch_{\De}(x)$.
Let $\rh_{\{\bx=x\}}$ be the output state given $\bx=x$ for input
state $\rh$.  Then, from \Eq{3622a}, the apparatus $\bA(\bx)$ 
satisfies the repeatability hypothesis if and only if we have
\beqa\label{eq:3622b}
\Pr\{\by\in\De\|\rh_{\{\bx=x\}}\}=\ch_{\De}(x)
\eeqa 
for any Borel set $\De$.

Now, we shall consider the case where apparatus $\bA(\bx)$
precisely measures an observable $A$. 
From \Eq{3622b}, in this case $\bA(\bx)$ satisfies the repeatability
hypothesis if and only if we have
\begin{equation}\label{eq:122g}
\Tr[E^{A}(\De)\rh_{\{\bx=x\}}]=\ch_{\De}(x)
\end{equation}
for any Borel set $\De$.
The last equality is equivalent to the condition
\beqa
E^{A}(\De)\rh_{\{\bx=x\}}E^{A}(\De)=
\ch_{\De}(x)\rh_{\{\bx=x\}}.
\eeqa
 
Suppose that $A$ has purely discrete nondegenerate
spectrum $a_{1},a_{2},\ldots$ with corresponding
orthonormal  basis $\ph_{1},\ph_{2},\ldots$ of
eigenvectors. Then the repeatability hypothesis holds if
and only if 
\begin{equation}\label{eq:122h}
\rh_{\{\bx=a_{n}\}}=|\ph_{n}\>\<\ph_{n}|
\end{equation}
for all $n=1,2,\ldots$.
In this case, the operational distribution $\cI$ of $\bA(\bx)$
is determined uniquely by
\begin{equation}\label{eq:122i}
\cI(\De)\rh
=\sum_{a_{n}\in\De}|\ph_{n}\>\<\ph_{n}|\rh|\ph_{n}\>\<\ph_{n}|
\end{equation}
for any $\rh\in\tc(\cH)$ and $\De\in\cB(\R)$.
Thus for any observable with purely discrete
nondegenerate spectrum the repeatability hypothesis
determines an apparatus uniquely up to statistical equivalence.

If $A$ has, however, purely discrete 
but degenerate spectrum then the repeatability hypothesis no
longer determines the state after the measurement.
In fact, in this case $\rh_{\{\bx=a_{n}\}}$ can be one of 
any eigenstates $\ket{\ph}\bra{\ph}$ with $A\ph=a_{n}\ph$
or even mixtures of them.
In order to determine the state after the measurement in this case,
L\"{u}ders \cite{Lud51} proposed the following 
requirement.

{\bf Projection Postulate.}
{\em For any input state $\rh$ for a precise measurement of
a purely discrete observable $A$, the output state 
$\rh_{\{\bx=a\}}$ is given by
\begin{equation}\label{eq:2.6}
\rh_{\{\bx=a\}}
=\frac{E^{A}\{a\}\rh\,E^{A}\{a\}}{\Tr[E^{A}\{a\}\rh]}
\end{equation}
for any $a\in\R$ with $\Pr\{\bx=a\|\rh\}>0$.}

According to the projection postulate, if the input state is a
vector state $\ps$, i.e., $\rh=\ket{\ps}\bra{\ps}$, then the
output state $\rh_{\{\bx=a\}}$ is represented by the projection
$E^{A}(\{a\})\ps$ of 
$\ps$ on the eigenspace corresponding to the output $a$,
i.e.,
\beqa
\rh_{\{\bx=a\}}
=
\frac{\ket{E^{A}\{a\}\ps}\bra{E^{A}\{a\}\ps}}
{\|E^{A}\{a\}\ps\|^{2}}.
\eeqa

It is obvious that the projection postulate implies 
the repeatability hypothesis.
The projection postulate yields the following 
operational distribution
\begin{equation}
\cI(\De)\rh=\sum_{a\in\De}E^{A}\{a\}\rh\,E^{A}\{a\}
\end{equation}
for all $\rh\in\tc(\cH)$.

\subsection{Discreteness of repeatable instruments}

Now we shall consider the general case where $A$ may 
have a continuous spectrum
or even $\bA(\bx)$ makes no precise measurement of an
observable.
Let us assume that a measurement using $\bA(\bx)$ is 
immediately followed by a measurement using another
statistically equivalent apparatus $\bA(\by)$.
Let $\cI$ be the common operational distribution of those
apparatuses.   It follows from \Eq{3624a} and \eq{122a} that
the repeatability  hypothesis holds if and only if
\begin{equation}\label{eq:122c}
\Pr\{\bx\in\De,\by\in\De'\|\rh\}=\Pr\{\bx\in\De\cap\De'\|\rh\}
\end{equation}
where $\De,\De'\in\cB(\R)$.
Thus, from Eqs.~\eq{DL1} and
\eq{JOPD1} 
we conclude that apparatus $\bA(\bx)$
satisfies the repeatability hypothesis if and only  if the
operational distribution $\cI$ satisfies
\begin{equation}\label{eq:122f}
\Tr[\cI(\De')\cI(\De)\rh]
=
\Tr[\cI(\De\cap\De')\rh]
\end{equation}
or equivalently
\begin{equation}
\cI(\De')^{*}\cI(\De)^{*}I=\cI(\De\cap\De')^{*}I
\end{equation}
for any input state $\rh$ and $\De,\De'\in\cB(\R)$.
The above conditions are also restated as
$\bA(\bx)$  satisfies the repeatability hypothesis if and
only if the operational distribution $\cI$ and the POVM
of $\bA(\bx)$ satisfies
\beqa
\cI(\De)^{*}\Pi (\De')=\Pi (\De\cap\De').
\eeqa

Motivated by the above argument, any DL instrument
satisfying \Eq{122f} for all $\De,\De'\in\cB(\La)$
is said to be {\em repeatable};
note that Davies and Lewis \cite{DL70} called originally 
such DL instruments  as ``weakly repeatable''.

Contrary to the fact that there can be many repeatable 
DL instruments corresponding to the same purely discrete
observables, the following theorem, conjectured in
Ref.~\cite{DL70} and proved in Ref.~\cite[Theorem
5.1]{85CA} shows that there are no repeatable DL instruments
corresponding to any observables with continuous spectrum.

\begin{Theorem}
Every repeatable DL instrument is discrete in 
the sense that there is a countable subset $\La_{0}$ of 
$\R$ such that $\cI(\R\setminus \La_{0})=0$.
\end{Theorem}

%The above theorem was generalized in Ref.~\cite[Theorem 5.1]{85CA}
%so as to conclude that 
%non-discrete repeatable normalized positive map valued measures
%(i.e., operational measures without requiring complete positivity)
%for a von Neumann algebra $cM$ and a standard Borel space $\La$ 
%have no families of posterior states.

It is concluded, therefore, that in order to model repeatable
measurements of continuous observables 
it is necessary to describe them approximately with arbitrary closeness
or to extend the formulation of quantum mechanics to include the limit
of those approximate models \cite{Sri80,88MR}.
In Ref.~\cite{93CA} it was shown that we have still satisfactory models of 
approximately repeatable measurement of 
continuous observables within arbitrarily
small error limit in the standard formulation of quantum mechanics.

\subsection{Approximate repeatability}

Whereas von Neumann considers only precise measurements
of observables and introduced the repeatability hypothesis
for canonical description of state changes caused by measurements,
the von Neumann model does not satisfy the preciseness
nor the repeatability.
One of the characteristic features of our $(1,-2,2)$ model
is that it precisely measures position, but our model does not satisfy
the repeatability hypothesis either.  
Thus, it is tempting to understand that the $(1,-2,2)$ model circumvent
Heisenberg's  noise-disturbance uncertainty relation
by paying the price of failing the repeatability.
In what follows we shall show that such a view cannot be supported.  

In the first place, as discussed in Subsection \ref{se:RH},
the repeatability hypothesis can be satisfied only by measurements   
of purely discrete observables.  
Thus, no precise position measurements satisfy 
the repeatability hypothesis.

Secondly, if we consider the approximate repeatability, our model
satisfies any stringent requirement of approximate repeatability.
In order to show this, we need the measure of approximate
repeatability introduce by Ref.~\cite{93CA}.

Let $\cI$ be a DL instrument. 
We define the {\it root-mean-square repetition error\/}
of $\cI$ on input state $\rh$, denoted by $R(\cI,\rh)$,
as follows.
\begin{equation}
R(\cI,\rh)
= \left(\iint _{\R^{2}}(x - y)^{2}\,\Tr[d\cI(x)\,d\cI(y)\rh]\right)^{1/2}.
\label{(3.5)}
\end{equation}
We shall write $R(\cI,\ps)
=R(\cI,\ket{\ps}\bra{\ps})$.
Since $\Tr[\cI(\De)\cI(\De')\rh]$ represents 
the joint probability
distribution of  the ouputs of the repeated measurements of 
statistically equivalent apparatuses with operational distribution  
$\cI$, the interpretation of the above error is obvious.
Then we have the following.

\begin{Theorem}\label{th:4.4}
A DL instrument $\cI$ is repeatable if and only if
$\cI$ satisfies
$$ 
R(\cI,\rh)=0
$$
for any density operator $\rh$ .
\end{Theorem}

For the proof, we refer to Ref.~\cite{93CA}.

A DL instrument $\cI$ is said to be {\it $\ve$-repeatable\/} if
$\cI$ satisfies  
$R(\cI,\rh)\leq\sqrt{2}\varepsilon$ for any
density operator $\rh$.  
Now, it is natural to say that 
an apparatus or an indirect measurement
model is said to be {\it $\ve$-repeatable\/} if the corresponding 
operational measure is {\it $\ve$-repeatable\/}.

Suppose that we measure the position  of mass $\hat{x}$
in succession using two apparatuses described by the 
identical indirect measurement models with operational 
distribution $\cI$.    
Suppose that the first apparatus with probe
$\hat{q}$ interacts with $\hat{x}$ in $(t,t+\Delta t)$ 
and the second apparatus with probe $\hat{q'}$ 
interacts with $\hat{x}$ in $(t+\Delta t, t+2\Delta t)$. 
Then, the root-mean-square repetition error $R$ of the above
apparatus is the root-mean-square difference between the
first output $\hat{q}(t+\De t)$ and 
the second output $\hat{q}(t+2\De t)$, i.e., 
\begin{equation}
R^{2}=\bracket{\ps\otimes\xi\otimes\xi|
[\hat{q'}(t+2\De t)-\hat{q}(t+\De)]^{2}|\ps\otimes\xi\otimes\xi}.
\end{equation}

If the apparatuses are described by the von Neumann model, 
we have 
\beqas
\hat{q'}(t+2\De t)&=&\hat{x}(t+\De t)+\hat{q'}(t+\De t)\\
&=&\hat{x}(t)+\hat{q'}(t+\De t)\\
\hat{q}(t+\De t)&=&\hat{x}(t)+\hat{q}(t).
\eeqas
Thus, we have
\beqas
\lefteqn{[\hat{q'}(t+2\De t)-\hat{q}(t+\De t)]^{2}}\quad\\
&=&
\hat{q'}(t+\De t)^{2}-2\hat{q'}(t+\De t)\hat{q}(t)
+\hat{q}(t)^{2}.
\eeqas
Since $\hat{q'}(t+\De t)$ and $\hat{q}(t)$ are
statistically independent and identically distributed we have
\beqas
\lefteqn{
\bracket{\ps\otimes\xi\otimes\xi|
[\hat{q}'(t+2\De t)-\hat{q}(t+\De)]^{2}|\ps\otimes\xi\otimes\xi}}
\qquad\qquad\nn\\ &=&
2(\bracket{\xi|\hat{q}(t)^{2}|\xi}-\bracket{\xi|\hat{q}(t)|\xi}^{2})\\
&=&
2\si(\hat{q}(t))^{2}.
\eeqas
Thus, we have
\beqa
R=\sqrt{2}\si(\hat{q}(t)).
\eeqa

If the apparatuses are described by the $(1,-2,2)$model,  we have
\beqa
\hat{q'}(t+2\De t)&=&\hat{x}(t+\De t)=\hat{x}(t)-\hat{q}(t),\\
\hat{q}(t+\De t)&=&\hat{x}(t),
\eeqa
and hence
\beqa
\hat{q'}(t+2\De t)-\hat{q}(t+\De)=-\hat{q}(t).
\eeqa
Thus, we have
\beqa
R=\bracket{\hat{q}(t)^{2}}^{1/2}.
\eeqa

Thus, for the probe preparation $\xi$ such that 
$\bracket{\hq(t)}=0$, 
the von Neumann model has 
\beqa
\ep(\hat{x})&=&\si(\hat{q}(t))\\
R&=&\sqrt{2}\si(\hat{q}(t)),
\eeqa
and the $(1,-2,2)$model has
\beqa
\ep(\hat{x})&=&0\\
R&=&\si(\hat{q}(t)).
\eeqa
Thus, for the identical preparation of the probe,
the $(1,-2,2)$ model is concluded to be
a $\si(\hat{q(t)})/\sqrt{2}$-repeatable
precise position measurement, 
whereas
the von Neumann model is a $\si(\hat{q}(t))$-repeatable
$\si(\hat{q}(t))$-precise position measurement.

Therefore, we conclude for any small $\ve>0$ 
we have an $\ve$-repeatable precise position measurement 
that violates Heisenberg's noise-disturbance uncertainty 
relation \eq{Heisenberg}.
This suggests that how stringent conditions on preciseness
and repeatability might be
posed for a class of position measurements, 
we can find in that class at least one position measurement  
that violates Heisenberg's noise-disturbance
uncertainty relation. 

\section{Concluding remarks}

In Ref.~\cite{03UVR}, we have obtained the universally 
valid noise-disturbance uncertainty relation \Eq{UVUR1} and
the generalized noise-disturbance uncertainty relation 
\Eq{GUR}, and also derived Theorems \ref{th:HNDUR}, \ref{th:URND},
and \ref{th:URPM} in the model dependent formulation.
However, the following problems have been remained open
concerning the foundations of the model dependent approach.
(I) Can every measuring apparatus be described by an indirect measurement
model?
(II) Are the root-mean-square noise and disturbance uniquely determined 
independent of the model?

Indirect measurement models, originally introduced by von Neumann \cite{vN32}
and generally formalized in Ref.~\cite{83CR,84QC}, are powerful tool to
study measuring processes, since the interaction between the measured object
and the apparatus is described purely by quantum mechanics.
This merit is strongly contrasted with a conventional view that the measuring
interaction involves the macroscopic part of the apparatus.
Although some measuring apparatuses, especially in the attempts for quantum
nondemolition measurements \cite{BK92}, allow indirect measurement 
model descriptions, it is still difficult to convince any schools of measurement
theory of the affirmative answer to question (I) above.
However, the present paper has shown that in order to establish uncertainty
relations for noise and disturbance the use of indirect measurement models
is justified regardless of the answer to question (I).

The strategy taken in the present paper is as follows.
We have started with listing up properties that obviously every measuring 
apparatus obeys, and then proven that every apparatus satisfying those 
properties is statistically equivalent to an apparatus described by an
indirect measurement model.
In the next step, we have proven that the root-mean-square noise and
disturbance are determined 
by the POVM and the nonselective operation, respectively, of the
apparatus, so that question (II) above has been answered affirmatively.
This means that if two apparatuses are statistically equivalent, they have
the same root-mean-square noise and disturbance.
Thus, if a formula for root-mean-square noise and disturbance is proven for 
one apparatus with an indirect measurement model,
every apparatus statistically equivalent to that apparatus obeys the same
formula.  In this way, we have justified the assertion of Ref.~\cite{03UVR}
that those formulas obtained for apparatuses with indirect measurement model
are universally true for every apparatus irrespective of the model that describes
the apparatus.

As properties that obviously every measuring apparatus obeys,
we have proposed the following axioms for general measuring apparatuses.

(i) Mixing law: If two apparatuses
are applied to a single system in succession, the joint probability 
distribution of outputs from those two  apparatuses 
depends affinely on the input state.

(ii) Extendability axiom: Every apparatus measuring one
system can be trivially extended to an apparatus measuring
a larger system including the original system without
changing the statistics. 

(iii) Realizability postulate:  Every indirect measurement
model corresponds to an apparatus whose measuring process
is described by that model.

From axioms (i) and (ii), we have demonstrated that
statistical properties of any apparatus is described by
a normalized completely positive map valued measure, 
called a CP instrument.
Then, it has been shown that two apparatus are statistically 
equivalent if and only if they corresponds to the same
CP instrument.
Thus, the set of the statistical equivalence classes of
all apparatuses are considered to be a subset of the set of
all CP instruments.
From the realization theorem of CP instruments (Theorem 
\ref{th:realization}) and axiom (iii), we have further concluded 
that the statistical
equivalence class of apparatuses are in one-to-one 
correspondence with the CP instruments.
Thus, we can conclude that every apparatus is statistically
equivalent to at least 
one apparatus which is described by an indirect measurement model,
in which the measuring interaction is simply described by a quantum
mechanical interaction between two quantum mechanical
systems, the object and the probe.  

There have been many attempts to define the root-mean-square noise
for some special classes of measurements.
In Section \ref{se:Noise} we have shown that all those convincing attempts
are equivalent to our notion of the distance of a POVM from an observable,
based on which we define the root-mean-square noise of an arbitrary 
measurement in the model independent formulation.
The empirical adequacy of our definition can be supported by the
following reasons.
(i) Our definition satisfies the requirement that if the 
measured observable has a definite value in the input state,
the root-mean-square noise be the root-mean-square of the
difference between the true value and the measured value
 (\Eq{definite2}).
(ii) Our definition satisfies the requirement that 
measurements with uniformly zero root-mean-square
noise coincide with precise measurements (Theorem \ref{th:noise-1}).
(iii) The difference between the standard deviations of the measured
observable and of the measured value is bounded from above by
the root-mean-square noise plus the bias, namely, the difference
of their means (\Eq{triangular5-A}). 
(iv) The root-mean-square noise in any input state can be statistically 
estimated from the experimental data (\Eq{data}).
(v) The root-mean-square noise defined through the noise operator 
has a clear geometric interpretation  (\Eq{geometric1}).
(iv) Even if another observer describes the same apparatus by a different
indirect measurement model and identify the noise operator in a 
different way, the root-mean-square noises for both observers are 
equal (\Eq{noise}).

In Ref.~\cite{03HUR}, we have discussed two distinct types of measurements
in which Heisenberg's noise-disturbance uncertainty relation is violated for
position measurement noise and momentum disturbance uniformly for any input state.  
These cases are generalized in Theorem \ref{th:URND} for type I violation
and Theorem \ref{th:URPM} for type II violation.
These relations clearly reveals possibilities of measurements beyond Heisenberg's
relation such as Yuen's contractive state measurement \cite{Yue83} 
and clarifies the new constraints for measurements beyond Heisenberg's relation.  
An experimental realization of a measurement with type II violation for
optical quadrature measurement is proposed in Ref.~\cite{03UVR}.
This measuring interaction is equivalent to the  (1,-2,2) model,
discussed in Section \ref{se:MBHUP}, which realizes Yuen's contractive 
state measurement as demonstrated in Ref.~\cite{88MS}, 
so that the realization of this measurement with required accuracy 
will open a way to the new technology for supersensitive sensors.
\bigskip

\begin{acknowledgments}
This work was supported by the 
Strategic Information and Communications R\&D Promotion Scheme
of the MPHPT of Japan,  by the CREST
project of the JST, and by the Grant-in-Aid for Scientific Research of
the JSPS.
\end{acknowledgments}

\end{document}